\pdfoutput=1
\documentclass[usenatbib]{mn2e}
\usepackage{apjfonts}
\usepackage{amssymb}
\usepackage{amsmath}
\usepackage{ctable}
\usepackage{fixltx2e} %% forces figure order to remain fixed (otherwise figure*'s can move out-of-order)

\newcommand{\Zsun}{Z_{\odot}}
\newcommand{\Mstar}{M_{\ast}}
\newcommand{\Rstar}{r_{\ast}}
\newcommand{\gas}{_{\rm g}}
\newcommand{\Sigmagas}{\Sigma\gas}
\newcommand{\hgas}{H\gas}
\newcommand{\rhogas}{\rho\gas}
\newcommand{\meanrhogas}{\langle\rhogas\rangle_{0}}
\newcommand{\cs}{c_{s}}
\newcommand{\grain}{_{\rm d}}
\newcommand{\Sigmagrain}{\Sigma\grain}
\newcommand{\hgrain}{h\grain}
\newcommand{\rhograin}{\rho\grain}
\newcommand{\meanrhograin}{\langle\rhograin\rangle_{0}}
\newcommand{\Zgrain}{Z\grain}
\newcommand{\rhoratio}{\tilde{\rho}}
\newcommand{\scalevar}{\lambda}
\newcommand{\Lmax}{\scalevar_{\rm max}}
\newcommand{\Ltilde}{\tilde{\scalevar}}
\newcommand{\Rgrain}{R\grain}
\newcommand{\Rgraincm}{{R\grain}_{\rm ,\,cm}}
\newcommand{\rhograininternal}{\bar{\rho}\grain}
\newcommand{\rhoratiocrit}{\rhoratio_{\rm crit}}
\newcommand{\deltarho}{\delta_\rho}
\newcommand{\betafit}{b_{d}}

\newcommand{\Minsolar}{M_{\ast,\,\sun}}
\newcommand{\Linsolar}{L_{\ast,\,\sun}}

\newcommand{\tstop}{t_{\rm s}}
\newcommand{\taustop}{\tau_{\rm s}}
\newcommand{\eddy}{_{\rm e}}
\newcommand{\Teddy}{t\eddy}
\newcommand{\Leddy}{\scalevar\eddy}
\newcommand{\fturb}{f_{t}}
\newcommand{\gturb}{g_{t}}
\newcommand{\tcollapse}{t_{\rm grav}}
\newcommand{\rau}{r_{\rm au}}
\newcommand{\SigmaMMSN}{\Sigma_{0}}
\newcommand{\rhocrit}{\rho_{\rm crit}}

\newcommand\plotonesize[2]
 {\centering \leavevmode \includegraphics[width={#2\columnwidth}]{#1}}
\newcommand{\plotsidesize}[2]
 {\centering \leavevmode \includegraphics[width={#2\textwidth}]{#1}}
\newcommand{\acknowledgments}{\begin{small}\section*{Acknowledgments}\end{small}}
\newcommand\altaffilmark[1]{$^{#1}$}
\newcommand\altaffiltext[1]{$^{#1}$}
\title[Jumping the Gap: Pebble-Pile Planetesimals]{Jumping the Gap: The Formation Conditions and Mass Function of ``Pebble-Pile'' Planetesimals\vspace{-0.5cm}}

\author[Hopkins]{
\parbox[t]{\textwidth}{ 
Philip F. Hopkins\altaffilmark{1,2}\thanks{E-mail:phopkins@caltech.edu}} 
\vspace*{6pt} \\
\altaffiltext{1}{TAPIR, Mailcode 350-17, California Institute of Technology, Pasadena, CA 91125, USA\vspace{-1.1cm}} \\
}

\date{Submitted to MNRAS, January, 2014\vspace{-0.6cm}}
\begin{document}
\maketitle
\label{firstpage}

\begin{abstract}
In a turbulent proto-planetary disk, dust grains undergo large density fluctuations and under the right circumstances, grain overdensities can collapse under self-gravity (forming a ``pebble pile'' planetesimal). Using a simple model for fluctuations predicted in simulations, we estimate the rate-of-formation and mass function of self-gravitating planetesimal-mass bodies formed by this mechanism. This depends sensitively on the grain size, disk surface density, and turbulent Mach numbers. However, when it occurs, the resulting planetesimal mass function is broad and quasi-universal, with a slope ${\rm d}N/{\rm d}M\propto M^{-(1-2)}$, spanning size/mass range $\sim10-10^{4}\,$km ($\sim 10^{-9}-5\,M_{\oplus}$). Collapse to planetesimal through super-Earth masses is possible. The key condition is that grain density fluctuations reach large amplitudes on large scales, where gravitational instability proceeds most easily (collapse of small grains is suppressed by turbulence). This leads to a new criterion for ``pebble-pile'' formation: $\taustop\gtrsim 0.05\,\ln{(Q^{1/2}/Z_{d})}/\ln{(1+10\,\alpha^{1/4})}\sim 0.3\,\psi(Q,\,Z,\,\alpha)$ where $\taustop=\tstop\,\Omega$ is the dimensionless particle stopping time. In a MMSN, this requires grains larger than $a=(50,\ 1,\ 0.1)\,{\rm cm}$ at $r=(1,\ 30,\ 100)\,{\rm au}$. This may easily occur beyond the ice line, but at small radii would depend on the existence of large boulders. Because density fluctuations depend strongly on $\taustop$ (inversely proportional to disk surface density), lower-density disks are more unstable. Conditions for pebble-pile formation also become more favorable around lower-mass, cooler stars.
\end{abstract}

\begin{keywords}
planets and satellites: formation --- protoplanetary discs --- accretion, accretion disks --- hydrodynamics --- instabilities --- turbulence
\vspace{-1.0cm}
\end{keywords}

\vspace{-1.1cm}
\section{Introduction}
\label{sec:intro}

It is widely believed that dust grains form the fundamental building blocks of planetesimals. But famously, models which attempt to form planetesimals via the growth of dust grains in proto-planetary disks face the ``meter barrier'' or ``gap'': planetesimals of sizes $\gtrsim$\,km are required before gravity allows them to grow further via accretion, but grains larger than $\sim\,$cm tend to shatter rather than stick when they collide, preventing further grain growth. 

Therefore, alternative pathways to planetesimal formation have received considerable attention. If a dust grains -- even small ones -- could be strongly enough concentrated at, say, the disk midplane, their density would be sufficient to cause the region to collapse directly under its own self-gravity, and ``jump the gap'' to directly form a km-sized planetesimal from much smaller grains -- what we call a ``pebble pile planetesimal'' (\citealt{goldreich:1973.planetesimal.formation.sedimentation}; for reviews see \citealt{chiang:2010.planetesimal.formation.review,johansen:2014.planetesimal.formation.review}). In general, though, turbulence in the disk sets a ``lower limit'' to the degree to which grains can settle into a razor-thin sub-layer, and this has generally been regarded as a barrier to the ``pebble pile'' scenario described above \citep[but see][and references therein]{lyra:2009.semianalytic.planet.form.model.grain.settling,lee:2010.grain.settling.vs.grav.instability,chiang:2010.planetesimal.formation.review}. 

However, it is also well-established that the number density of solid grains can fluctuate by multiple orders of magnitude when ``stirred'' by turbulence, even in media where the turbulence is highly sub-sonic and the gas is nearly incompressible -- it has therefore been proposed that in some ``lucky'' regions, this turbulent concentration might be sufficient to trigger ``pebble pile'' formation \citep[see e.g.][]{bracco:1999.keplerian.largescale.grain.density.sims,cuzzi:2001.grain.concentration.chondrules,johansen:2007.streaming.instab.sims,carballido:2008.large.grain.clustering.disk.sims,lyra:2008.dust.traps.low.mass.disks,lyra:2009.planet.imf.from.sims,lyra:2009.semianalytic.planet.form.model.grain.settling,bai:2010.grain.streaming.sims.test,bai:2010.streaming.instability,bai:2010.grain.streaming.vs.diskparams,pan:2011.grain.clustering.midstokes.sims}. These studies -- mostly direct numerical simulations -- have un-ambiguously demonstrated that this is possible, {\em if} grains are sufficiently large, abundant, and the disks obey other conditions on e.g.\ their gas densities and sound speeds. The concentration phenomenon (including so-called ``vortex traps''; \citealt{barge:1995.vortex.trap.idea,zhu:2014.non.ideal.mhd.vortex.traps}) can occur via self-excitation of turbulent motions in the ``streaming'' instability \citep{johansen:2007.streaming.instab.sims}, or in externally driven turbulence, such as that excited by the magneto-rotational instability (MRI), global gravitational instabilities, convection, or Kelvin-Helmholtz/Rossby instabilities \citep{dittrich:2013.grain.clustering.mri.disk.sims,jalali:2013.streaming.instability.largescales,hendrix:2014.dust.kh.instability}. The direct numerical experiments have shown that the magnitude of these fluctuations depends on the parameter $\taustop=\tstop\,\Omega$, the ratio of the gas ``stopping'' time (friction/drag timescale) $\tstop$ to the orbital time $\Omega^{-1}$, with the most dramatic fluctuations around $\taustop\sim1$. These experiments have also demonstrated that the magnitude of clustering depends on the volume-averaged ratio of solids-to-gas ($\rhoratio\equiv \rhograin/\rhogas$), and basic properties of the turbulence (such as the Mach number). These have provided key insights and motivated considerable work studying these instabilities; however, the parameter space spanned by direct simulations is always limited. Moreover, it is not possible to simulate the full dynamic range of turbulence in these systems: the ``top scales'' of the system are $\Lmax\sim$\,AU, while the viscous/dissipation scales $\scalevar_{\eta}$ of the turbulence are $\scalevar_{\eta}\sim$\,m-km (Reynolds numbers $Re\sim10^{6}-10^{9}$). Clearly, some analytic models for these fluctuations -- even simplistic ones -- are needed for many calculations. 

Fortunately, the question of ``preferential concentration'' of aerodynamic particles is actually much more well-studied in the terrestrial turbulence literature. There both laboratory experiments \citep{squires:1991.grain.concentration.experiments,fessler:1994.grain.concentration.experiments,rouson:2001.grain.concentration.experiment,gualtieri:2009.anisotropic.grain.clustering.experiments,monchaux:2010.grain.concentration.experiments.voronoi} and numerical simulations \citep{cuzzi:2001.grain.concentration.chondrules,yoshimoto:2007.grain.clustering.selfsimilar.inertial.range,hogan:2007.grain.clustering.cascade.model,bec:2009.caustics.intermittency.key.to.largegrain.clustering,pan:2011.grain.clustering.midstokes.sims,monchaux:2012.grain.concentration.experiment.review} have long observed that very small grains, with stokes numbers $St\equiv \tstop/\Teddy(\scalevar_{\eta})\sim1$ (ratio of stopping time to eddy turnover time at the viscous scale) can experience order-of-magnitude density fluctuations at small scales (at/below the viscous scale). Considerable analytic progress has been made understanding this regime: demonstrating, for example, that even incompressible gas turbulence is unstable to the growth of inhomogeneities in grain density \citep{elperin:1996:grain.clustering.instability,elperin:1998.grain.clustering.instability.rotation}, and predicting the behavior of the small-scale grain-grain correlation function using simple models of gaussian random-field turbulence \citep{sigurgeirsson:2002.grain.markovian.concentration.toymodel,bec:2007.grain.clustering.markovian.flow}. 

These studies have repeatedly shown that grain density fluctuations are tightly coupled to the local vorticity field: grains are ``flung out'' of regions of high vorticity by centrifugal forces, and collect in the ``interstices'' (regions of high strain ``between'' vortices).\footnote{It is sometimes said that anti-cyclonic, large-scale vortices ``collect'' dust grains. But it is more accurate to say that grains preferentially avoid regions with high magnitude of vorticity $|\boldsymbol{\omega}|$. If sufficiently large vortices are anti-cyclonic and aligned with the disk plane, they represent a local {\em minimum} in $|\boldsymbol{\omega}|$, so grains concentrate by being dispersed out of higher-$|\boldsymbol{\omega}|$ regions.}
%This is somewhat misleading: grains always preferentially avoid regions with high absolute value of vorticity $|\boldsymbol{\omega}|\sim |{\bf v}_{e}\,\Leddy^{-1} + {\boldsymbol{\Omega}}|$. It is simply that very large eddies ($\Teddy \gtrsim\Omega^{-1}$, with $\Teddy=\Leddy/|{\bf v}_{e}|$) which are locally anti-cyclonic and in-plane ($\hat{\bf v}_{e} = -\hat{\boldsymbol{\Omega}}$) have lower $|\boldsymbol{\omega}|$ than the mean (${\bf v}_{e}=0$) Keplerian flow; so grains concentrate there by being dispersed out of higher-$|\boldsymbol{\omega}|$ regions.}
 Studies of the correlation functions and scaling behavior of higher Stokes-number particles suggest that, in the inertial range (ignoring gravity and shear), the same dynamics apply, but with the scale-free replacement of a ``local Stokes number'' $\tstop/\Teddy$, i.e.\ what matters for the dynamics on a given scale are the vortices of that scale, and similar concentration effects can occur whenever the eddy turnover time is comparable to the stopping time \citep[e.g.][]{yoshimoto:2007.grain.clustering.selfsimilar.inertial.range,bec:2008.markovian.grain.clustering.model,wilkinson:2010.randomfield.correlation.grains.weak,gustavsson:2012.grain.clustering.randomflow.lowstokes}. Several authors have pointed out that this critically links grain density fluctuations to the phenomenon of intermittency and discrete, time-coherent structures (vortices) on scales larger than the Kolmogorov scale in turbulence \citep[see][and references therein]{bec:2009.caustics.intermittency.key.to.largegrain.clustering,olla:2010.grain.preferential.concentration.randomfield.notes}. In particular, \citet{cuzzi:2001.grain.concentration.chondrules} demonstrate convincingly that grain density fluctuations behave in a multi-fractal manner: multi-fractal scaling is a key signature of well-tested, simple geometric models for intermittency \citep[e.g.][]{sheleveque:structure.functions}. In these models, the statistics of turbulence are approximated by regarding the turbulent field as a hierarchical collection of ``stretched'' singular, coherent structures (e.g.\ vortices) on different scales \citep{dubrulle:logpoisson,shewaymire:logpoisson,chainais:2006.inf.divisible.cascade.review}. 

Such statistical models have been well-tested as a description of the {\em gas} turbulence statistics \citep[including gas density fluctuations; see e.g.][]{burlaga:1992.multifractal.solar.wind.density.velocity,sorriso-valvo:1999.solar.wind.intermittency.vs.time,budaev:2008.tokamak.plasma.turb.pdfs.intermittency,shezhang:2009.sheleveque.structfn.review,hopkins:2012.intermittent.turb.density.pdfs}. More recently, first steps have been taken to link them to grain density fluctuations: for example, in the cascade model of \citet{hogan:2007.grain.clustering.cascade.model}, and the hierarchical model of \citet{hopkins:2013.grain.clustering}. These models ``bridge'' between the well-studied regime of small-scale turbulence and that of large, astrophysical particles in shearing, gravitating disks. The key concepts are based on the work above: we first assume that grain density fluctuations are driven by coherent velocity ``structures,'' for which we can solve analytically the perturbation owing to a single structure of a given scale. Building on \citet{cuzzi:2001.grain.concentration.chondrules} and others, we then attach this calculation to a well-tested, simple, cascade model for the statistics of velocity structures. In \citet{hogan:1999.turb.concentration.sims,cuzzi:2001.grain.concentration.chondrules,hogan:2007.grain.clustering.cascade.model,teitler:2009.turbulent.concentration.tests} and \citet{hopkins:2013.grain.clustering}, these models are shown to give a good match to both direct numerical simulations and laboratory experiments. 

In this paper, we combine these analytic approximations with simple criteria for gravitational collapse, to calculate the conditions under which ``pebble pile'' planetesimal formation may occur, and in that case, to estimate the mass function of planetesimals formed.

\begin{figure}
    \centering
    \plotonesize{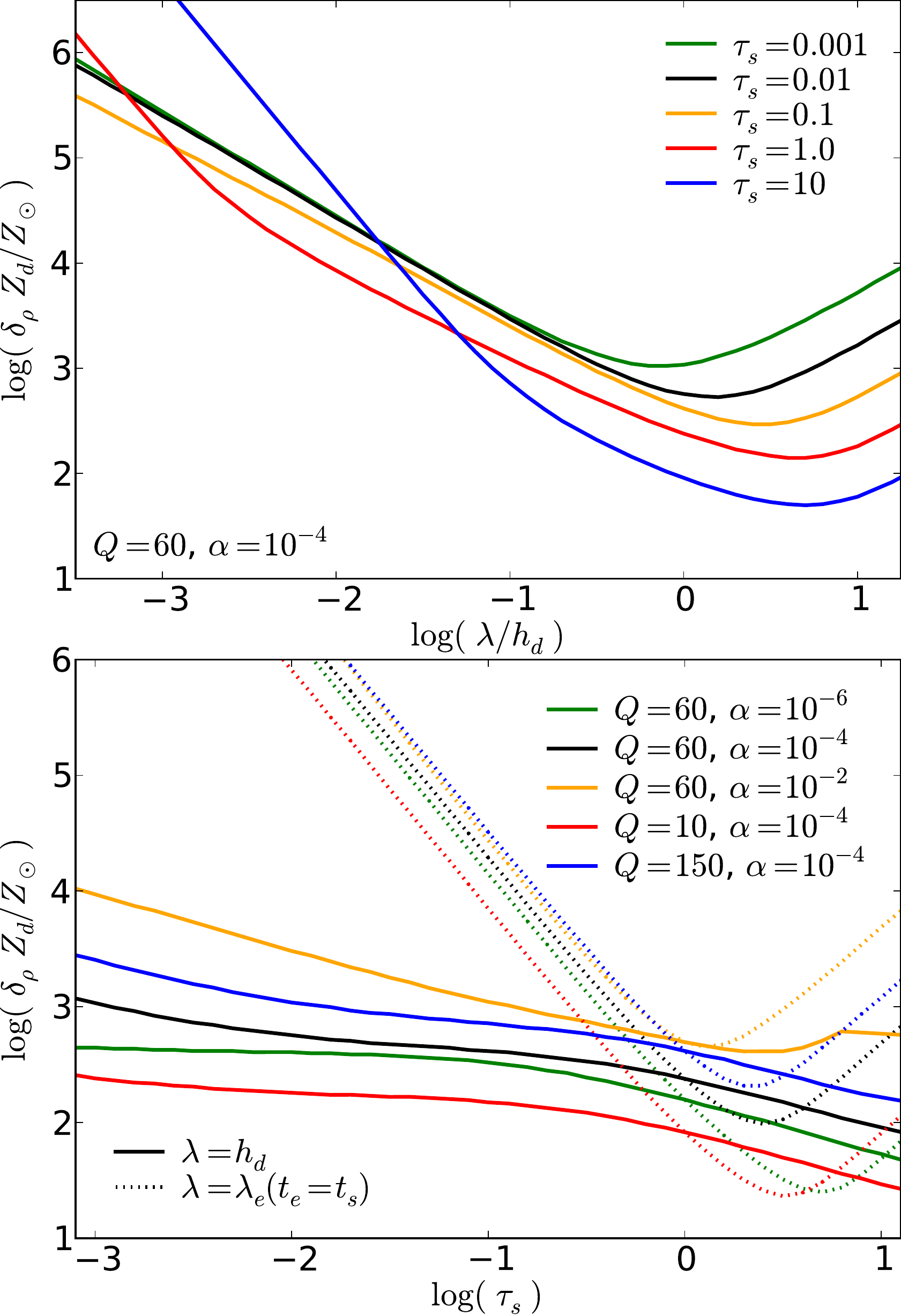}{0.99}
    \vspace{-0.2cm}
    \caption{Critical grain overdensity $\deltarho \equiv \rhograin(\scalevar)/\langle\rhograin\rangle$ for dynamical collapse under self-gravity (Eq.~\ref{eqn:stability}). We plot $\deltarho\,\Zgrain/Z_{\odot}$, since it is the combination $\deltarho\,\Zgrain$ that must exceed some critical value as a function of scale ($\scalevar/\hgrain$), Toomre $Q$, stopping time $\taustop=\tstop\,\Omega$, and turbulence strength $\alpha$. 
    {\em Top:} Critical density vs.\ scale, for different grain sizes ($\taustop$) in a disk with ``standard'' MMSN properties at $\sim1\,$au ($Q=60$, $\alpha=10^{-4}$). On most scales, larger grains require smaller fluctuations to collapse, because the initial dust disk is thinner (higher-density) and resistance from gas pressure is weaker. 
    {\em Bottom:} Critical density vs.\ $\taustop$, evaluated either at the disk scale height ($h_{d}$, near where the collapse overdensity $\deltarho(\scalevar)$ is minimized; solid) or the characteristic scale where density fluctuations are maximized ($\Leddy(\Teddy=\tstop)$; dotted). Both generally decrease with $\taustop$ until $\taustop\sim1$. For small grains ($\taustop\ll1$), the critical overdensity near $\Leddy(\Teddy=\tstop)\ll h_{d}$ is large because of turbulent support. We vary $Q$ and $\alpha$; the critical overdensities increase with $Q$, as expected, and with $\alpha$ (since turbulent support vs.\ gravity is larger), though the latter effect is weak.
    \label{fig:rhocrit}}
\end{figure}

\begin{figure}
    \centering
    \plotonesize{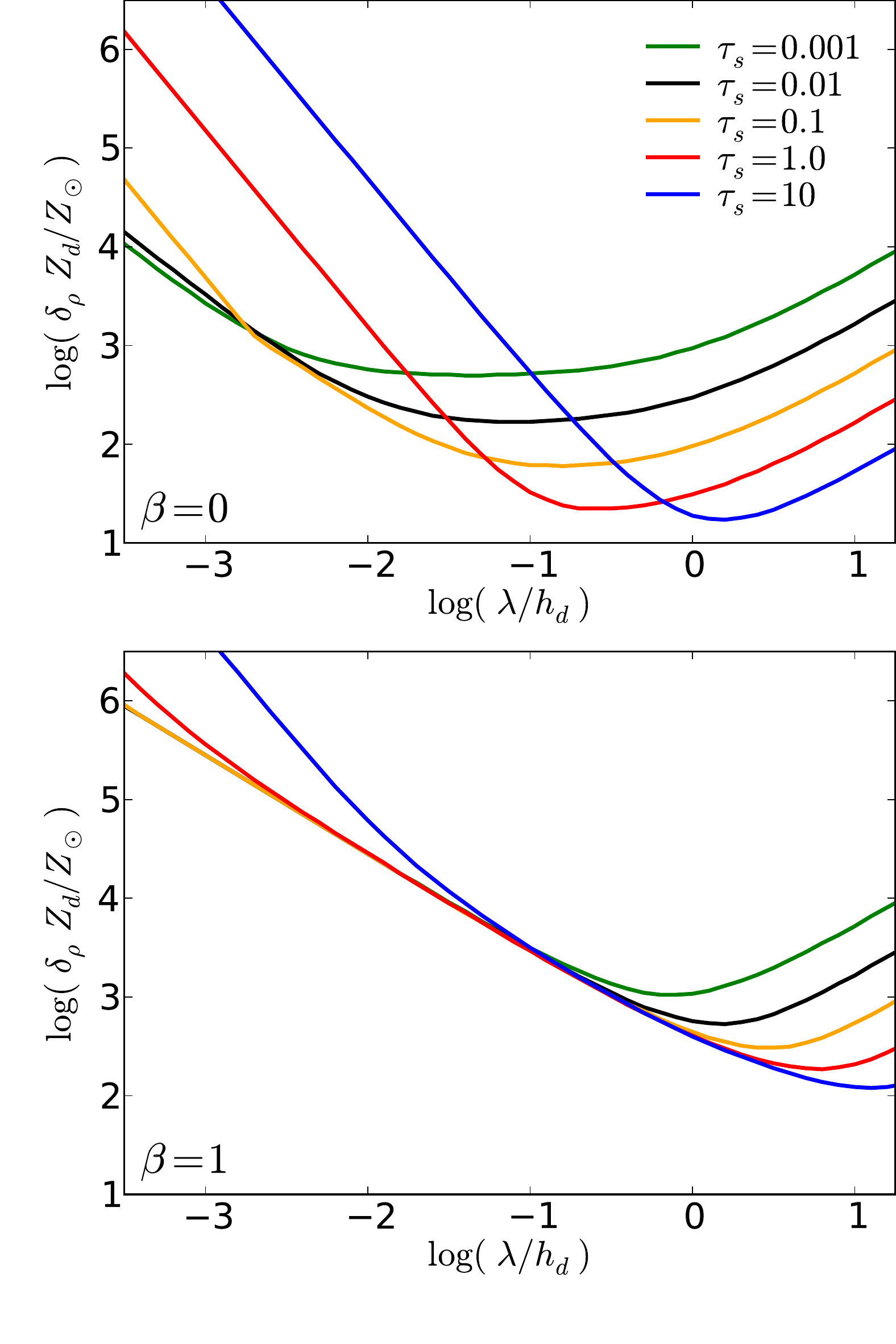}{0.99}
    \vspace{-0.6cm}
    \caption{As Fig.~\ref{fig:rhocrit} ({\em top}), but simply forcing $\beta=0$ (no gas pressure) or $\beta=1$ (treating the gas as perfectly-coupled). For $\beta=1$, the predicted thresholds are not dramatically different from our full calculation except at intermediate scales or for large grains on small scales. Taking $\beta=0$, however, would lead one to infer much smaller collapse densities (by an order of magnitude or more) on many scales. One must account for gas pressure resisting the collapse of grains on large scales. Note, though, that even neglecting gas pressure entirely, collapse on small scales $\lesssim 10^{-3}\,\hgrain$ requires enormous density fluctuations $\deltarho\gtrsim10^{4}$.
    \label{fig:rhocrit.beta}}
\end{figure}

\vspace{-0.5cm}
\section{The Model}
\label{sec:model}

\subsection{Overview and Basic Parameters}
\label{sec:model:overview}

Consider a grain-gas mixture in a Keplerian disk, at some (midplane) distance $\Rstar$ from the central star of mass $\Mstar$. Assume that the grains are in a disk with surface density $\Sigmagrain$ and exponential vertical scale-height $\hgrain$ ($\langle\rhograin(z)\rangle\propto \exp{(-|z|/\hgrain)}$), so the mid-plane density $\meanrhograin \equiv \langle\rhograin(z=0)\rangle = \Sigmagrain/(2\,\hgrain)$. This is embedded in a gas disk with corresponding $\Sigmagas$, $\hgas$, $\meanrhogas$, and sound speed $\cs$ and the global grain-to-gas mass ratio is defined by $\Zgrain \equiv \Sigmagrain/\Sigmagas$. Being Keplerian, the disk has orbital frequency $\Omega\equiv (G\,\Mstar/\Rstar^{3})^{1/2}$ and epicyclic frequency $\kappa\approx\Omega$ (Keplerian circular velocity $V_{K}=\Omega\,R$). In the regime of interest in this paper, the Mach numbers of gas turbulence within the mid-plane dust layer are small ($\ll1$; \citealt{voelk:1980.grain.relative.velocity.calc,laughlin:1994.protostar.disk.instabilities,gammie:2001.cooling.in.keplerian.disks,hughes:2011.turb.protoplanetary.disk.obs}), so the gas density fluctuations are much smaller than the grain density fluctuations \citep{passot:1988.proof.lognormal,vazquez-semadeni:1994.turb.density.pdf,scalo:1998.turb.density.pdf} and we can treat the gas as approximately incompressible $\rhogas \approx\langle\rhogas(z)\rangle$. This also gives the gas scale height $\hgas=\cs/\Omega$, and the usual Toomre $Q$ parameter $Q\equiv \cs\,\kappa/(\pi\,G\,\Sigmagas) = \Omega^{2}/(2\pi\,G\,\meanrhogas)$. We can define the usual turbulent $\alpha\equiv \langle v\gas^{2}\rangle / \cs^{2}$, where $\langle v^{2}\gas\rangle$ is the rms turbulent velocity of the gas averaged on the largest scales of the system.\footnote{We stress that $\alpha =  \langle v\gas^{2}\rangle / \cs^{2}$ is here purely to function as a useful parameter defining the turbulent velocities. We are {\em not} specifically assuming a \citet{shakurasunyaev73}-type viscous ``$\alpha$-disk'' nor a \citet{gammie:2001.cooling.in.keplerian.disks}-type gravito-turbulent disk.}

We will focus on a monolithic grain population with size $\Rgrain = \Rgraincm\,{\rm cm}$, internal density $\rhograininternal\approx2\,{\rm g\,cm^{-3}}$ \citep{weingartner:2001.dust.size.distrib}, mass fraction $\Zgrain$. The mid-plane stopping time is 
\begin{align}
\tstop &= \frac{\rhograininternal\,\Rgrain}{\meanrhogas\,\cs}\,\times
\begin{cases}
      {\displaystyle 1}\ \ \ \ \ \hfill {\tiny (\Rgrain\le 9\,\scalevar_\sigma/4)} \\ 
      {\displaystyle {(4\,\Rgrain)}/{(9\,\scalevar_\sigma)}}\ \ \ \ \ \hfill {\tiny (\Rgrain>9\,\scalevar_\sigma/4)} \
\end{cases}
\end{align}
where $\scalevar_\sigma = 1/(n\gas\,\sigma({\rm H}_{2})) = \mu\,m_{p}/(\meanrhogas\,\sigma(H_{2}))$ is the mean-free path in the gas ($n\gas$ is the gas number density, $m_{p}$ the proton mass, $\mu$ the mean molecular weight, and $\sigma(H_{2})$ the cross section for molecular collisions). We can then define $\taustop \equiv \tstop\,\Omega$. This and $\alpha$ determine the dust scale height, $\hgrain = \sqrt{\alpha/(\alpha+\taustop)}\approx\sqrt{\alpha/\taustop}\,\hgas$, a general result that holds for both large and small $\taustop$ \citep{carballido:2006.grain.sedimentation}.

Now allow a fluctuation $\rhograin(k) = \deltarho\, \langle \rhograin\rangle$ ($\deltarho\ne1$) of the mean grain density averaged on the scale $k$ centered near the midplane, where $k \equiv 1/\scalevar$ is the wavenumber ($\scalevar$ the wavelength) of the fluctuation. For the incompressible (Kolmogorov) turbulent cascade, we expect an rms turbulent velocity on each scale $\langle v\gas^{2}(k)\rangle \equiv \alpha\,\cs^{2}\,\fturb(\scalevar/\Lmax)$ with $\fturb\sim(\scalevar/\Lmax)^{2/3}$ where $\Lmax$ is the top/driving scale of the cascade (we take $\Lmax\approx \hgas$).\footnote{More accurately, we can correctly include the energy-containing range ($\scalevar>\Lmax$) by taking the isotropic turbulent power spectrum $E(k)\propto k^{-5/3}\,(1+|k\,\Lmax|^{-2})^{-(2-[-5/3])/2}$ \citep{bowman:inertial.range.turbulent.spectra}. This gives
\begin{align}
\langle v\gas^{2}(\scalevar)\rangle &= \alpha\,\cs^{2}\,\fturb(\scalevar/\Lmax) \\ 
\fturb(x) &\equiv \frac{4\,\Gamma[11/6]}{\sqrt{\pi}\,\Gamma[1/3]}\,\int_{0}^{x}\,t^{-1/3}\,(1+t^{2})^{-11/6}\,{\rm d}t\\
&\approx 1.189\,x^{2/3}\,(1 + 1.831\,x^{7/3})^{-2/7}
\end{align}
We use this (the approximate form is accurate to $\sim1\%$ at all $k$) in our full numerical calculations. Some cutoff is necessary at large scales or else a power-law cascade contains a divergent kinetic energy (and we do not expect $\Lmax\gg \hgas$). However, we do not include an explicit model for the dissipation range (scales below Kolmogorov $\scalevar_\eta$) -- i.e.\ we assume infinite Reynolds number -- since all quantities in this paper are converged already on much larger scales. For the conditions of interest, $\scalevar_\eta\sim 0.1$\,km \citep{cuzzi:2008.chondrule.planetesimal.model.secular.sandpiles}.} We can define the corresponding eddy turnover time $\Teddy(k) = \scalevar/\langle v\gas^{2}(k)\rangle^{1/2}$. Note that by assuming a Kolmogorov spectrum, we are implicitly assuming that grains do not modify the gas velocity structures. This is true when the density fluctuations are weak, but less clear when they are large (so that the local grain density becomes large compared to the gas density). However, preliminary simulation results suggest the power spectra of the gas turbulence are not much different in this limit \citep[see e.g.][]{johansen:2007.streaming.instab.sims,pan:2011.grain.clustering.midstokes.sims,hendrix:2014.dust.kh.instability}. 

The grains will also have a scale-dependent velocity dispersion following the turbulent cascade, for which we can define $\langle v\grain^{2}(k)\rangle \equiv \alpha\,\cs^{2}\,\gturb(\scalevar/\Lmax)$. However, grains are partially-coupled to gas, so $\gturb$ is in general a non-trivial function which we derive in Appendix~\ref{sec:appendix:graindisp}. On large scales (for small grains) where $\tstop\ll \Teddy(k)$, the grains are well-entrained by the gas, so we expect $\gturb\approx \fturb$, but on small scales where $\tstop\gg \Teddy(k)$, the grains are effectively collisionless, so have a constant (scale-independent) minimum velocity dispersion. 

In order to calculate the mass function of ``pebble piles'' -- the number density or probability of forming ``interesting'' fluctuations -- we require four things:
\begin{itemize}
\item{{\bf (1)} A model for the proto-planetary disk. This is given by the description above and \S~\ref{sec:disk.model}}

\item{{\bf (2)} A model for the statistics of grain density fluctuations. This is outlined in \S~\ref{sec:grain.density.fluct.model}, and is based on the direct numerical simulations and experiments described in \S~\ref{sec:intro}.}

\item{{\bf (3)} A criterion for an ``interesting'' fluctuation. We take this to be a fluctuation which is sufficiently large that it can undergo dynamical collapse under self-gravity (overcoming resistance to collapse from gas drag and pressure, turbulent kinetic energy, shear and angular momentum). We derive this criterion in \S~\ref{sec:collapse.crit} and Appendices~\ref{sec:stability.dustgas.fluid}-\ref{sec:turb.fluct.fx.on.collapse.criterion}.}

\item{{\bf (4)} A mathematical method to ``count'' the interesting fluctuations, given the assumptions above. This is provided by the excursion-set formalism, as summarized in \S~\ref{sec:excursion.sets}}
\end{itemize}

\vspace{-0.5cm}
\subsection{Physical Disk Models}
\label{sec:disk.model}

In order to attach physical values to the dimensionless quantities above, we require a disk model. We will adopt the following, motivated by the minimum-mass solar nebula (MMSN), for a disk of arbitrary surface density around a solar-type star:\footnote{We assume a stellar mass $M_{\ast} \approx M_{\odot}$; using a size $R_{\ast}\approx 1.8\,R_{\odot}$ and effective temperature $T_{\ast}\approx 4500\,$K for a young star, or $R_{\ast}\approx R_{\odot}$, $T_{\ast}\approx 6000\,$K for a more mature star give identical results in our calculations. Variations in the assumed stellar age lead to percent-level corrections to the model, much smaller than our other uncertainties.}
\begin{align}
\Omega &= \sqrt{\frac{G\,M_{\ast}}{\Rstar^{3}}} \approx 6.3\,\rau^{-3/2}\,{\rm yr^{-1}} \\ 
\Sigmagas &= \SigmaMMSN\,1000\,\rau^{-3/2}\,{\rm g\,cm^{-2}} \\ 
T_{\rm eff,\,\ast} &= {\Bigl(}\frac{(0.05\,\rau^{2/7})\,R_{\ast}^{2}}{4\,\Rstar^{2}} {\Bigr)}^{1/4}\,T_{\ast} \approx 140\,\rau^{-3/7}\,{\rm K} 
\end{align}
where $T_{\rm eff,\,\ast}$ is the effective temperature of the disk \citep{chiang:1997.protostellar.disk.sed}; $R_{\ast}$ and $T_{\ast}$ are the effective size and temperature of the star. 

Following \citet{chiang:2010.planetesimal.formation.review}, Eqs.~3-15, these choices determine the parameters
\begin{align}
\cs &= \sqrt{\frac{k_{B}\,T_{\rm mid}}{\mu\,m_{p}}} \approx 0.64\,\rau^{-3/14}\,{\rm km\,s^{-1}}\\
\frac{\hgas}{\Rstar} &= \frac{\cs}{V_{K}} \approx 0.022\,\rau^{2/7}\\
\meanrhogas &= \frac{\Sigmagas}{2\,\hgas} \approx 1.5\times10^{-9}\,\SigmaMMSN\,\rau^{-39/14}\,{\rm g\,cm^{-3}}\\ 
Q &= \frac{\cs\,\Omega}{\pi\,G\,\Sigmagas} \approx 61\,\SigmaMMSN^{-1}\,\rau^{-3/14} \\ 
\Pi &= \frac{1}{2\,\meanrhogas\,V_{K}\,\cs}\frac{\partial (\meanrhogas\,\cs^{2})}{\partial \ln{r}} \approx 0.035\,\rau^{2/7}\\
\scalevar_\sigma &= \frac{1}{n\gas\,\sigma(H_{2})} \approx \frac{1.2\,\SigmaMMSN^{-1}\,\rau^{39/14}}{1+(\rau/3.2)^{3/7}}\,{\rm cm}\\
\taustop &\approx {\rm MAX}
\begin{cases}
      {\displaystyle 0.004\,\SigmaMMSN^{-1}\,\rau^{3/2}\,\Rgraincm} \\ 
      {\displaystyle 0.0014\,\Rgraincm^{2}\,\rau^{-9/7}\,(1+(\rau/3.2)^{3/7})}
\end{cases}
\end{align}
with $\mu\approx 2.3$ (appropriate for a solar mixture of molecular gas) and we take the molecular cross-section $\sigma(H_{2})\approx2\times10^{-15}\,(1+(T/70\,{\rm K})^{-1})$ \citep{chapman:1970.gas.dynamics.book}. Here $T_{\rm mid,\,\ast}$ is the disk mid-plane temperature, and $\Pi$ defines the offset between the mean gas circular velocity and the Keplerian circular velocity $V_{K} - \langle V_{\rm gas} \rangle \equiv \eta\,V_{K}$, where $\Pi \equiv \eta V_{K}/\cs$; this is related to the grain drift velocity $v_{\rm drift} \equiv {2\,\eta\,V_{K}\,\taustop\,[{(1+\rhoratio)^{2}+\taustop^{2}/4}}]^{1/2}\,[{\taustop^{2}+(1+\rhoratio)^{2}}]^{-1}$ \citep{nakagawa:1986.grain.drift.solution}.

The expression we use for $T_{\rm mid,\,\ast}$ is the approximate expression for the case of a passive flared disk irradiated by a central solar-type star, assuming the disk is optically thick to the incident and re-radiated emission (in which case the external radiation produces a hot surface dust layer which re-radiates $\sim1/2$ the absorbed light back into the disk, maintaining $T_{\rm mid,\,\ast}^{4}\approx T_{\rm eff,\,\ast}^{4}/2$; see \citealt{chiang:1997.protostellar.disk.sed}).\footnote{More accurately, we can take the effective temperature from illumination to be: $T_{\rm eff,\,\ast}^{4} = T_{\ast}^{4}\,\alpha_{T}\,R_{\ast}^{2}/(4\,\Rstar^{2})$ with $\alpha_{T} \approx 0.005\,\rau^{-1} + 0.05\,\rau^{2/7}$ \citep{chiang:1997.protostellar.disk.sed}. Since we allow non-zero $\alpha$, this implies an effective viscosity and accretion rate $\dot{M}\approx3\pi\,\alpha\,\cs^{2}\,\Sigmagas\,\Omega^{-1}$ \citep{shakurasunyaev73}, which produces an effective temperature $T_{\rm eff,\,acc}^{4}\approx 3\,\dot{M}\,\Omega^{2}/(8\pi\,\sigma_{B})$ ($\sigma_{B}$ is the Boltzmann constant). Note this depends on the term $\cs^{2} = \sqrt{k_{B}\,T_{\rm mid}/(\mu\,m_{p})}$. A more accurate estimate of $T_{\rm mid}$ is then given by solving the implicit equation $T_{\rm mid}^{4} = (3/4)\,[\tau_{V}+4/3 + 2/(3\,\tau_{V})]\,T_{\rm eff,\,acc}^{4} + [1+\tau_{V}^{-1}]\,T_{\rm eff,\,\ast}^{4}$, where $\tau_{V}=\tau_{V}(T_{\rm mid}) = \kappa_{R}(T_{\rm mid})\,\Sigmagas/2$ is the vertical optical depth from the midplane. Here $\kappa_{R}$ is the Rosseland mean opacity, which we can take from the tabulated values in \citet{semenov:2003.dust.opacities} (crudely, $\kappa_{R}\sim 5\,{\rm cm^{2}\,g^{-1}}$ at $T_{\rm mid}>160\,$K and $\kappa_{R}\sim2.4\times10^{-4}\,T^{2}\,{\rm cm^{2}\,g^{-1}\,K^{-2}}$ at lower $T_{\rm mid}$. We use this more detailed estimate for our full numerical calculation, however it makes almost no difference for the parameter space we consider, compared to the simple scalings above.}

%sec:stability.dustgas.fluid
%sec:turb.fluct.fx.on.collapse.criterion

\vspace{-0.5cm}
\subsection{Criteria for Dynamical Gravitational Collapse}
\label{sec:collapse.crit}

Now, to define the mass function of ``interesting'' grain density fluctuations. Here, we will define ``interesting'' as those fluctuations which exceed some critical density $\rhocrit$, above which they can collapse under self-gravity on a dynamical timescale. This is not the only channel by which dust overdensities can form planetesimals! There are secular instabilities \citep[e.g.][]{youdin:2011.grain.secular.instabilities.in.turb.disks,shariff:2011.secular.grain.instability}, and grain overdensities could promote grain growth; but these require different considerations (see \S~\ref{sec:discussion}), and are outside the scope of our calculation here. 

For an grain overdensity or mode with size/wavelength $\scalevar = 1/k$, the critical density $\rhocrit$ will be a function of that wavelength, $\rhocrit = \rhocrit(\scalevar = k^{-1})$. It is convenient to define the local gas-to-dust mass ratio averaged on a scale $\scalevar$ around a point ${\bf x}$ (e.g.\ averaged in a sphere of radius $\scalevar$ about the point ${\bf x}$)
\begin{align}
\rhoratio = \rhoratio({\bf x},\,k) \equiv 1 + \frac{\rhograin({\bf x},\,k)}{\rhogas} 
\end{align}

If we consider grains which are purely collisionless (no grain-gas interaction), then a Toomre analysis gives the following criterion for gravitational instability of a mid-plane perturbation of wavenumber $k$:  
\begin{align}
\label{eqn:critrho.collisionless}
0 > \omega^{2} &= \kappa^{2} + \langle v\grain^{2}(k)\rangle\,k^{2} - 4\pi\,G\,\rhogas\,\rhoratio\,\frac{|k\,\hgrain|}{1+|k\,\hgrain|}
\end{align}
The $\omega$ here is the frequency of the assumed (linear) perturbations ($\propto \exp{(-\imath\,\omega\,t)}$; see Appendix~\ref{sec:stability.dustgas.fluid}); for $\omega^{2}<0$, the mode is unstable. Note that this is identical to the criterion for a stellar galactic disk \citep{binneytremaine}. We show in Appendix~\ref{sec:stability.dustgas.fluid} that a systematic dust settling/drift velocity does not change this criterion significantly, so long as the drift velocity $v_{\rm drift}\sim \taustop\,\eta\,V_{K} \ll V_{K}$. Here the $\kappa$ term represents the contribution of angular momentum resisting collapse, and $\langle v^{2}\grain(k) \rangle$ is the rms turbulent velocity of grains on the scale $k$; for a derivation of the turbulent term here see \citet{chandrasekhar:1951.turb.jeans.condition,chavanis:2000.vortex.trapping.disk.instability}.\footnote{More exactly, for grains on small scales -- where they are locally collisionless -- we should combine the turbulent velocity and density terms, taking instead $\rho\grain\rightarrow \rho\grain\,\mathcal{F}(\omega/\kappa,\,k^{2}\,\langle v\grain^{2}(k) \rangle/\kappa^{2}$) where $\mathcal{F}$ is the reduction factor determined by integration over the phase-space distribution. However, the relevant stability threshold comes from evaluating $\mathcal{F}$ near $\omega\approx0$; in this regime we can Taylor expand $\mathcal{F}$ (assuming a Maxwellian velocity distribution), and to leading order we recover the solution in Eq.~\ref{eqn:stability}. The exact solution can be determined for the purely collisional limit (identical to Eq.~\ref{eqn:stability}) or the purely collisionless limit (identical to a stellar disk, where the minimum density for collapse $\rhoratio$ is smaller by a factor $=0.935$. Given the other uncertainties in our calculation, this difference is negligible. Moreover, in Appendix~\ref{sec:turb.fluct.fx.on.collapse.criterion}, we show that the form of the turbulent terms in Eq.~\ref{eqn:critrho.collisionless} accounts for the non-linear, time-dependent and stochastic behavior of the turbulence (i.e.\ accounts for fluctuations in the velocity dispersions, and represents the criterion for a region where the probability of successful collapse is large).}
The negative term in $G$ represents self-gravity, and de-stabilizes the perturbation at sufficiently large $\rhoratio$. The terms in $|k\,\hgrain|$ on the right are the exact solution for an exponential vertical disk and simply interpolate between the two-dimensional (thin-disk) case on scales $\gtrsim \hgrain$ and three-dimensional case on scales $\lesssim \hgrain$ \citep[see][for derivations]{elmegreen:1987.cloud.instabilities,kim:2002.mhd.disk.instabilities}. 

In the opposite, perfectly-coupled ($\tstop\rightarrow0$) limit, we have a single fluid, so the dispersion relation is identical to that of a pure, single collisional fluid, in which we simply define ``dust'' and ``gas'' sub-components
\begin{align}
0 > \omega^{2} &= \kappa^{2} + \frac{1}{\rhoratio}\,(\cs^{2}+\langle v^{2}\gas(k) \rangle)\,k^{2}  \\
\nonumber & + \frac{\rhoratio-1}{\rhoratio}\,\langle v\grain^{2}(k)\rangle\,k^{2} - 4\pi\,G\,\rhogas\,\rhoratio\,\frac{|k\,\hgrain|}{1+|k\,\hgrain|}\\
\nonumber &= \kappa^{2} + \frac{\cs^{2}\,k^{2}}{\rhoratio} + \langle v^{2}(k) \rangle\,k^{2} - 4\pi\,G\,\rhogas\,\rhoratio\,\frac{|k\,\hgrain|}{1+|k\,\hgrain|}
\end{align}
Here $\cs$ and $\langle v^{2}\gas(k) \rangle$ represent gas pressure and turbulent support (and we used $\langle v^{2}\gas(k) \rangle=\langle v^{2}\grain(k) \rangle$ for the perfectly-coupled case). Note that the terms describing the gas pressure/kinetic energy density have a pre-factor $1/\rhoratio = \rhogas/(\rhogas + \rhograin)$, since what we need for the mixed-grain-gas perturbation is the energy density {per unit mass} in the perturbation. Likewise, the grain kinetic energy density term has a pre-factor $(\rhoratio-1)/\rhoratio = \rhograin/(\rhogas+\rhograin)$. Since both sit in the same external potential and self-gravitate identically, the $\kappa$ and $G$ terms need no pre-factor. In this limit we can think of the $1/\rhoratio$ factor as simply an enhanced ``mean molecular weight'' from the perfectly-dragged gas grains (so the effective sound speed of the gas $\cs^{\rm eff}\rightarrow \cs/\sqrt{1 + \rhograin/\rhogas} = \cs/\rhoratio^{1/2}$). In other words, in the perfectly-coupled limit, the system behaves as gas which must ``carry'' some extra weight in dust \citep[see also][]{youdin:2011.grain.secular.instabilities.in.turb.disks,shariff:2011.secular.grain.instability,shariff:2014.collapse.of.tightly.coupled.dust.gas.mixture}.

We can interpolate between these cases by writing 
\begin{align}
\nonumber 0 > \omega^{2} &\equiv \kappa^{2} + \frac{\beta}{\rhoratio}\,(\cs^{2}+\langle v^{2}\gas(k) \rangle)\,k^{2}  \\
& + \frac{\rhoratio-1}{\rhoratio}\,\langle v\grain^{2}(k)\rangle\,k^{2} - 4\pi\,G\,\rhogas\,\rhoratio\,\frac{|k\,\hgrain|}{1+|k\,\hgrain|}\label{eqn:stability}
\end{align}
The only important ambiguity in the above is the term $\beta$, which we introduce (in a heuristic and admittedly ad-hoc manner) to represent the strength of coupling between grains and gas ($\beta=0$ is un-coupled/collisionless; $\beta=1$ is perfectly-coupled). In general, $\beta$ is some unknown, presumably complicated function of all the parameters above, which can only be approximated in the fully non-linear case by numerical simulations (and the exact criterion for intermediate cases between these limits may require terms beyond those which can be approximated by the $\beta$ here). However, the limits are straightforward: if a perturbation collapses on a free-fall time $\tcollapse$, and $\tcollapse \ll \tstop$, we expect $\beta\rightarrow0$ (since there is no time for gas to decelerate grains). Conversely if $\tcollapse\gg\tstop$, $\beta\rightarrow1$. Therefore in this paper we make the simple approximation\footnote{This approximation is motivated by the \citet{maxey:1987.grav.settling.aerodynamic.particles} linear expansion of the solution for the de-celeration of dust grains by molecular collisions for times $t\ll \tstop$ in a symmetric, homogeneous sphere. The ratio of the dust de-celeration term to the term from gas pressure if the sphere were pure gas ($\rho^{-1}\,\nabla P$) in this limit is the same as $\beta$ in Eq.~\ref{eqn:beta.of.rho}.}
\begin{align}
\label{eqn:beta.of.rho}
\beta \approx \frac{\tcollapse}{\tstop + \tcollapse} = (1+\tstop/\tcollapse)^{-1} = {\Bigl(}1 + \frac{4\,\taustop}{\pi}\sqrt{\frac{\rhoratio}{3\,Q}} {\Bigr)}^{-1}
\end{align} 
where we assume $\tcollapse \approx (3\pi/32\,G\,\rho)^{1/2}$, {\em for regions which meet our dynamical collapse criterion} (i.e.\ this does not apply to secularly-sedimenting regions, or regions where self-gravity is not stronger than all other forces including the support from gas drag). 

Alternative derivations of these scalings from the linear equations for coupled gas-dust fluid, and including the non-linear stochastic effects of turbulence, are presented in Appendices~\ref{sec:stability.dustgas.fluid}-\ref{sec:turb.fluct.fx.on.collapse.criterion}, respectively. If anything, we have chosen to err on the side of caution and define a strict criterion for collapse -- almost all higher-order effects make collapse slightly easier, not harder. This criterion is sufficient to ensure that (at least in the initial collapse phase) a pebble pile is gravitationally bound (including the thermal pressure of gas being dragged with dust and turbulent kinetic energy of gas and dust), and gravitational collapse is sufficiently strong to overcome gas pressure forces, tidal forces/angular momentum/non-linear shearing of the overdensity, turbulent vorticity and ``pumping'' of the energy and momentum in the region, and ram-pressure forces from the ``headwind'' owing to radial drift. Similarly, when this criterion is met, the gravitational collapse timescale is faster than the orbital time, the grain drift timescale, the effective sound-crossing time of the clump, and the eddy turnover time. 

Using the definitions in \S~\ref{sec:model:overview}, Eq.~\ref{eqn:stability} can be re-written: 
\begin{align}
0 &> 1 + {\Bigl(}\frac{\hgas}{\hgrain}{\Bigr)}^{2}\,\frac{1}{\Ltilde^{2}\,\rhoratio}\,
{\Bigl [} \beta\,{\Bigl(}1 + \alpha\,\fturb{\bigl[}\frac{\scalevar}{\Lmax}{\bigr]} {\Bigr)} \\ 
& + \alpha\,(\rhoratio-1)\,\gturb{\bigl[}\frac{\scalevar}{\Lmax}{\bigr]} {\Bigr]}
\nonumber -\frac{2\,Q^{-1}}{1+\Ltilde}\,\rhoratio \\ 
&= 1 + \frac{\taustop}{\Ltilde^{2}\,\rhoratio}\,
{\Bigl [} \beta\,{\Bigl(}\alpha^{-1} + \fturb {\Bigr)}  + (\rhoratio-1)\,\gturb {\Bigr]}
-\frac{2\,Q^{-1}}{1+\Ltilde}\,\rhoratio
\end{align}
where $\Ltilde \equiv \scalevar/\hgrain$ and we abbreviate $\fturb = \fturb(\scalevar/\Lmax)$.
This has the solution
\begin{align}
\rhoratio &> \rhoratiocrit(\scalevar) \equiv \psi_{0}\,(1 + \sqrt{1 + \psi_{1}/\psi_{0}}) \\ 
\psi_{0} &\equiv \frac{Q}{4}\,(1+\Ltilde)\,{\Bigl[}1 + \taustop\,\Ltilde^{-2}\,\gturb {\Bigr]} \\ 
\psi_{1} &\equiv \frac{2\,\taustop}{\Ltilde^{2}}\,{\Bigl[}\beta(\alpha^{-1}+\fturb) - \gturb {\Bigr]}\,{\Bigl [} 1 + \taustop\,\Ltilde^{-2}\,\gturb {\Bigr]}^{-1}
\end{align}
(Note, if $\beta$ itself is a function of $\rhoratio$, then this is an implicit equation for $\rhoratiocrit$ which must be solved numerically). 
Recall the dimensionless grain density fluctuation $\deltarho = \rhograin/\meanrhograin$, so $\rhoratio = 1 + \deltarho\,(\meanrhograin/\meanrhogas) = 1 + \deltarho\,(\Sigmagrain/\Sigmagas)\,(\hgas/\hgrain) = 1 + \deltarho\,\Zgrain\,\sqrt{\taustop/\alpha}$. 
So in terms of $\deltarho$, the criterion becomes
\begin{align}
\deltarho\,\Zgrain > \sqrt{\frac{\alpha}{\taustop}}\,{\Bigl(}\rhoratiocrit(\scalevar) - 1 {\Bigr)} \label{eqn:deltarhocrit}
\end{align}

\vspace{-0.5cm}
\subsection{A Simple Representation of Grain Density Fluctuations in Incompressible Gas}
\label{sec:grain.density.fluct.model}

\citet{hopkins:2013.grain.clustering} present some simple, analytic expressions for the statistics of grain density fluctuations in a turbulent proto-planetary disk. For our purposes here, what is important is that these expressions provide a reasonable ``fitting function'' to the results of direct numerical simulations and laboratory experiments studying grain density fluctuations resulting from a variety of underlying mechanisms (e.g.\ driven turbulence, zonal flows, streaming-instability and Kelvin-Helmholtz instabilities; \citealt{johansen:2007.streaming.instab.sims,bai:2010.grain.streaming.vs.diskparams,dittrich:2013.grain.clustering.mri.disk.sims,jalali:2013.streaming.instability.largescales,hendrix:2014.dust.kh.instability}). We should note that this directly builds on several previous analytic models for grain clustering around the Kolmogorov (turbulent dissipation) scale and in the inertial range \citep[e.g.][]{elperin:1996:grain.clustering.instability,elperin:1998.grain.clustering.instability.rotation,hogan:2007.grain.clustering.cascade.model,bec:2008.markovian.grain.clustering.model,wilkinson:2010.randomfield.correlation.grains.weak,gustavsson:2012.grain.clustering.randomflow.lowstokes}. We have experimented with some of these models and find, for small grains where the clustering occurs on small scales (where $\Teddy \ll \Omega^{-1}$, and shear/rotation can be neglected), they give qualitatively similar results; we discuss this further in \S~\ref{sec:compare.alt.calcs}. However these previous models did not consider the case of a rotating disk with large grains where $\tstop \sim \Omega^{-1}$, in which case the behavior can differ dramatically (see e.g.\ \citealt{lyra:2008.dust.traps.low.mass.disks}). 

We briefly describe the model in \citet{hopkins:2013.grain.clustering} here, but interested readers should see that paper. Fundamentally, it follows \citet{hogan:2007.grain.clustering.cascade.model} in assuming grain density fluctuations on different scales can -- like gas density fluctuations in supersonic turbulence \citep{hopkins:2012.intermittent.turb.density.pdfs} -- be represented by a multiplicative random cascade. At any instant, a group of grains ``sees'' a gas vorticity field which can be represented as a superposition of coherent velocity structures or ``eddies'' with a wide range of characteristic spatial scales $\scalevar$ and timescales $\Teddy$. If we consider a single, idealized eddy or vortex, we can analytically solve for the effect it has on the grain density distribution in and around itself (assuming the vortex survives for some finite timescale $\sim \Teddy$). When $\tstop$ is much smaller than $\Teddy$, the grains are tightly-coupled to the gas, so the vortex has no effect on the average grain density distribution (since the gas is incompressible); when $\tstop$ is much larger than $\Teddy$, the vortex is unable to perturb the grains. But when $\tstop\sim \Teddy$, the vortex imprints large (order-unity) changes in the density field. These changes are multiplicative, so to the extent that the vorticity field can be represented by hierarchical cascade models, the grain density distribution on various scales behaves as a multiplicative random cascade. Assuming the turbulence obeys a Kolmogorov power spectrum, and assuming some ``filling factor'' of structures which each behave as scaled versions of the ideal vortex (constrained to match that power spectrum), with each grain encountering a random Gaussian field of vortex structures over time, and adopting a simple heuristic correction for the ``back-reaction'' of grains on gas, this leads to a prediction for a lognormal-like (random multiplicative) distribution of local grain densities.

Quantitatively, for a given set of {\em global}, dimensionless properties of the turbulent disk: $\taustop$, $\alpha$, and $\Pi$, the model predicts the distribution of grain density fluctuations (relative to the mean), $\delta_{\rho}({\bf x},\,\scalevar) = \rhograin({\bf x},\,\scalevar)/\meanrhograin$. Recall, this is the grain density averaged within a radius $\scalevar$ (as $\rho\grain({\bf x},\,\scalevar) = M\grain(|{\bf x^{\prime}-x}|<\scalevar)/(4\pi\,\scalevar^{3}/3) = \delta(\scalevar)\,\langle \rhograin \rangle$) around a random point in space ${\bf x}$. Using these parameters, we obtain $P(\delta_{\rho},\,\scalevar)$, the probability that any point in space lives within a region, averaged on size scale $\scalevar$, with grain over-density between $\delta_{\rho}$ and $\delta_{\rho} + d\delta_{\rho}$. This is approximately log-normal, with a variance that depends on scale (where most of the power, or contribution to the variance, comes from scales where the ``resonant condition'' $\tstop\sim \Teddy$ is satisfied).

\vspace{-0.5cm}
\subsection{Counting ``Interesting'' Density Fluctuations}
\label{sec:excursion.sets}

For {\em any} model for the statistics of grain density fluctuations, and any threshold criterion for an ``interesting'' fluctuation (both as a function of scale), there is a well-defined mathematical framework for calculating the predicted mass function, size distribution, correlation function, and related statistics of the objects/regions which exceed the threshold. This is the ``excursion set formalism,'' well-known in cosmology as the ``extended Press Schechter'' method by which dark matter halo mass functions, clustering, and merger histories can be analytically calculated \citep[there, the statistics are given by the initial Gaussian random field and cosmological power spectrum, and ``interesting'' regions are those which turn around from the Hubble flow; see][]{bond:1991.eps}. Recently, the same framework has been applied to predict the mass function of structures (e.g.\ giant molecular clouds and voids) formed on galactic scales by super-sonic interstellar turbulence \citep{hopkins:excursion.ism}; the initial mass function \citep{hennebelle:2008.imf.presschechter,hennebelle:2009.imf.variation,hopkins:excursion.imf,hopkins:excursion.imf.variation} and correlation functions/clustering \citep{hopkins:excursion.clustering} of cores and young stars inside molecular clouds; and the mass spectrum of planets which can form via direct collapse in turbulent, low-$Q$ disks \citep{hopkins:2013.turb.planet.direct.collapse}. For reviews, see \citet{zentner:eps.methodology.review,hopkins:frag.theory,offner:2013.imf.review}.

There are many ways to apply this methodology. Probably the simplest, and what we use here, is a Monte Carlo approach. Consider some annulus in the disk at radius $\Rstar$. Select some arbitrarily large number of Monte Carlo ``sampling points''; each of these represents a different random location ${\bf x}$ within the annulus (i.e.\ they randomly sample the volume) -- really, each is a different random realization of the field, given its statistics. Now, we can ask what the density of dust grains is, averaged in spheres of size $\scalevar$, around each of these points ($\rhograin({\bf x},\,\scalevar)$). If this ``initial'' $\scalevar = \scalevar_{0}$ is sufficiently large: 
\begin{align}
\scalevar &= \scalevar_{0} \gg \hgas \\ 
\rhograin({\bf x},\,\scalevar_{0}) & \rightarrow \langle \rhograin({\bf x},\,\scalevar_{0}) \rangle
\end{align}
for all ${\bf x}$, i.e.\ all points have the same mean density around them on sufficiently large scales (this is just the definition of the mean density, after all). Now, take a differential step ``down'' in scale -- this corresponds to shrinking the smoothing sphere by some increment $\Delta \scalevar$, and ask what the mean density inside each sphere is. For the sphere around each point ${\bf x}$, this is given by the appropriate conditional probability distribution function $P(\delta_{\rho},\,\scalevar_{0} - \Delta \scalevar\,|\,\delta_{\rho}[\scalevar_{0}],\,\scalevar_{0})$ (essentially the probability of a given {\em change} in the mean density, between two volumes separated by some differentially small smoothing size), or:
\begin{align}
\scalevar &\rightarrow \scalevar_{0} - \Delta \scalevar \\ 
\delta_{\rho}({\bf x},\,\scalevar_{0}) &\rightarrow \delta_{\rho}({\bf x},\,\scalevar_{0} - \Delta \scalevar) = \delta_{\rho}({\bf x},\,\scalevar_{0}) + \Delta \delta_{\rho}({\bf x})\\ 
P(\Delta \delta_{\rho}) &= P[\Delta \delta_{\rho}\,|\,\delta_{\rho}(\scalevar_{0}),\,\scalevar_{0},\,\Delta \scalevar]
\end{align}
The conditional probability distribution function $P[\Delta \delta_{\rho}\,|\,\delta_{\rho}(\scalevar_{0}),\,\scalevar_{0},\,\Delta \scalevar]$ is directly related to the power spectrum of density fluctuations (see \citealt{hopkins:frag.theory}, Eq.~2-12), and is determined by the model for $P(\delta_{\rho},\,\scalevar)$ described in \S~\ref{sec:grain.density.fluct.model}. Knowing that distribution, we draw a random value of $\Delta \delta_{\rho}$ for each Monte Carlo point ${\bf x}$, determining $\delta_{\rho}({\bf x},\,\scalevar_{0}-\Delta\scalevar)$. We then repeat this until we reach $\scalevar\rightarrow 0$ (making sure to take small enough steps $\Delta\lambda$ so that the statistics are converged). 

For each random point ${\bf x}$, we now have the value of the density field smoothed on all scales, $\delta_{\rho}({\bf x},\,\scalevar)$ -- in the excursion-set language, this is referred to as its ``trajectory.'' We can now simply compare this to the predicted collapse threshold on each scale $\rhoratiocrit(\scalevar)$ (Eq.~\ref{eqn:deltarhocrit}), to ask whether the region is ``interesting'' (exceeds the critical density for dynamical collapse). To avoid the ambiguity of ``double-counting'' or ``clouds in clouds'' (i.e.\ trajectories which exceed $\rhoratiocrit(\scalevar_{1})$ but also have some $\scalevar_{2}>\scalevar_{1}$ where they exceed $\rhoratiocrit(\scalevar_{2})$, which represents smaller scales that are independently self-gravitating/collapsing embedded in larger collapsing regions), we specifically consider the ``first crossing distribution'' \citep[see][]{bond:1991.eps,hopkins:excursion.imf,hopkins:excursion.ism}. Namely, if a trajectory exceeds $\rhocrit(\scalevar)$ anywhere, we uniquely identify the {\em largest} size/mass scale $\scalevar = \scalevar_{\rm first}$ on which $\rhoratio({\bf x},\,\scalevar) > \rhoratiocrit(\scalevar)$ as the ``total'' collapsing object. Since the trajectory $\delta_{\rho}({\bf x},\,\scalevar)$ is continuous in $\scalevar$, $\rhoratio({\bf x},\,\scalevar_{\rm first}) = \rhoratiocrit(\scalevar_{\rm first})$, and there is actually a one-to-one mapping between the first-crossing scale and mass enclosed in a first-crossing, given by the integral over volume in an exponential disk (since that is the vertical profile we assumed): 
\begin{align}
M(\scalevar_{\rm first}) &\equiv 4\,\pi\,\rhoratio_{\rm crit}(\scalevar_{\rm first})\,\meanrhogas\,\hgrain^{3} \nonumber \\ 
\times &{\Bigl[}\frac{\scalevar_{\rm first}^{2}}{2\,\hgrain^{2}} + {\Bigl(}1+\frac{\scalevar_{\rm first}}{\hgrain}{\Bigr)}\,\exp{{\Bigl(}-\frac{\scalevar_{\rm first}}{\hgrain}{\Bigr)}}-1 {\Bigr]}
\end{align}
\citep{hopkins:excursion.ism}. It is easy to see that on scales $\scalevar_{\rm first}<\hgrain$, this is just $M=(4\pi/3)\,\rhocrit\,\scalevar_{\rm first}^{3}$, on scales $\scalevar_{\rm first}>\hgrain$, just $M=\pi\,\Sigma_{\rm crit}\,\scalevar_{\rm first}^{2}$ (where $\Sigma_{\rm crit} = 2\,\hgrain\,\rho_{\rm crit}$).

Finally, we can use our ensemble of trajectories to define the function $f = f(\scalevar_{\rm first})$, where $f$ is the fraction of Monte Carlo ``trajectories'' that have a first-crossing on scales $\scalevar > \scalevar_{\rm first}$. Since $M(\scalevar_{\rm first})$ is a function of $\scalevar_{\rm first}$, we can just as well write this as a function of mass, $f = f(M)$, where $M\equiv M(\scalevar_{\rm first})$. Now, since each Monte Carlo trajectory represents the probability that a random point in space -- i.e.\ a random differential volume element -- is embedded in such a region, the differential value $|{\rm d}f(M)/{\rm d}\ln{M}|\,{\rm d}\ln{M}$ represents the differential volume fraction embedded inside of regions of with masses $M=M(\lambda_{\rm first})$ between $\ln{M}$ and $\ln{M} + {\rm d}\ln{M}$. Since these first-crossing regions have mean internal mass density (by definition) $\rho = \rhocrit(\lambda_{\rm first})$, the number of independent ``regions'' or ``objects'' (per unit volume) must be  
\begin{align}
\label{eqn:mf.general}
\frac{{\rm d}n_{\rm first}(M)}{{\rm d}\ln{M}} \equiv \frac{\rhocrit(\scalevar_{\rm first}[M])}{M}\,{\Bigl |}\frac{{\rm d}\,f(M)}{{\rm d}\ln{M}} {\Bigr|} 
\end{align}
And this is the desired mass function of collapsing objects. To turn this into an absolute number (instead of a number density), we simply need to integrate over the ``effective'' volume (differential volume in a radial annulus ${\rm d}R$ is just $(2\,\hgrain)\,(2\pi\,R\,{\rm d}R)$); or we can directly convert from volume fraction to absolute number based on the argument above, for an assumed disk size. 

For our calculations here, we typically use $\sim 10^{9}$ Monte Carlo ``trajectories'' to sample the statistics, and sample those trajectories in logarithmically-spaced steps $\Delta \lambda \approx 0.001\,\lambda$. Most of the results here converge at much coarser sampling, but the mass function at the lowest masses requires a large number of trajectories to be properly represented.

\begin{figure*}
    \centering
    \plotsidesize{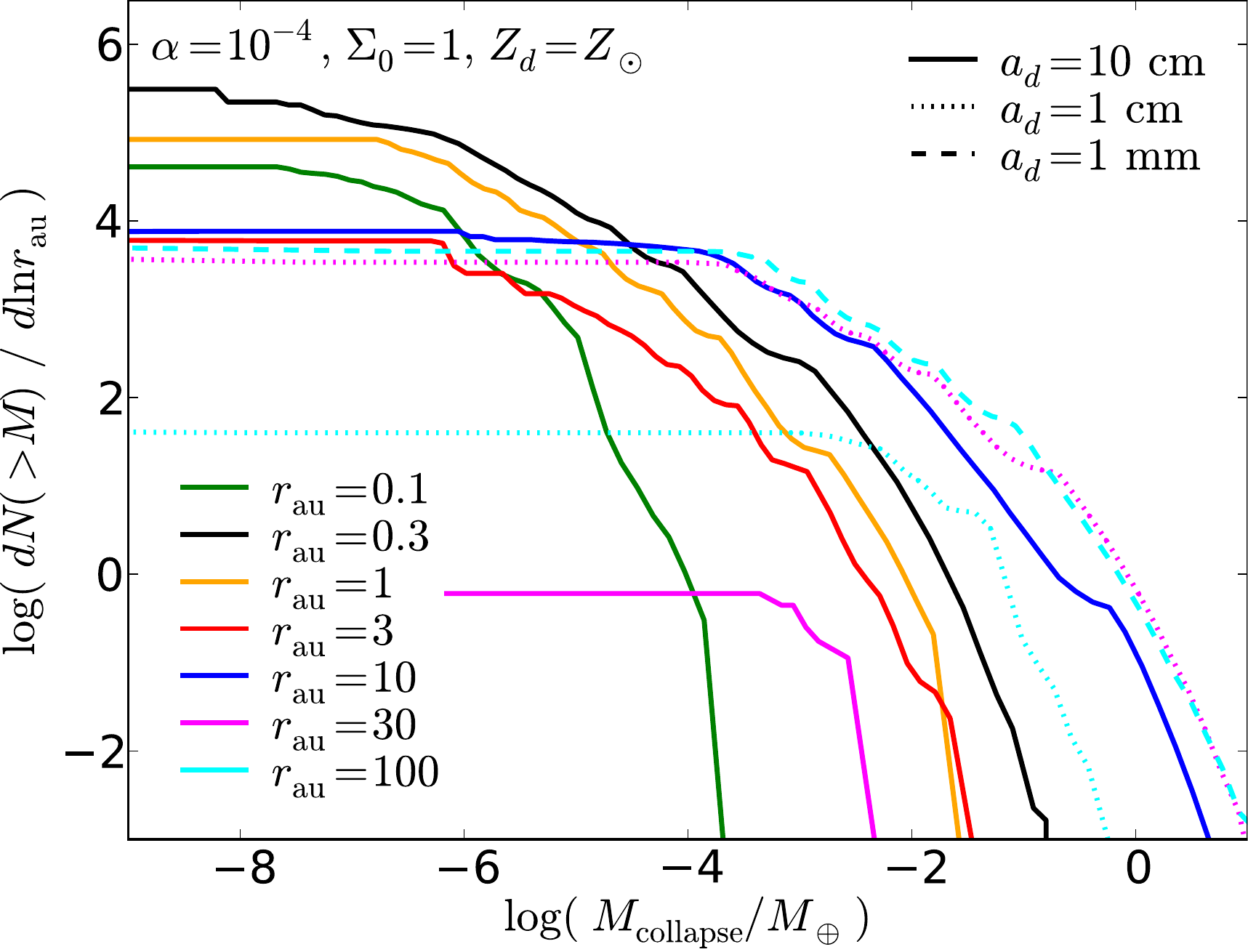}{0.9}
    \vspace{-0.2cm}
    \caption{Predicted mass function of collapsing (self-gravitating) pebble-pile planetesimals formed by turbulent grain density concentrations. We plot the cumulative number formed at various radial distances from the star (per unit orbital distance: ${\rm d}N/{\rm d}\ln{r_{\rm au}}$), as a function of mass (in Earth masses). The disk is our standard MMSN model ($Z=Z_{\odot}$, $\SigmaMMSN=1$; see \S~\ref{sec:disk.model}), with $\alpha=10^{-4}$. Different line types assume the grain mass is concentrated in grains of different sizes (as labeled). If the grains are large ($10\,$cm), then pebble piles can collapse directly to masses from $\sim10^{-8}-1\,M_{\oplus}$ over a range of orbital radii $\sim 0.1-20\,$au. If grains only reach $1\,$cm, the lower $\taustop$ super-exponentially suppresses this process at smaller radii, and it can only occur at large radii $\gtrsim 20-30\,$au, where $\taustop\gtrsim0.1$ (however the range of masses at these radii is large, from $\sim10^{-4}-10\,M_{\oplus}$). For maximum grain sizes $=1\,$mm, this is pushed out to $\gtrsim100\,$au. 
    \label{fig:mf.demo}}
\end{figure*}

\vspace{-0.5cm}
\section{Approximate Expectations}
\label{sec:approx}

We now have everything needed to calculate the detailed statistics of collapsing regions. Before we do so, however, we can gain considerable intuition using the some simple approximations. 

\vspace{-0.5cm}
\subsection{Small Grains}
\label{sec:smallgrain.criterion}

For small grains ($\taustop\ll1$), density fluctuations on scales $\sim \hgrain$ are weak (since the grains are well-coupled to gas on these scales). Large density fluctuations are, however, still possible on small scales, where $\Teddy\sim \tstop$. Consider this limit. In this regime, in a large Reynolds-number flow, the fluctuations are approximately self-similar, because all grains ``see'' a large, scale-free (power-law) turbulent cascade at both larger scales ($\Teddy\gg \tstop$) and smaller scales ($\Teddy\ll \tstop$). As shown in many studies \citep{cuzzi:2001.grain.concentration.chondrules,yoshimoto:2007.grain.clustering.selfsimilar.inertial.range,hogan:2007.grain.clustering.cascade.model,pan:2011.grain.clustering.midstokes.sims,hopkins:2013.grain.clustering}, the maximum local density fluctuations in this limit saturate at values $\deltarho^{\rm max}\sim300-1000$.

Since $\Teddy(\scalevar < \Lmax) \propto \scalevar^{2/3}$, this ``resonance'' will occur at scales $\scalevar\approx\Lmax\,\taustop^{3/2}\,\alpha^{3/4}\,(\hgas/\Lmax)^{3/2} \ll \Lmax$ 
%(so $\Ltilde\sim\taustop^{2}$). %% not Lmax=H
(so $\Ltilde\sim\alpha^{1/4}\,\taustop^{2}$). 
We can, on these scales, also approximate 
%$\fturb\approx\gturb\approx (\scalevar/\Lmax)^{2/3} \approx \taustop$, 
$\fturb\approx\gturb\approx (\scalevar/\Lmax)^{2/3} \approx \alpha^{1/2}\,\taustop$, 
and drop higher-order terms in $\scalevar/\Lmax$ or $\Ltilde$. If we take either the tightly coupled ($\beta=1$) or un-coupled ($\beta=0$) limits, we obtain
\begin{align}
\rhoratiocrit &\sim
\begin{cases}
%      {\displaystyle {\Bigl(}{\frac{Q}{2\,\alpha\,\taustop^{3}}}{\Bigr)}^{1/2}}\ \ \ \ \ \hfill {\tiny (\beta=1)} \\ 
      {\displaystyle {\Bigl(}{\frac{Q}{2\,\alpha^{3/2}\,\taustop^{3}}}{\Bigr)}^{1/2}}\ \ \ \ \ \hfill {\tiny (\beta=1)} \\ 
      \\
      {\displaystyle \frac{Q}{2}\,{\Bigl(} 1 + \taustop^{-2} {\Bigr)}}\ \ \ \ \ \hfill {\tiny (\beta=0)} \
\end{cases}
\end{align}
or
\begin{align}
\deltarho &\gtrsim
\begin{cases}
%      {\displaystyle \Zgrain^{-1}\,\taustop^{-2}\,({Q/2})^{1/2}}\ \ \ \ \ \hfill {\tiny (\beta=1)} \\ 
      {\displaystyle \Zgrain^{-1}\,\taustop^{-2}\,\alpha^{-1/4}\,({Q/2})^{1/2}}\ \ \ \ \ \hfill {\tiny (\beta=1)} \\ 
      \\
      {\displaystyle \Zgrain^{-1}\,\taustop^{-5/2}\,\alpha^{1/2}\,(Q/2)}\ \ \ \ \ \hfill {\tiny (\beta=0)} \
\end{cases}
\end{align}
This requires extremely large density fluctuations: for $\Zgrain\sim \Zsun$, and $Q\sim 60$ (MMSN at $\Rstar\sim1\,$au), this gives minimum $\deltarho$ of $\sim3\times10^{5}\,(\taustop/0.1)^{-2}\,(\alpha/10^{-4})^{-1/4}$ and $\sim 5000\,(\taustop/0.1)^{-5/2}\,(\alpha/10^{-4})^{1/2}$, respectively. 

Physically, even if we ignore gas pressure, and the density fluctuation is small-scale (so shear can be neglected), grains must still overcome their turbulent velocity dispersion in order to collapse. A simple energy argument requires $G\,M\grain^{2}(<\scalevar)/\scalevar \gtrsim M\grain(<\scalevar)\,\langle v\grain^{2}(\scalevar)\rangle$ (where $M\grain$ is the dust mass inside the region of size $\scalevar$); using $M\grain(<\scalevar) \sim \rhograin\,\scalevar^{3}$ and $\langle v\grain^{2}(\scalevar)\rangle\sim (\scalevar/\Teddy^{\grain})^{2}$, this is just $G\,\rhograin \gtrsim (\Teddy^{\grain})^{-2}$. In other words, the collapse time $\tcollapse\sim(G\,\rhograin)^{-1/2}$ must be shorter than the eddy turnover time (within the grains) $\Teddy^{\grain}$ on the same scale. But recall, the clustering occurs characteristically on a scale where for the gas, $\Teddy\sim\tstop$. Thus, the grains are at least marginally coupled, and the grain $\Teddy^{\grain}\sim\Teddy\sim\tstop$ -- the same eddies that induce strong grain clustering {\em necessarily} induce turbulent grain motions with eddy turnover time on the same scale $\sim \tstop$ \citep[see][]{bec:2009.caustics.intermittency.key.to.largegrain.clustering}. So collapse of even a ``collisionless'' grain population requires $\tcollapse \lesssim \tstop$. Using $Q\sim \Omega^{2}/(G\,\rhogas)$ and $\rhograin\sim\rhogas\rhoratio$ (for $\rhoratio\gg1$), we see this is equivalent to the $\beta=0$ criterion above. Since, in this limit, $\tcollapse<\tstop$, taking $\beta=0$ is in fact a good approximation (and since the $\beta=1$ criterion requires a still higher density, so $\tcollapse\ll\tstop$, it is not the relevant case limit here). 

Thus even with no gas pressure effects ($\beta=0$), collapse ($\deltarho^{\rm max}\gtrsim \deltarho^{\rm collapse}$) requires $\taustop \gtrsim 0.2\,(\alpha/10^{-4})^{1/5}\,(\deltarho^{\rm max}\,\Zgrain/1000\,Z_{\odot})^{-2/5}\,(Q/60)^{2/5}$ -- unless the disks are extremely quiescent ($\alpha \ll 10^{-7}$), we are forced to consider large grains (where $\taustop\ll1$ is not true). 

Before going on, however, note that the arguments we make above apply only to {\em dynamical} collapse of small grains. {\em Secular} collapse of small grains, through the slow, nearly incompressible ``sedimentation'' mode described in e.g.\ \citep{youdin:2011.grain.secular.instabilities.in.turb.disks,shariff:2011.secular.grain.instability} may still be possible in this regime. As noted above, this requires a different treatment entirely and is outside the scope of this paper; however, it may present an alternative channel for planetesimal formation if only small grains are present.

\vspace{-0.5cm}
\subsection{Large Grains}
\label{sec:approx.largegrain.criterion}

For large grains, fluctuations are possible on large scales. For a flat perturbation spectrum, the most unstable scale is $\scalevar\sim\hgrain$ \citep{goldreich:1965.spiral.stability,toomre77,lau:spiral.wave.dispersion.relations,laughlin:1994.protostar.disk.instabilities}, so take this limit now. In this case 
%$\fturb\approx\gturb\approx1$ 
$\fturb\approx\gturb\approx(\alpha/\taustop)^{1/3}$ 
and $\Ltilde\approx1$, giving
\begin{align}
\rhoratiocrit &\sim
\begin{cases}
      {\displaystyle {\Bigl(}\frac{Q\,\taustop}{\alpha} {\Bigr)}^{1/2}}\ \ \ \ \ \hfill {\tiny (\beta=1)} \\ 
      \\
%      {\displaystyle Q\,{(} 1 + \taustop {)}}\ \ \ \ \ \hfill {\tiny (\beta=0)} \
      {\displaystyle Q\,{(} 1 + \taustop^{2/3}\,\alpha^{1/3} {)}}\ \ \ \ \ \hfill {\tiny (\beta=0)} \
\end{cases}
\end{align}
or
\begin{align}
\deltarho &\gtrsim
\begin{cases}
      {\displaystyle \Zgrain^{-1}\,Q^{1/2}}\ \ \ \ \ \hfill {\tiny (\beta=1)} \\ 
      \\
      {\displaystyle \Zgrain^{-1}\,Q\,\sqrt{\alpha/\taustop}}\ \ \ \ \ \hfill {\tiny (\beta=0)} \
\end{cases}
\end{align}
Even at $\Zgrain\sim\Zsun$ and $Q\sim60$, this gives a minimum $\deltarho$ of $\sim400$ and $\sim100\,(\taustop/0.1)^{-1/2}\,(\alpha/10^{-4})^{1/2}$, respectively. Collapse is far ``easier'' when grains can induce fluctuations on large scales.

In this limit, the $\beta=0$ criterion is just the a Roche criterion, $\tcollapse\lesssim\Omega^{-1}$ (the turbulence is sub-sonic, so its support is not dominant on large scales). The $\beta=1$ criterion is more subtle: recall that the ``effective'' sound speed of the coupled fluid is $\sim \cs^{\rm eff}/\sqrt{\rhoratio}$ \citep{safronov:1969.largest.body.planetesimal.before.instabilities,marble:1970.dust.gas.fluid.dynamics,sekiya:1983.secular.grav.instability.pebbles.in.disk}, and that $\taustop/\alpha = (\hgas/\hgrain)^{2}$, $\cs\sim\Omega\,\hgas$, and $Q\sim\Omega^{2}/G\,\rhogas$. Then we see this criterion is equivalent to $\tcollapse\lesssim t_{\rm cross} \equiv \hgrain/\cs^{\rm eff}$, i.e.\ that the collapse time is shorter than the effective sound-crossing time on the scale $\hgrain$. For $\taustop\lesssim1$, these generally do allow $\tcollapse \gtrsim \tstop$, so $\beta\sim1$ is the more relevant limit -- but importantly, collapse of the two-fluid medium even on timescales $\gg \tstop$ is allowed, provided a large overdensity can form on sufficiently large scales (i.e.\ collapse is ``slow'' compared to the stopping time, but ``fast'' compared to the dynamical and effective sound-crossing times).  In other words, the important criterion is the ``effective'' Jeans number of the coupled dust-gas fluid, 
\begin{align}
\label{eqn:jeans.number}
J \equiv \frac{\rho\,G}{c_{s,\,{\rm eff}}^{2}\,k^{2}} &=  \frac{(\rhograin+\rhogas)^{2}\,G}{\rhogas\,\cs^{2}\,k^{2}} \sim \frac{\rhoratio^{2}\,\meanrhogas\,G\,\scalevar^{2}}{\cs^{2}} \sim \frac{t_{\rm cross}^{\rm eff}}{\tcollapse} \lesssim 1\\ 
\rhoratio_{\rm crit}(\scalevar) &\sim \frac{\cs}{\scalevar\,\sqrt{G\,\meanrhogas}} \sim Q^{1/2}\, \frac{\cs}{\scalevar\,\Omega} \\ 
\rhoratio_{\rm crit}(\scalevar & \sim \hgrain) \sim Q^{1/2}\, \frac{\cs}{\sqrt{\alpha/\taustop}\,\hgas\,\Omega} \sim \left( \frac{Q\,\taustop}{\alpha} \right)^{1/2}
\end{align}
(where we have dropped the order-unity prefactors). This was first proposed as a key revision to the \citet{goldreich:1973.planetesimal.formation.sedimentation} midplane density by \citet{safronov:1969.largest.body.planetesimal.before.instabilities} and \citet{marble:1970.dust.gas.fluid.dynamics}, and appreciated by \citet{sekiya:1983.secular.grav.instability.pebbles.in.disk} \citep[see also][]{youdin:2011.grain.secular.instabilities.in.turb.disks,shariff:2011.secular.grain.instability}; recently, direct numerical simulations by \citet{shariff:2014.collapse.of.tightly.coupled.dust.gas.mixture} following the full non-linear collapse of a dusty sphere (no angular momentum or turbulent terms) in the perfect-coupling ($\taustop\rightarrow0$) limit have explictly confirmed that for $J\gtrsim 0.4$, collapse will occur and proceeds on the dynamical (free-fall) time, while for smaller $J$ only the ``secular'' sedimentation mode survives \citep[see also][who obtain consistent results in the perfectly un-coupled limit]{wahlberg-jansson:2014.pebble.pile.collapse.process}. This agrees well with our criterion, if we take the appropriate limits (and keep the relevant pre-factors). 

In \citet{hopkins:2013.grain.clustering}, approximate expressions are given for the maximum density fluctuations seen in simulations of large grains on large scales (Table~2 therein), which provide a good fit to the results of numerical simulations of MHD (zonal-flow), streaming-instability, and driven turbulence \citep{hogan:2007.grain.clustering.cascade.model,bai:2010.grain.streaming.sims.test,dittrich:2013.grain.clustering.mri.disk.sims,hendrix:2014.dust.kh.instability}. These are $\ln{\deltarho^{\rm max}}\sim 6\,\delta_{0}\,(1+\delta_{0})^{-1}\ln{[1+\betafit\,(1+\delta_{0})+\betafit^{3/2}\,(\sqrt{1+\delta_{0}^{2}}-1)]}$ where $\delta_{0}\approx 3.2\,\taustop/(1+\taustop^{2})$ and $\betafit\sim1$ depends on the ratio of drift to turbulent velocities. For $\taustop\lesssim1$, this becomes $\deltarho^{\rm max}\sim \exp{[20\,\ln{(1+\betafit)}\,\taustop]}$; comparing this to the above ($\beta=1$) criterion requires $\taustop\gtrsim0.05\,\ln{(Q^{1/2}/\Zgrain)}/\ln{(1+\betafit)} \sim 0.3$. So sufficiently large grains can indeed achieve these fluctuations.

\begin{figure}
    \centering
    \plotonesize{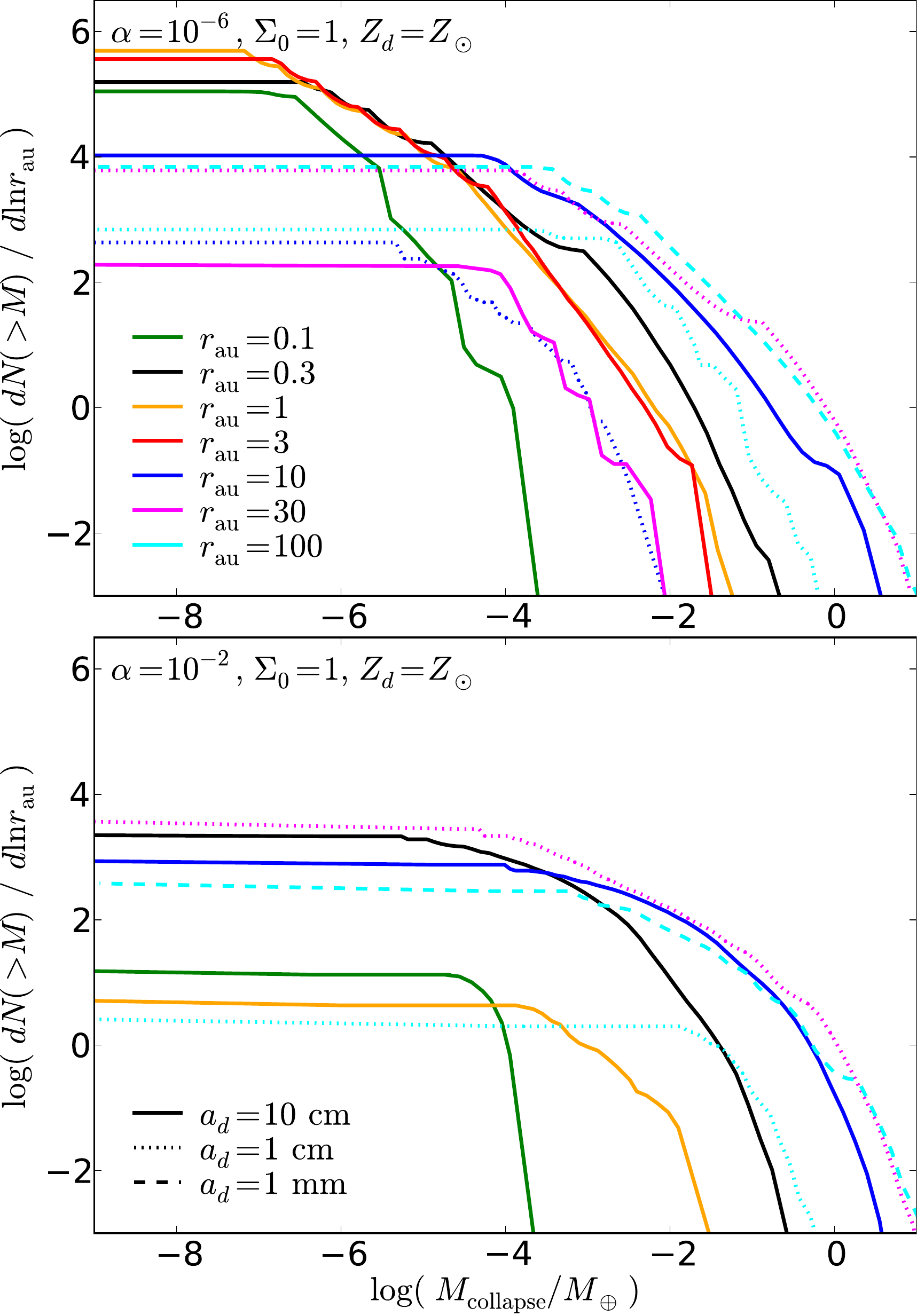}{0.99}
    \vspace{-0.2cm}
    \caption{Predicted pebble pile mass function, as Fig.~\ref{fig:mf.demo}, for varied $\alpha=10^{-6}-10^{-2}$ ({\em top} and {\em bottom}). At large masses, $\alpha$ has little effect on the MF. At low masses, increasing $\alpha$ means larger turbulent support on small scales, suppressing low-mass pebble pile formation. For high-$\alpha$, this flattens the MF slope and eliminates pebble pile formation at some smaller radii even for large grains. However, most of our conclusions about the radii where pebble pile formation can occur, as a function of grain size, are not changed.
    \label{fig:mf.alpha}}
\end{figure}

\begin{figure*}
    \centering
    \plotsidesize{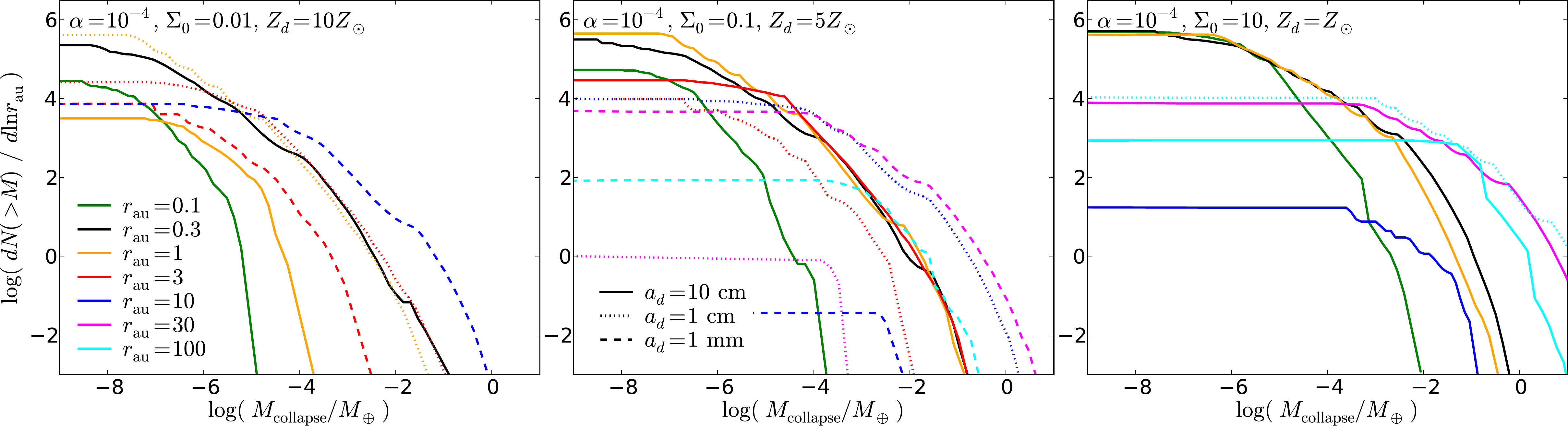}{0.99}
    \vspace{-0.2cm}
    \caption{Predicted pebble pile mass function, as Fig.~\ref{fig:mf.demo}, for varied protoplanetary disk properties. Here we fix $\alpha=10^{-4}$, but vary the disk mass/surface density (proportional to $\SigmaMMSN\equiv \Sigma/\Sigma_{\rm MMSN}$) and metallicity $\Zgrain$. 
    {\em Left:} Very low-density (but high-$\Zgrain$ disk); this corresponds to a MMSN which has lost $\sim 99\%$ of its gas but only $\sim 90\%$ of its large grains. Such a disk is expected to form collapsing pebble piles at $\sim1-10$\,au even with $\sim$mm-cm sized grains!
    {\em Middle:} Low density disk (MMSN after losing $\sim90\%$ of its gas and $\sim50\%$ of its large grains). Intermediate grains 
    {\em Right:} High-density disk ($\sim10\times$ MMSN), with solar abundances. Only large grains can form pebble piles.
    Although the mass and mean density increase, and Toomre $Q$ decreases, with increasing $\SigmaMMSN$, the parameter $\taustop\propto \SigmaMMSN^{-1}$ (at fixed grain size) decreases. Since the maximum amplitude of grain density fluctuations scales super-exponentially with $\taustop$ (while the threshold for collapse is only linear in $Q$), this means smaller grains can preferentially form pebble piles in {\em lower} density disks (where $\taustop\sim1$). 
    \label{fig:mf.disk}}
\end{figure*}

\vspace{-0.5cm}
\section{Numerical Results}

Now we show the results of the full numerical model described in \S~\ref{sec:model}, for specific choices of the disk parameters.

\vspace{-0.5cm}
\subsection{The Collapse Threshold}

\subsubsection{Dependence on Spatial Scale: Large Scales are Favored}

In Fig.~\ref{fig:rhocrit} we illustrate how the threshold for self-gravity derived in \S~\ref{sec:collapse.crit} scales as a function of various properties. Recall, the combination $\deltarho\,\Zgrain$ must exceed some value (Eq.~\ref{eqn:deltarhocrit}) which is a function only of ($\taustop$, $Q$, $\alpha$) in order for an over-density to collapse on a dynamical timescale. So the collapse threshold in dimensionless units of grain-density fluctuations ($\deltarho$) scales inversely with the dust-to-gas mass ratio $\Zgrain$. We see that, as is generic for Jeans/Toomre collapse and expected from the arguments in \S~\ref{sec:approx}, higher over-densities are required for collapse on small scales, with a minimum in $\deltarho$ around $\scalevar\sim\hgrain$. On small scales the thermal pressure term in Eq.~\ref{eqn:stability} ($\propto \scalevar^{-2}/\rhoratio$) dominates the support vs.\ gravity ($\propto \rhoratio$), giving $\rhoratiocrit \propto \scalevar^{-1}$. On large scales $\scalevar\gg \hgrain$ angular momentum dominates Eq.~\ref{eqn:stability} and, just as in the Toomre problem, $\rhoratiocrit \propto \scalevar$. 

\vspace{-0.5cm}
\subsubsection{Dependence on Grain Properties}

We also see that, generically, larger grains (larger $\taustop$) require smaller $\deltarho$ for collapse. This is because (with other disk properties fixed) the initial dust disk settles to a smaller scale height (larger density), and because the resistance by gas pressure is weaker. The change in this behavior for large grains $\taustop\gtrsim1$ on small scales owes to the fact that the velocity dispersions of large grains de-couple from the gas and become scale-independent (do not decrease with $\scalevar$) on small scales. 

If we focus on $\deltarho$ around scales $\scalevar\sim\hgrain$ or $\scalevar\sim\Leddy(\Teddy=\tstop)$, as in \S~\ref{sec:approx}, we confirm our approximate scalings above. Near $\sim \hgrain$, collapse requires modest over-densities $\sim100-1000$, weakly dependent on $\taustop$ or $\alpha$ (for small $\alpha \lesssim 10^{-3}$) and $\propto Q^{1/2}$, confirming our approximate scaling for $\beta=1$ (since in this limit, $\tcollapse \gtrsim \tstop$, $\beta\sim1$). Around $\scalevar\sim\Leddy(\Teddy=\tstop)$, we see, as expected, a strong scaling $\deltarho \propto \taustop^{-5/2}$ with weak residual dependence on $\alpha$ (and also $\propto Q$), as expected from our derivation above.

\vspace{-0.5cm}
\subsubsection{Importance of Gas Pressure}

In Fig.~\ref{fig:rhocrit.beta} we repeat this exercise but simply force $\beta=1$ or $\beta=0$. We can see that either approximation fails to match our usual ``hybrid'' interpolated model, at some range of scales, by about an order of magnitude. This indicates that it is clearly necessary to understand the non-linear behavior of collapsing objects, when $\taustop\sim \tcollapse$. However, assuming $\beta=1$ does not much change the criteria for large-scale collapse, and while the change at small scales is large we require such large values of $\deltarho$ that dynamical collapse is unlikely in any any case. But assuming $\beta=0$ gives lower collapse thresholds by an order-of-magnitude or more on large scales (the most interesting range for our calculation). In this regime the collapse thresholds are such that the collapse time is longer than the stopping time, so it is probably not a good approximation to neglect gas pressure. The collapsing dust will drag gas with it, increasing the gas pressure \citep{youdin:2011.grain.secular.instabilities.in.turb.disks,shariff:2011.secular.grain.instability,shariff:2014.collapse.of.tightly.coupled.dust.gas.mixture}, hence our assumption of $\beta=1$, but the full non-linear behavior in this regime remains poorly understood.

\vspace{-0.5cm}
\subsection{The Mass Function of Resulting Pebble-Pile Planetesimals}

Given our assumptions, we can now estimate the mass function of collapsing dust density fluctuations. 
Fig.~\ref{fig:mf.demo} shows the results for our ``default'' MMSN model ($\SigmaMMSN=1$, $\Zgrain=Z_{\odot}$, $\alpha=10^{-4}$), at various radii, assuming different grain sizes.

\vspace{-0.5cm}
\subsubsection{Dependence on Orbital Distance and Grain Size}

As expected, if the grains are sufficiently large ($\taustop\sim1$), the model predicts that self-gravitating pebble piles will form over a range of orbital radii, with a wide range of self-gravitating masses. For $\Rgrain\sim10\,$cm, all radii $\rau\sim0.1-10$ have $\taustop\sim1$ and form pebble piles. At still smaller radii, large $Q$ values imply sound speeds sufficient to suppress collapse; at larger radii, $\taustop\gg1$, and so grain-density fluctuations are actually suppressed because the grains are approximately collisionless (large density fluctuations cannot be generated by the gas, for $\taustop\gtrsim 3-5$). For smaller grains, we must go to larger radii before $\taustop\sim1$, and collapse becomes possible. For $\Rgrain\sim1\,$cm, pebble pile formation at $\ll10\,$au is completely suppressed -- we stress that because the density fluctuations depend exponentially on $\taustop$, the predicted number density is $<10^{-10}$ here! We see this rapid threshold behavior set in between $\taustop\sim0.1-0.3$, a parameter space we explore further below.

\vspace{-0.5cm}
\subsubsection{The Minimum and Maximum Masses}

Where possible, these collapse events form objects with a range of masses $\sim 10^{-8}-10\,M_{\oplus}$. 

The {\em maximum} mass is given by the behavior of the largest velocity structures. Recall, in this model, grains are essentially {passive}, so if structures of non-zero vorticity exist with $\Leddy\gtrsim \hgrain$ (with the appropriate $\Teddy\sim \tstop$), they still drive grain density fluctuations in the midplane dust layer (so long as the eddy intersects the midplane somewhere) on scales $\sim \Leddy$, even if we take the dust layer to be infinitely thin.\footnote{Note that, even if the grains themselves drive the turbulence, as in the streaming instability case, such large velocity structures form via the shearing of smaller structures, albeit with relatively limited power in their associated velocity fluctuations \citep[see e.g.][]{johansen:2007.streaming.instab.sims}.} Indeed, this is just one of the toy-model cases considered in \citet{hopkins:2013.grain.clustering} (see also \citealt{lyra:2013.max.density.in.vortex.trap.calc}): a large in-plane vortex in a disk perturbing a razor-thin (two-dimensional) dust distribution in the midplane; for which the same scalings apply as the three-dimensional case. It simply becomes dust {\em surface density} fluctuations that are driven by the large vortices trapping or expelling dust, rather than three-dimensional density fluctuations. So surface density fluctuations can, in principle, form over a wide range of scales; for a large structure with $\Leddy\gtrsim \hgrain$, the enclosed grain mass in the perturbation becomes $M\approx \pi\,\Sigma_{\rm crit}\,\scalevar^{2} = 2\pi\,\meanrhogas\,\rhoratiocrit\,\hgrain^{3}\,\tilde{\scalevar}^{2}$. Based on the arguments in \S~\ref{sec:approx}, we expect $\tcollapse\gtrsim\tstop$ (so $\beta\sim1$) on these scales, so $\rhoratiocrit\sim (Q\,\taustop/2\,\alpha\,\tilde{\scalevar})^{1/2}$. If we assume $\taustop\sim1$, and that the largest {\em possible} structures reach $\sim \hgas$, then we obtain 
\begin{align}
\label{eqn:maxmass} M_{\rm collapse}^{\rm max}(\taustop\sim1) &\sim \sqrt{2}\pi\,\frac{\alpha^{1/4}\,Q^{1/2}}{\taustop^{1/4}}\,\meanrhogas\,\hgas^{3} \\
&\sim 0.03\,{\Bigl(}\frac{\alpha}{10^{-4}}{\Bigr)}^{1/4}\,\SigmaMMSN^{1/2}\,\rau^{27/28}\,M_{\oplus}
\end{align}
So the maximum mass is only weakly-dependent on $\alpha$, while it increases with disk surface density and is nearly proportional to radius (because the disk mass increases with $\rau$).\footnote{Interestingly, if we had very large-scale fluctuations, $\scalevar\gtrsim \hgas$, then the shear/angular momentum term would be the dominant term resisting collapse and we would obtain $M_{\rm collapse}^{\rm max} \sim \pi\,Q\,\meanrhogas\,\hgas^{3}$, i.e.\ just the standard Hill mass.}
Note that this is not the ``typical'' object or structure -- we do not expect the driving scale for turbulence, for example, to reach $\hgas$. Rather it is where we expect a very sleep cutoff in the largest possible object sizes (in Fig.~\ref{fig:mf.demo}, note that this corresponds to a number of such objects a factor of $\sim 10^{7}$ lower than the ``typical'' objects). If we want to estimate a more typical mass, where, say, the mass function begins to turn over more steeply, we should take a size scale of $\sim \hgrain$ or of the velocity structures with turnover times $\sim \Omega^{-1}$ (which should have sizes $\sim \alpha^{1/2}\,\hgas$). For $\taustop\sim1$, these are the same, and this gives us a characteristic mass 
\begin{align}
M_{\rm collapse}^{\rm intermediate}(\taustop\sim1) & \sim \sqrt{2}\pi\,\alpha\,Q^{1/2}\,\meanrhogas\,\hgas^{3} \\
\nonumber & \sim 3\times10^{-4}\,\left(\frac{\alpha}{10^{-4}}\right)\,\SigmaMMSN^{1/2}\,\rau^{27/28}\,M_{\oplus}
\end{align}

The {\em lower} mass limit is also predicted because on sufficiently small scales, $t_{\rm cross} = \Leddy/v_{\rm drift} \ll \Teddy$ (so grains do not have time to interact with eddies) and/or $\tstop\gg \Teddy$ (so turbulent eddies do not significantly perturb the dust). For $\taustop\lesssim1$, $\tstop\gg \Teddy$ occurs on scales $\scalevar/H \lesssim \alpha^{3/4}\,\taustop^{3/2}\ll \hgrain$ (so $M\approx(4\pi/3)\,\rho_{\rm crit}\,\scalevar^{3}$), where a combination of gas pressure and turbulence form the dominant source of support ($\rho_{\rm crit}\propto\scalevar^{-1}$; see Fig.~\ref{fig:rhocrit}). Plugging in this scale to get values of $\rhocrit$, after some algebra we obtain
\begin{align}
M_{\rm collapse}^{\rm min}(\taustop\sim1) &\sim \frac{2\sqrt{2}\pi}{3}\,\frac{\alpha^{3/2}\,Q^{1/2}}{\taustop^{3}}\,\meanrhogas\,\hgas^{3} \\
&\sim 2\times10^{-7}\,{\Bigl(}\frac{\alpha}{10^{-4}}{\Bigr)}^{3/2}\,\SigmaMMSN^{1/2}\,\rau^{27/28}\,M_{\oplus}
\end{align}

For these simplifying cases, we predict that the ``dynamic range'' of the mass function is 
\begin{align}
\frac{M_{\rm collapse}^{\rm min}}{M_{\rm collapse}^{\rm max}} &\sim \alpha^{5/4}
\end{align}

\vspace{-0.5cm}
\subsubsection{Dependence on Turbulence Strength}

Fig.~\ref{fig:mf.alpha} shows how the MF depends on $\alpha$. As expected from our simple calculation above, the ``maximum'' masses and top-end of the MF depend weakly on $\alpha$, but the ``minimum'' mass and low-mass end depend strongly on $\alpha$. Increasing $\alpha$ truncates the MF at higher minimum masses, because collapse is more difficult both owing to the thicker grain disk (so it is harder to collapse on scales $\lesssim \hgrain$) and increased local turbulent kinetic energy resisting collapse. At  high $\alpha\gtrsim 10^{-2}$, this eliminates entirely collapse at some orbital radii $\lesssim 1\,$au (though for the most part the criteria for collapse at high masses are unchanged). 

\begin{figure*}
    \centering
    \plotsidesize{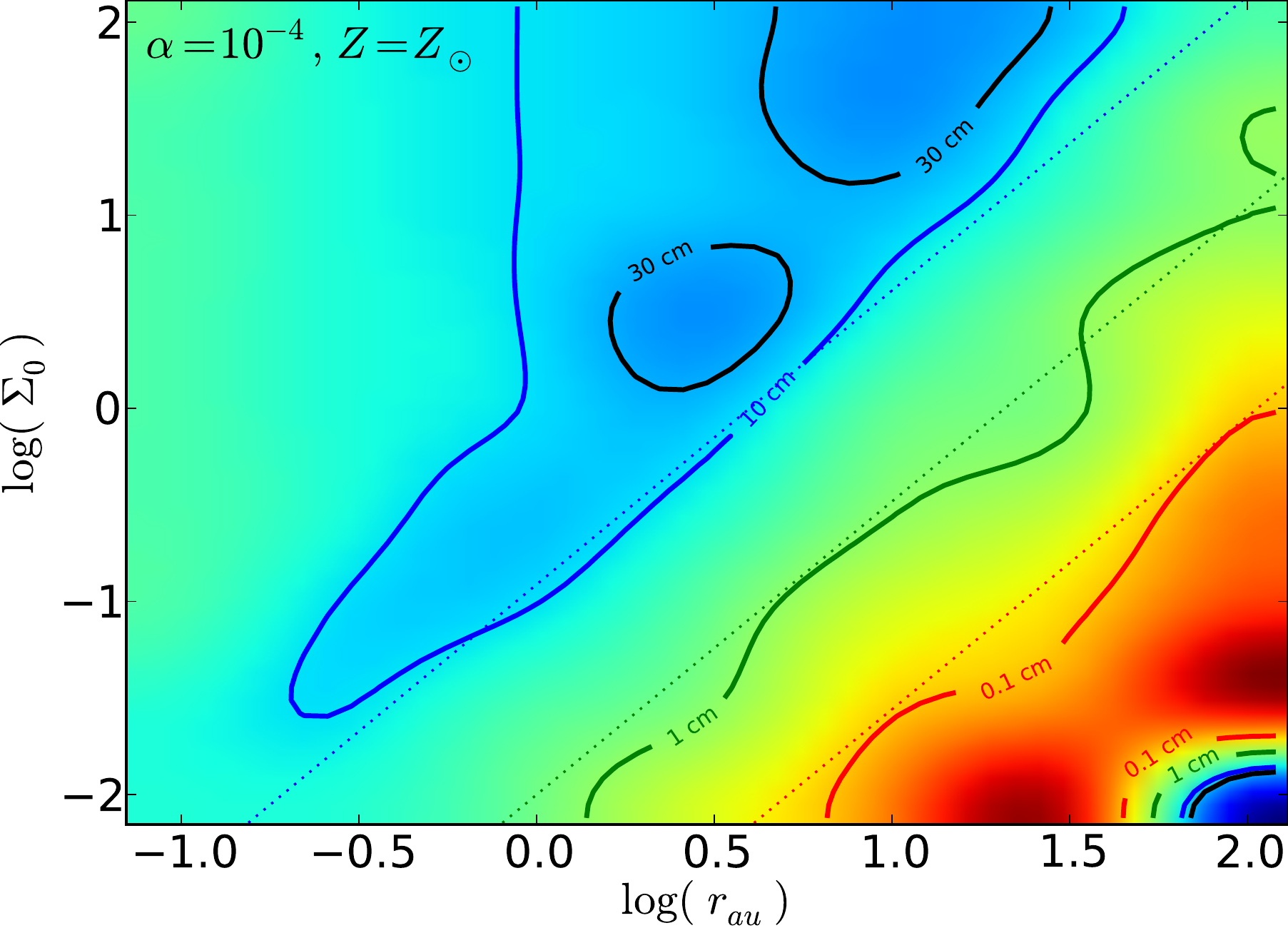}{0.99}
    \vspace{-0.2cm}
    \caption{Minimum grain size needed for pebble-pile formation, as a function of orbital distance from a solar-type star and disk surface density. Distance is in au, and surface density $\SigmaMMSN\equiv \Sigma(r)/(1000\,\rau^{-3/2})$ is the density normalized to the MMSN. In all cases we take $\Zgrain=\Zsun$ and $\alpha=10^{-4}$. Color encodes the minimum grain size above which formation and collapse of pebble pile planetesimals will occur, increasing from red-green-blue (lines show the contours for specific values of $\Rgraincm=0.1,\,1,\,10,\,30$). Dotted lines of the corresponding color show our simple analytic threshold estimate for the same grain size. In the MMSN ($\log{\SigmaMMSN}=0$), small grains with $\gtrsim 1\,{\rm cm}$ ($0.1\,$cm) can form pebble piles at $r\gtrsim 30$\,au ($\gtrsim100\,$au), but large $\sim10-30\,$cm ``boulders'' are required to trigger the process at $\sim1-3$\,au. However, the process is strongly sensitive to surface density, and {\em lower} density disks will, at the same $\Rgraincm$, form pebble piles more easily. At $\Sigma\sim0.1\,\Sigma_{\rm MMSN}$, $\sim1\,$cm grains can trigger pile formation at $\sim 3\,$au.
    \label{fig:min.grain.size}}
\end{figure*}

\begin{figure}
    \centering
    \plotonesize{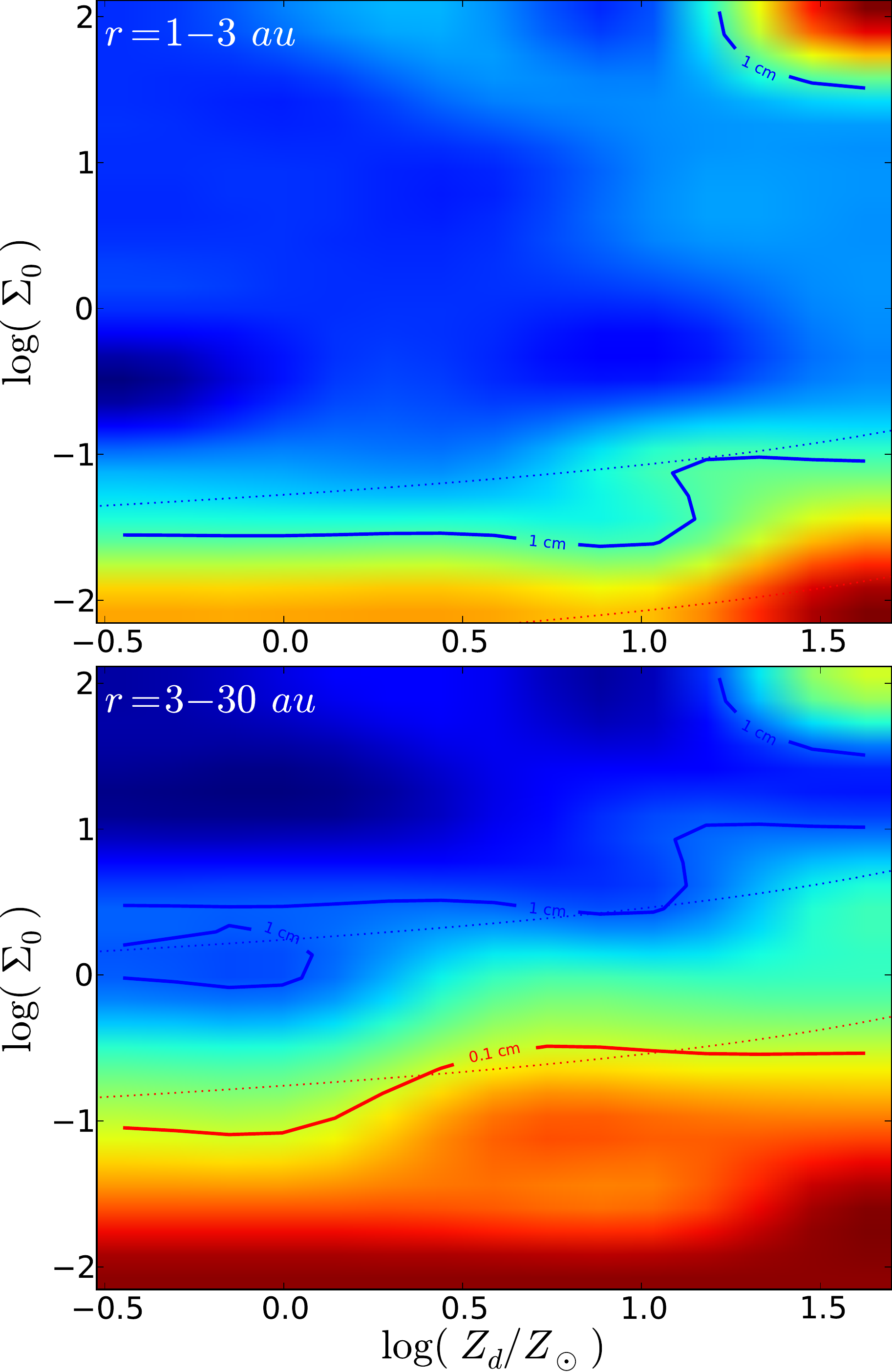}{0.99}
    \vspace{-0.2cm}
    \caption{Minimum grain size needed for pebble pile formation, as a function of disk metallicity and surface density integrated over two ranges of orbital distance. The figure style is as Fig.~\ref{fig:min.grain.size}. Higher-$\Zgrain$ disks require smaller fluctuations to collapse and have more ``seed material,'' so require smaller grains to seed planetesimal formation, but the dependence is weak (logarithmic).  
    \label{fig:min.grain.size.zsurvey}}
\end{figure}

\vspace{-0.5cm}
\subsubsection{The Mass Function Slope}

We also see the MF becomes flatter as $\alpha$ increases. Qualitatively, this follows from the same argument, that higher $\alpha$ suppresses small-scale collapse. Quantitatively, we can understand the slope as follows. The MF is given by Eq.~\ref{eqn:mf.general}; the exact solution must be evaluated numerically for the non-Gaussian statistics and complicated collapse threshold here. However, if the density fluctuations are distributed approximately as a log-normal, and the dependence of the {\em logarithmic} collapse threshold ($\ln{\delta_{\rho}}$) on scale is weak (logarithmic), then we can roughly approximate the MF by the \citet{pressschechter} solution for the mass function (number density $n$ of density fluctuations above a fixed threshold $B$ in a log-normal Gaussian random field\footnote{In \citealt{hopkins:frag.theory}, we derive this for more generic random fields, and include a detailed discussion of the accuracy of the approximation for different collapse thresholds (dependence of $\delta_{\rho}^{\rm crit}$ on scale) and statistics (Gaussian, lognormal, log-Poisson, etc.). For our purposes here, it is adequate -- given the other assumptions in our model -- to approximate the slope of the predicted MF over regions where it is locally power-law like.})
\begin{align}
\frac{{\rm d}n}{{\rm d}\ln{M}} &\sim \frac{\rho_{\rm crit}(M)}{M}\,\frac{B_{0}}{\sqrt{2\pi\,S^{3}}}\,{\Bigl|}\frac{{\rm d}S}{{\rm d}\ln{M}} {\Bigr|}\,\exp{{\Bigl(}-\frac{B^{2}}{2\,S}{\Bigr)}} 
\end{align}
Here $B$ is the ``barrier'' -- a variable which represents the critical density of collapse, which for a lognormal distribution of the dimensionless density fluctuation $\delta_{\rho} = \rhograin / \langle \rhograin \rangle$ is just $B \equiv \ln \delta_{\rho}^{\rm crit} + S/2$ \citep[see][]{hennebelle:2008.imf.presschechter,hopkins:excursion.imf}. And $S$ is the log-normal variance in the dust density distribution $\rhograin$ on a smoothing scale $\scalevar_{\rm first}$ corresponding to $M=M(\scalevar_{\rm first})$ (see \S~\ref{sec:excursion.sets}). Expanding this out and dropping the numerical pre-factors (since we want to isolate just the logarithmic slope), we obtain 
\begin{align}
\label{eqn:mf.eps}
M\,\frac{{\rm d}n}{{\rm d}\ln{M}} &\propto \deltarho\,\frac{|\ln{\deltarho}+S/2|_{\scalevar=\Lmax}}{S^{3/2}}\, \deltarho^{-\frac{1}{2}-\frac{\ln{\deltarho}}{2\,S}}\,\exp{(-S/8)}
\end{align}

Both numerical simulations \citep{johansen:2007.streaming.instab.sims,johansen:2012.grain.clustering.with.particle.collisions} and analytic estimates \citep{hopkins:2013.grain.clustering} show that for large grains ($\taustop\sim1$), the power in logarithmic density fluctuations on large scales ($\Teddy\gtrsim\Omega^{-1}$) is approximately scale-free: ${\rm d}S/{\rm d}\ln{\scalevar}\approx S_{0}=$\,constant, with $S_{0}\approx C_{\infty}\,|\delta_{0}|^{2} = 2\,|(5/2)\,\taustop/(1+\taustop^{2})|^{2}$. This comes simply from the fact that the centrifugal force in large eddies is dominated by the Keplerian/orbital term ($\Omega$), which is scale-independent. This is directly verified in numerical simulations in \citet{johansen:2007.streaming.instab.sims,zhu:2014.dust.power.spectra.mri.turbulent.disks}. And over most of the dynamic range of interest, the critical density depends on scale as $\delta_{\rho}\propto \scalevar_{\rm first}^{-1}$ (see \S~\ref{sec:approx}). 

On small scales (well below the grain disk scale height), we also have $M_{\rm collapse}\propto \rho_{\rm crit}\,\scalevar_{\rm first}^{3}\propto \delta_{\rho}\,\scalevar_{\rm first}^{3}$. Combining these power-law approximations with Eq.~\ref{eqn:mf.eps}, and ignoring the factors that are either constant or slowly (logarithmically) varying in $\scalevar_{\rm first}$ (such as the $S^{3/2}$ term), we obtain 
\begin{align}
M\,\frac{{\rm d}n}{{\rm d}\ln{M}} &= M^{2}\,\frac{{\rm d}n}{{\rm d}M} \propto f(\ln{M})\,M^{q} \\
\label{eqn:mf.pwrlaw} q &\sim \frac{\ln{\deltarho}}{4\,S_{0}} + {\Bigl(}\frac{S_{0}}{16}-\frac{1}{4}{\Bigr)} \\
%&\sim \frac{0.6}{1-0.1\,\ln{(\alpha/10^{-4})}}
&\sim \frac{1.1}{1-0.2\,\ln{(\alpha/10^{-2})}}\ \ \ \ \ (\scalevar \ll \hgrain)
\end{align}
where the latter equality comes from noting that the ($S_{0}/16-1/4$) term is small for all $\taustop\sim1$ of interest, and evaluating $\deltarho$ as in \S~\ref{sec:approx} for $\taustop\sim1/3$ (the approximate threshold where we see the MF rise, though variations $\taustop\sim1/3-3$ have weak effects here). 

On somewhat larger scales, the structures become comparable in size to $\sim \hgrain$; so we must modify this for effectively ``two-dimensional'' structures, $M_{\rm collapse}\propto \rho_{\rm crit}\,\hgrain\,\scalevar_{\rm first}^{2}\propto \delta_{\rho}\,\scalevar_{\rm first}^{2}$, and $\delta_{\rho} \sim $\,constant (see Fig.~\ref{fig:rhocrit}; this is the ``turnover'' or ``trough'' in the Figure). With our other assumptions, this just gives 
\begin{align}
q &\rightarrow 0\ \ \ \ \ (\scalevar \gtrsim \hgrain)
\end{align}
This is the turnover see see in Figs.~\ref{fig:mf.demo}-\ref{fig:mf.disk}, where ${\rm d}N/{\rm d}M\propto M^{-2}$ and the number of planetesimals predicted falls rapidly (while at smaller scales it falls slowly). 

We can understand this as follows. Since the density fluctuations are approximately scale-free over some range, if the ``collapse threshold'' were also scale-free, then the entire system would be scale-free and we would expect self-similar structure, or $q\approx0$ (equal mass over each logarithmic interval in mass). And indeed we do see this on scales close to $\sim \hgrain$. This is a generic consequence of many very different processes, for example supersonic gas turbulence \citep{hopkins:excursion.ism} or cosmologically-seeded dark matter density fluctuations \citep{pressschechter}. However, as we push to smaller scales, the threshold is not scale free; collapse is ``more difficult'' (requires larger $\deltarho$) on small scales, so the MF is biased towards higher-mass objects (larger-scale fluctuations). To leading order, a threshold which grows ``steeply'' below $\sim \hgrain$ leads to $q\sim1$; the logarithmic correction for $\alpha$ reflect the fact that as $\alpha$ is lowered, collapse on small scales becomes ``easier'' (for the reasons discussed above), so the MF is less biased towards higher-masses.

\vspace{-0.5cm}
\subsubsection{Dependence on Metallicity \&\ Disk Densities}

Fig.~\ref{fig:mf.disk} repeats our MF calculation, this time varying the nebula properties (surface density $\SigmaMMSN$ and metallicity $\Zgrain$). At otherwise fixed conditions, increasing the metallicity does not have much effect on the predicted mass function, when planetesimals form (as expected from our simple derivation above). However, it does increase the range of orbital radii over which pebble piles can form at all -- we discuss this further below.

Varying the disk surface density -- with otherwise fixed properties -- has a more dramatic predicted effect on pebble pile formation. Once again though, {\em most} of this effect is in controlling whether piles form at all, not changing the mass function when they do form. Increasing $\SigmaMMSN$ does (weakly) shift the maximum in the MF to higher masses, in line with our expectation for $M_{\rm collapse}^{\rm max}$. 

\begin{figure}
    \centering
    \plotonesize{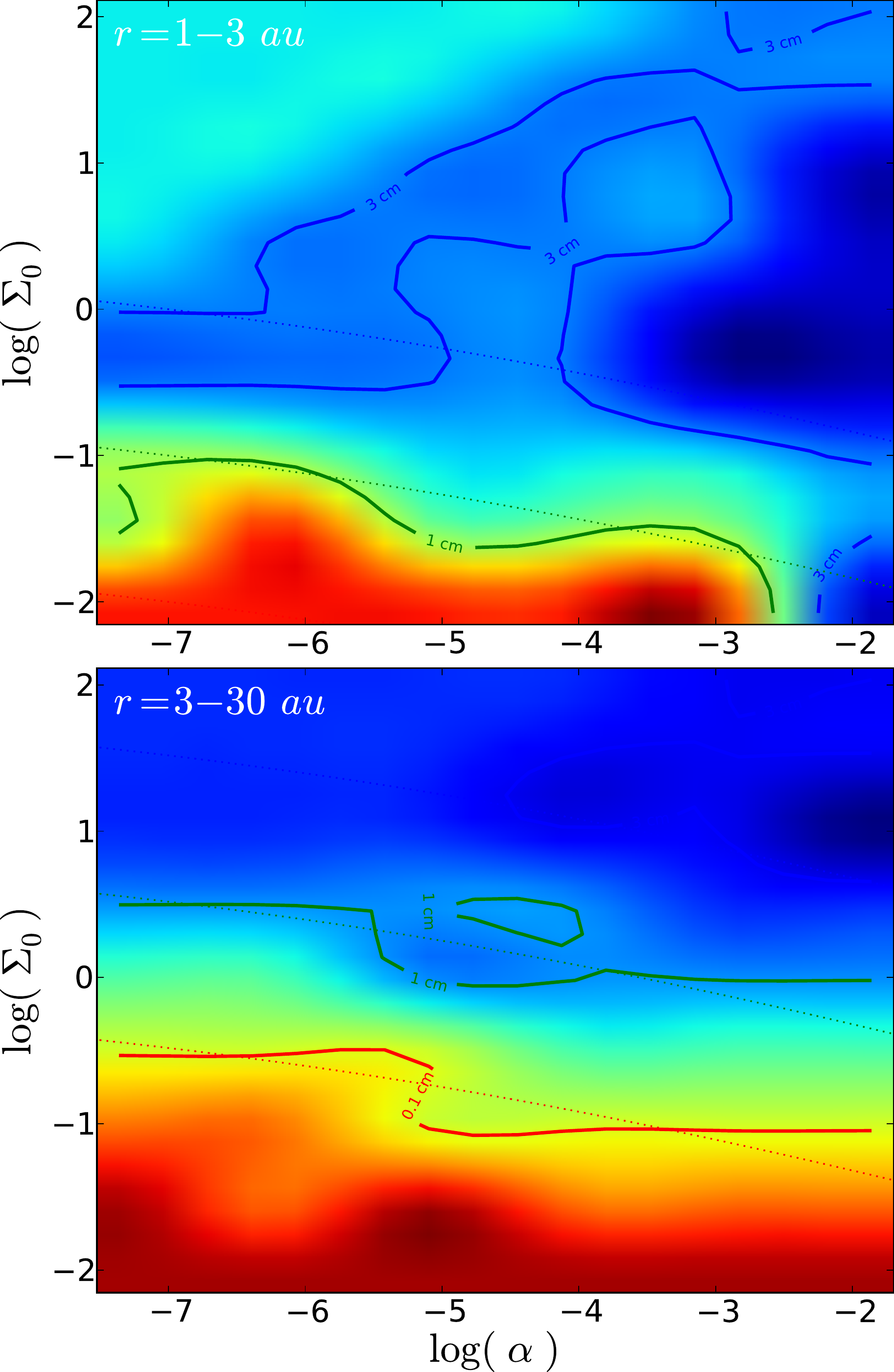}{0.99}
    \vspace{-0.2cm}
    \caption{As Fig.~\ref{fig:min.grain.size.zsurvey}, but here showing the minimum grain size for pebble pile formation as a function of the disk turbulent velocities ($\langle v\gas^{2}(\sim H)\rangle = \alpha\,\cs^{2}$) and surface density. Again, the dependence is weak. Weaker turbulence (lower $\alpha$) leads to denser midplane disks and less turbulent resistance to collapse, so promotes planetesimal formation, but also drives weaker turbulent clustering of grains, so the net effect is both weak and somewhat complicated (non-monotonic).
    \label{fig:min.grain.size.asurvey}}
\end{figure}

\vspace{-0.5cm}
\subsection{General Conditions for Collapse}

Our numerical calculation allows us to map the parameter space in which dynamical grain collapse may occur. As we noted above, the solutions are essentially Boolean: depending on the parameters of a given disk, either pebble pile formation is common, or it is exceptionally rare/impossible. Therefore we treat this as a binary process and ask under which parameters we recover an interesting probability of pebble pile formation.\footnote{Formally we define ``an interesting probability of pebble pile formation'' as a mean predicted $\langle N(>M) \rangle > 1$ in the mass function, integrated down to a mass $M_{\rm min}=10^{-8}\,M_{\oplus}$. But because the scaling of the predicted MF is super-exponential in the important quantities above, modest changes to the exact threshold we choose makes only a tiny difference to our calculation.} At a given radius, for a solar-type star, the key parameters are the grain size $\Rgrain$ and the disk parameters: $\alpha$, $\SigmaMMSN$, and $\Zgrain$.\footnote{Implicitly, $\Zgrain$ is the metallicity of grains with size about $\sim \Rgrain$. For grains with a ``normal'' size distribution, most of the mass $\Zgrain$ is in grains with the largest sizes. But this does not always have to be true.}

Figs.~\ref{fig:min.grain.size}-\ref{fig:min.grain.size.asurvey} map the {\em minimum} grain size $\Rgrain$ needed for the formation of collapsing pebble piles (by our definition), as a function of $\alpha$, $\SigmaMMSN$, and $\Zgrain$, at different radii in a protoplanetary disk. 
Fig.~\ref{fig:min.grain.size} shows the dependence of this grain size on orbital radius and surface density (relative to the MMSN) in a disk around a solar-type star, with our ``default'' $\alpha=10^{-4}$, $\Zgrain=\Zsun$. 

First, we confirm that our approximate estimate of the critical $\taustop \gtrsim 0.05\,{\ln(Q^{1/2}/\Zgrain)}/{\ln{(1+\betafit)}}$ (\S~\ref{sec:approx.largegrain.criterion}) provides a reasonably good approximation to the full numerical calculation. We also can read off that for the MMSN ($\log{(\SigmaMMSN)}=0$), grains with $\Rgrain>(10,\ 1,\ 0.1)\,{\rm cm}$ are required to form pebble piles at $r=(3,\ 30,\ 100)\,{\rm au}$, as we saw from our MF predictions in Figs.~\ref{fig:mf.demo}-\ref{fig:mf.disk}.

\vspace{-0.5cm}
\subsubsection{Dependence on Disk Densities: Lower-Density Disks Promote Collapse}

Given this, we see that at fixed $\rau$, varying the disk surface density -- with otherwise fixed properties -- has a dramatic effect on pebble pile formation. First recall that since $\meanrhogas\,\cs = \meanrhogas\,\hgas\,\Omega$, $\taustop \propto \Rgraincm/\Sigmagas(r)$ depends only on the grain size and disk surface density for any equilibrium disk. Combining that with the simple analytic criterion on $\taustop$ we derived above for large fluctuations, we require a minimum $\Rgraincm \propto \Sigmagas(r)$ for pebble pile formation (or in more detail, $\Rgraincm \gtrsim100\,\psi(Q,\,\Zgrain,\,\alpha)\,(\Sigmagas/1000\,{\rm g\,cm^{-2}})$ where $\psi$ collects the logarithmic corrections; see \S~\ref{sec:discussion}). 

This means that for otherwise fixed grain sizes, {\em lower} surface density disks are more prone to pebble pile formation! Physically, if we keep $\Rgraincm$ fixed and decrease $\Sigma$, $\taustop$ increases. But the maximum amplitude of grain fluctuations then grows super-exponentially in $\taustop$ (for $\taustop\lesssim1$, because the ability of grains to concentrate particles is sensitive to this number, and there is a large ``multiplier'' effect from all turbulent eddies in the cascade; see \citealt{hogan:2007.grain.clustering.cascade.model,bec:2007.grain.clustering.markovian.flow}). The threshold for a density fluctuation to collapse does increase also, but this scales only linearly $\propto Q\propto \Sigma^{-1}$. So the increased clustering ``wins.''

Specifically, if we assume maximum sizes $\Rgraincm\sim1$, then pebble pile formation is only possible at $\gtrsim 30$\,au in a MMSN, but this radius moves in to $\gtrsim 3\,$au in a $\SigmaMMSN=0.1$ disk ($10x$ lower-density), and $\gtrsim1\,$au in a $\SigmaMMSN=0.01$ disk. 

Such low-density disks may be common. \citet{andrews:2013.mdisk.mstar.protoplanetary.compilation} recently compiled a large sample of protoplanetary disks; they found $M_{\rm disk} \propto M_{\ast}$, with a median disk-to-stellar mass ratio of $\approx 0.003$; for the MMSN profile out to $\sim100\,$au, this would give $\SigmaMMSN\sim0.2$; these are consistent with direct measurements of surface density profiles at large radii \citep{isella:2009.protoplanetary.disk.surface.densities}. So at least $\sim50\%$ of disks may be in this regime! If we interpret some of the observational scatter in $M_{\rm disk}/M_{\ast}$ or $\SigmaMMSN$ as an evolutionary effect, then most disks must spend a significant fraction of their lifetime in this lower-density state -- more than sufficient for pebble pile formation to occur in the model here. Indeed, at {\em some} point, disks must evaporate, so all disks pass through such a phase -- and because the collapse is dynamical (occurs on timescale $\sim \Omega^{-1}$), all disks should experience a phase where cm-sized grains have $\taustop\sim1$ even at small radii.

The real question is not whether such grains would cluster -- the simulations modeling clustering can be freely scaled to this parameter space and show large-amplitude fluctuations \citep[see][]{bai:2010.streaming.instability,johansen:2012.grain.clustering.with.particle.collisions,dittrich:2013.grain.clustering.mri.disk.sims,jalali:2013.streaming.instability.largescales}. The question is whether such low-density disks could contain or support cm-sized grains. Some models suggest the maximum grain size scales $\propto M_{\rm disk}$; the maximum size also depends on $\taustop$ for fixed $\alpha$, because the relative velocity of grains increases with $\taustop$ and grains will shatter in collisions at sufficiently high velocities. Detailed calculations suggest that a population of such large grains would be difficult to sustain in a low-density disk (see \S~\ref{sec:compare.expected.sizes}), so the existence of large-grains in a low-density disk would depend on their surviving from an earlier phase (which they can only do for the shorter of either the drift or shattering timescales). Whether they can survive long enough to trigger the instabilities described here is a question outside the scope of this paper, but of major importance for future study.

\vspace{-0.5cm}
\subsubsection{Dependence on Metallicity: Higher-Metallicity Helps, But Only Weakly}

As noted above, the metallicity $\Zgrain$ has a weak effect on the conditions where pebble piles can form. In agreement with the threshold we estimated $\taustop\gtrsim0.05\,\ln{(Q^{1/2}/\Zgrain)}/\ln{(1+\betafit)}$, the minimum $\taustop$ (hence minimum grain size) needed to trigger collapse decreases with increasing metallicity. But this dependence is only logarithmic; so for $\Rgrain\sim10\,{\rm cm}$ the range of pebble-pile forming radii in e.g.\ a $\SigmaMMSN\sim0.1$ disk shrinks from $\sim0.05-6$\,au when $\Zgrain\sim20\,\Zsun$ to $\sim 0.2-3\,$au when $\Zgrain\sim 1\,\Zsun$ and $\sim0.3-3$\,au when $\Zgrain\sim 0.1\,\Zsun$. For a higher-density disk the effects are slightly weaker; for a lower-density disk ($\SigmaMMSN\sim1$), pebble pile formation ceases even for large grains below $\Zgrain\lesssim0.1\,\Zsun$. However, higher metallicities can help promote grain growth to larger sizes, so in this (indirect) sense, may be important.

\vspace{-0.5cm}
\subsubsection{Dependence on Turbulent $\alpha$}

We can also examine the dependence on the turbulent $\alpha$ parameter. Higher-$\alpha$ increases the {\em relative} clustering amplitude of grains \citep[see][]{hogan:1999.turb.concentration.sims,hogan:2007.grain.clustering.cascade.model}, because it implies a larger dynamic range of the turbulent cascade; but the effect is weak because so long as any eddies exist with $\Teddy\sim \Omega^{-1}$, the ``added'' dynamic range is outside the resonant range. Lower-$\alpha$ implies a more-dense grain disk, hence a lower threshold for pebble pile formation; this enters logarithmically in the critical $\taustop$. Together, these effects mean that the net dependence of the minimum grain size on $\alpha$ is quite weak. 

However, we stress that some of this weak dependence stems from the {\em assumption} in our model that the characteristic timescale of large eddies is $\sim \Omega^{-1}$. Depending on the details of the mechanism driving the turbulence, long-lived ``zonal flows'' with coherence time $\gg \Omega^{-1}$ can form \citep[see][]{dittrich:2013.grain.clustering.mri.disk.sims}. As shown in \citet{hopkins:2013.grain.clustering}, these can individually strongly alter the local grain clustering (see Fig.~9 therein).

\vspace{-0.5cm}
\section{Comparison to Other Calculations of Pebble-Pile Planetesimal Formation}
\label{sec:compare.alt.calcs}

At this point, it is instructive to compare the results of our calculation to other calculations (both analytic and numerical) which attempted to estimate the formation rates and mass function of self-gravitating grain-piles. 

\vspace{-0.5cm}
\subsection{Recent Simulations}

Relevant to our models here are some recent direct numerical simulations which include self-gravity. Recent shearing-box simulations of grains undergoing the streaming instability with self-gravity, with $\taustop\approx 0.3$ (about equal to our estimate of the critical value for pebble-pile formation; see \S~\ref{sec:approx.largegrain.criterion}) have been able to confirm that dynamical collapse is possible \citep[see also][]{johansen:2009.particle.clumping.metallicity.dependence} and that the mass function of collapsing pebble-piles has a power-law slope $q\sim1$, in good agreement with our prediction from Eq.~\ref{eqn:mf.pwrlaw} (A.\ Johansen, private communication). Unfortunately the dynamic range in these simulations is not large enough to define the upper/lower limits of the mass function. 

More surprisingly, the predicted mass function shapes here agree well with those from direct numerical simulations in  \citet{lyra:2009.planet.imf.from.sims} (see their Figs.~5 \&\ 10). Those simulations followed the dynamics of particles in an initially laminar disk with a gap carved by the presence of a single Jupiter-mass planet, and found that grains piled up and dynamically collapsed at the Lagrange points. Consistent with our predictions, they find optimum pebble-pile formation for grains with $\taustop\sim 0.4$ (at the location where the grains actually are located); with the mass function rapidly falling off for factor $\sim 3-10$ smaller or larger grains. For the optimal case, they find a cutoff in the mass function of $\sim 0.1\,M_{\oplus}$ at $\rau \sim 3-5$, very similar to our prediction (Eq.~\ref{eqn:maxmass}), with a small ``tail'' reaching earth-to-super-earth masses (as we predict in Figs.~\ref{fig:mf.demo}-\ref{fig:mf.disk}). Finally, the mass function range they follow is entirely dominated by the large-scale (effectively two-dimensional) domain, where we expect a slope $q\sim0$, which also agrees well with the simulations. Given that the physics driving the grains is nominally different (``pressure bumps'' induced by the presence of a pre-existing planet), this agreement is perhaps surprising. However, as \citet{lyra:2009.planet.imf.from.sims} note, the gaps set up vortices at their edges which do much of the trapping; these vortices have characteristic scales in resonance with the dynamical time and grain sizes; so the actual grain-trapping physics is not so different in the end (moreover, \citet{hopkins:2013.grain.clustering} point out that the solution the growth of a grain overdensity in a ``pressure bump'' and an appropriately-aligned vortex are identical up to an order-unity prefactor). And of course the collapse threshold should be set by the same physics we describe here. 

\vspace{-0.5cm}
\subsection{Analytic Models in (Fundamentally) Different Physical Regimes}

\citet{cuzzi:2008.chondrule.planetesimal.model.secular.sandpiles,cuzzi:2010.planetesimal.masses.from.turbulent.concentration.model} consider a model which is similar in spirit to that here -- they represent grain density fluctuations via a hierarchical turbulent cascade model, and calculate the probability of those fluctuations exceeding some threshold which would enable collapse. They reach some of the same qualitative conclusions we do: in their model, grain density fluctuations in simulations are large enough, in principle, to lead to direct collapse, even accounting for gas pressure, turbulence, ram pressure, and angular momentum. They also find that larger values of $\alpha$ suppress the formation of smaller-mass pebble piles, and higher metallicities increase the probability of such events. 

However, there are fundamental differences in what we both model. The \citet{cuzzi:2008.chondrule.planetesimal.model.secular.sandpiles,cuzzi:2010.planetesimal.masses.from.turbulent.concentration.model} model for grain density fluctuations and planetesimal formation applies {\em only} to grains in resonance with the {\em smallest} eddies in a proto-planetary disk. Specifically, they model the dynamics of grains with microscopic Stokes number of unity ($St=1$) -- a stopping time $\tstop$ equal to the eddy turnover time ($\Teddy(\scalevar_{\eta})$) at the Kolmogorov or dissipation scale of turbulence ($\scalevar_{\eta}$). For the MMSN model here, these are eddies with sizes $\scalevar_{\eta}\sim 0.1\,$km, and turnover times less than an hour! In detail, at $\sim(1,\,30,\,100)\,$au, respectively, this corresponds to grain sizes $\Rgrain\sim (300,\,20,\,7)\,\mu{\rm m}\,\SigmaMMSN^{1/2}\,(\alpha/10^{-4})^{-1/2}$ -- i.e.\ micron-through-sub-mm sized grains -- and $\taustop(\Rgrain) = \Teddy(\scalevar=\scalevar_{\eta})\,\Omega \sim (10^{-4},\,10^{-3},\,3\times10^{-3})\,\SigmaMMSN^{-1/2}\,(\alpha/10^{-4})^{-1/2}$. In numerical simulations and laboratory experiments, the clustering of $St=1$ grains behaves {\em qualitatively} differently from ``inertial range'' grains\footnote{For example, small grains with $St=1$ ($\tstop=\Teddy(\scalevar_{\eta})$; resonant with the Kolmogorov scale) have stronger clustering determined by an interplay between molecular viscosity (irrelevant in larger eddies) and drag forces; this makes them the only grain type that can cluster significantly even when $\tstop \neq \Teddy$ \citep[see][]{squires:1991.grain.concentration.experiments,fessler:1994.grain.concentration.experiments,rouson:2001.grain.concentration.experiment,yoshimoto:2007.grain.clustering.selfsimilar.inertial.range,gualtieri:2009.anisotropic.grain.clustering.experiments,monchaux:2010.grain.concentration.experiments.voronoi}. Related to this, the shape of the correlation function/power spectrum of clustering for these grains is qualitatively different from any other (smaller or larger) grains \citep[see e.g.][]{pan:2011.grain.clustering.midstokes.sims,monchaux:2012.grain.concentration.experiment.review}. And $St=1$ grain overdensities are specifically associated with singular (Kolmogorov-scale) intermittent structures in the turbulence (e.g.\ stretched vortex tubes) which can (unlike inertial-range structures) persist for enormously long times relative to their internal turnover time \citep[lifetimes $\gg \Teddy(\scalevar_{\eta})$; see][]{marcu:1995.grain.burgers.vortex,bec:2009.caustics.intermittency.key.to.largegrain.clustering,olla:2010.grain.preferential.concentration.randomfield.notes}.} (those where $\tstop$ corresponds to $\Teddy(\scalevar)\gg \Teddy(\scalevar_{\eta})$ with $\scalevar \gg \scalevar_{\eta}$ within the turbulent inertial range). The clustering model we use from \citet{hopkins:2013.grain.clustering} is specific to the inertial-range (larger) grains.

Clearly, the physical case studied in \citet{cuzzi:2008.chondrule.planetesimal.model.secular.sandpiles,cuzzi:2010.planetesimal.masses.from.turbulent.concentration.model} is radically different from ours. For example, they argue that the $St=1$ grains with $\taustop=\Omega\,\Teddy(\scalevar_{\eta})\sim10^{-4}-10^{-3}$ can form coherent overdensities with $\rhograin/\meanrhograin > 100$ containing enough dust mass ($\sim 10^{22}$\,g) to form $\sim100\,$km planetesimals with marginal probability ($N(>M)\sim1$) at $\sim$\,au in a MMSN. We find this is impossible (in detail, the probability of such events we calculate is lower by a factor $>10^{10}$). The reason is simple: if clustering characteristically occurs on scales where $\Teddy(\scalevar)\sim \tstop$, then for these grains this is $\scalevar\sim0.1\,$km -- smaller than the final size of the planetesimal! Obviously such a ``fluctuation'' ($\rhograin/\meanrhograin > 10^{19}$) is impossible in any model. The key difference is the \citet{cuzzi:2010.planetesimal.masses.from.turbulent.concentration.model} model specifically assumes that for the $St=1$ grains, the power in fluctuations per logarithmic interval in scale is {\em scale-independent}, i.e.\ that $St=1$ grains exhibit comparably large density fluctuations smoothed on the driving scale $\sim H$ as they do on the Kolmogorov scale $\sim \scalevar_{\eta}$; in practice, the relevant overdensities they model have $\rhograin(\scalevar)/\meanrhograin \gtrsim100$ on scales $\scalevar\sim \alpha^{1/2}\,H \sim 10^{5}\,{\rm km}\sim 10^{5-6}\,\scalevar_{\eta}$ (so $\Teddy(\scalevar)\sim 10^{3-4}\,\tstop$). Physically, they argue this is justified for the $St=1$ grains (and {\em only} $St=1$ grains) because their clustering is driven by singular dissipation structures (e.g.\ vortex tubes) with short-axis sizes $\sim \scalevar_{\eta}$, which can persist for timescales $\gtrsim \Omega^{-1}$ and be stretched to uniformly cover all long-axis scales up to $\sim H$; however, this has yet to be tested in numerical simulations at the relevant scales. Regardless, what is clear from both simulations and experiments is that such scale-invariance does {\em not} apply to the inertial-range particles which we model here \citep{squires:1991.grain.concentration.experiments,bec:2007.grain.clustering.markovian.flow,yoshimoto:2007.grain.clustering.selfsimilar.inertial.range,bai:2010.streaming.instability,pan:2011.grain.clustering.midstokes.sims,johansen:2012.grain.clustering.with.particle.collisions,dittrich:2013.grain.clustering.mri.disk.sims}.

It is worth noting, however, that even assuming vastly larger clustering amplitudes of the small $St=1$ grains on large scales, \citet{cuzzi:2010.planetesimal.masses.from.turbulent.concentration.model} reach the same conclusion we do in \S~\ref{sec:smallgrain.criterion}: that dynamical collapse of small grains is essentially impossible. They therefore consider only the secular (sedimentation) mode of collapse -- so these small-grain piles require hundreds of disk dynamical times to collapse. This leads to a different set of criteria -- basically, one must ask whether the ``pile'' could survive so long (whereas the collapsing pebble-piles of interest here all, by definition, are rapidly collapsing on dynamical timescales $\lesssim \Omega^{-1}$). Long survival is very challenging -- one concern (which requires further exploration) is that, over such a long timescale, some turbulent eddy will eventually disrupt the slowly-contracting grain overdensity.

\vspace{-0.5cm}
\section{Discussion}
\label{sec:discussion}

%\subsection{Summary}

We use a recently-developed analytic approximation \citep{hopkins:2013.grain.clustering}, which describes the statistics of grain density fluctuations in a turbulent proto-planetary disk, to estimate the rate and probability of formation of ``pebble-pile'' planetesimals -- self-gravitating collections of (relatively large) grains, which could collapse rapidly (on a dynamical timescale) into $>$km-size planetesimals. 

%The analytic model can provide a good fit to a the grain density statistics in simulations where the turbulence is driven by a variety of different physics, including the streaming instability \citep{youdin.goodman:2005.streaming.instability.derivation}, MRI \citep{dittrich:2013.grain.clustering.mri.disk.sims}, and Kelvin-Helmholtz/Rossby-type instabilities \citep{bai:2010.grain.streaming.vs.diskparams,jalali:2013.streaming.instability.largescales}; the differences between these different cases manifests -- in the leading-order approximation -- as different parameters describing the turbulence itself (e.g.\ different values of $\alpha$, and different driving scales and turnover times of the largest velocity structures). 

\vspace{-0.5cm}
\subsection{Key Conclusions}

\begin{itemize}

\item{{\bf Dynamical Collapse is Possible for Large Grains:}} We argue that the most important parameter determining the collapse of grains is the ratio of stopping to orbital time, $\taustop\equiv \tstop\,\Omega$. Large grain density fluctuations occur on large scales in the disk when $\taustop\sim1$. We derive the criterion for the largest of these fluctuations to overcome tidal/centrifugal/coriolis forces, shear, gas pressure, and turbulent kinetic energy, and undergo rapid (dynamical) gravitational collapse. This can occur when 
\begin{align}
\label{eqn:taustop.crit.discussion}
\taustop \gtrsim 0.05\,\frac{\ln(Q^{1/2}/\Zgrain)}{\ln{(1+\betafit)}}
\end{align}
which we can write as  
\begin{align}
\taustop &\approx 0.004\,{\Bigl(}\frac{\Rgrain}{{\rm cm}}{\Bigr)}\,{\Bigl(}\frac{\Sigma_{\rm gas}(r)}{1000\,{\rm g\,cm^{-2}}}{\Bigr)}^{-1}\gtrsim0.4\,\psi\\
\psi &\equiv \frac{{}1 + 0.08\,\ln{{\Bigl[}(Q(r)/60)\,(\Zgrain/Z_{\odot})^{-2} {\Bigr]}} {}}
{{\ln{{\Bigl[}1+{(}{\alpha}/{10^{-4}}{)}^{1/4}{\Bigr]}}}}\sim1\nonumber
\end{align}
or
\begin{align}
\Rgraincm \gtrsim 100\,\psi\,{\Bigl(}\frac{\Sigma_{\rm gas}(r)}{1000\,{\rm g\,cm^{-2}}}{\Bigr)}
\end{align}

For a MMSN with plausible turbulent $\alpha$ values, this criterion translates to large ``boulders'' with $\Rgrain\gtrsim 10-30\,$cm at $1\,$au; but more plausible ``large grains'' or ``pebbles'' with $\Rgrain \sim1\,$cm at $\sim30\,$au (or $\sim1\,$mm at $\sim100$\,au). For the MMSN regime, which is well-sampled by simulations, this analytically-calculated threshold is in excellent agreement with the results of full numerical simulations \citep[see e.g.][]{bai:2010.streaming.instability,johansen:2012.grain.clustering.with.particle.collisions}. 

\item{{\bf Dynamical Collapse is Not Possible for Small Grains:}} Small grains also cluster strongly -- in fact they can, under the right circumstances, cluster just as strongly as large grains \citep[see][]{squires:1991.grain.concentration.experiments,cuzzi:2001.grain.concentration.chondrules,pan:2011.grain.clustering.midstokes.sims}. However, this clustering occurs on small scales, where $\tstop\sim\Teddy$ (the small-scale eddy turnover time). On these scales, {\em even if we ignore gas drag}, the local turbulent velocity dispersion (induced by the same eddies that generate the density fluctuations) means that the grain free-fall time must be shorter than the stopping time in order for {\em dynamical} collapse to proceed ($G\,\rhograin\gtrsim \tstop^{-2}$). For small grains, this requires an enormous overdensity which is not achieved in any calculations. However, we stress that this conclusion applies only to dynamical grain collapse -- there is a second, secular or slow mode of collapse, which occurs on a timescale $\sim (G\,\rhograin\,\tstop)^{-1} \gg (G\,\rhograin)^{-1/2}$, in which grains slowly sediment to the center of an overdensity. Small grains may be able to form planetesimals via this channel \citep[see e.g.][]{cuzzi:2010.planetesimal.masses.from.turbulent.concentration.model,youdin:2011.grain.secular.instabilities.in.turb.disks,shariff:2011.secular.grain.instability}, though avoiding disruption by turbulent velocity fluctuations is still a significant challenge. 

\item{{\bf Lower-Surface Density Disks are More Prone to Grain-Pile Collapse:} Lower-surface density disks are ``more stable'' in the Toomre sense, and require larger relative overdensities to overcome the Roche and other criteria and collapse. However, if we assume fixed physical grain sizes, then the parameter $\taustop \propto \Rgrain/\Sigma_{\rm gas}$ is inversely proportional to the disk surface density, and the relative magnitude of the maximum grain density fluctuations scales {\em super-exponentially} with $\taustop$ (for $\taustop\lesssim1$). So for reasonable densities $\Sigma_{\rm gas}\sim0.01-1$, the enhanced grain clustering ``wins,'' and the minimum grain size needed for fluctuations decreases with $\Sigma_{\rm gas}$ (although the maximum planetesimal size will also decrease). This has been confirmed in numerical simulations \citep{lyra:2008.dust.traps.low.mass.disks}.

For a disk which begins as a MMSN at $\sim1-3\,$au, if the maximum grain size can reach $\sim1-5$\,cm, then the grains are too-well coupled to collapse ``initially.'' But, as the gaseous disk is eventually dissipated, when more than $\sim90\%$ of its mass has been removed, then the grains will suddenly cross the threshold above (Eq.~\ref{eqn:taustop.crit.discussion}), and the density fluctuations will increase super-exponentially until collapse occurs. 
The key question is whether such large grains could survive (possible) or still be newly-made (unlikely) at this late stage in proto-planetary disk evolution.}

\item{{\bf We Predict a General ``Initial Mass Function'' of Planetesimals:} When this instability occurs, it leads to a mass function of collapsing grain overdensities with a quasi-universal form, which we can approximate as a power-law with a lognormal-like cutoff above/below some maximum/minimum mass:}
\begin{align}
\frac{{\rm d}N}{{\rm d}M} &\propto M^{q-2}\,\exp{{\Bigl(}-\ln^{2}{{\Bigl[}1+\frac{M}{M_{\rm max}}+\frac{M_{\rm min}}{M}{\Bigr]}} {\Bigr)}}\\
q &\approx \frac{1.1}{1-0.2\,\ln{(\alpha/0.01)}} \sim 1 \\
M_{\rm max} &\sim 0.03\,{\Bigl(}\frac{\alpha}{10^{-4}}{\Bigr)}^{1/4}\,\SigmaMMSN^{1/2}\,\rau^{27/28}\,M_{\oplus}\\
M_{\rm min} &\sim \alpha\,M_{\rm max}
\end{align}
Since $q\sim1>0$, this means that most of the {\em mass} in the new collapsing planetesimals is in relatively large objects, with mass $\sim M_{\rm max}$. The mass function then turns over, and on large scales (corresponding to spatial scales of the ``initial'' collapsing regions $\gtrsim \hgrain$) takes on a scale free ($q\approx 0$) mass spectrum. This also appears to agree well with early results from direct numerical simulations \citep{lyra:2009.planet.imf.from.sims,johansen:2012.grain.clustering.with.particle.collisions}. 

\item{{\bf Direct-Collapse to Earth Masses is Possible:} This characteristic maximum mass of pebble piles increases with disk surface density and distance from the star (approximately linearly), in the same qualitative manner as a Jeans mass, although they are not identical. At sufficiently large radii in dense disks -- e.g.\ $\rau\gtrsim30-100\,$au in a MMSN, direct collapse to Earth and super-Earth masses can become possible! Super-earth masses appear to constitute the maximum masses that can be achieved under realistic circumstances. Of course, the mass will continue to evolve as these objects collapse (with some material ejected while other material is accreted). Modeling the non-linear collapse of these systems is key to determine whether they would retain most of their mass, fragment into multiple planetesimals, and/or accrete massive gaseous atmospheres.

As the turbulence becomes weaker ($\alpha$ decreases), the characteristic mass decreases as well, and the mass function becomes more concentrated towards the low-mass end. These lower masses are still more than large enough to provide self-gravitating, $>\,$km-size planetesimal seeds. However, capturing this low mass behavior {\em is} potentially a problem for direct numerical simulations, given both the small mass and size resolution required to capture the relevant scales. }

\item{{\bf Only Modest Metallicities Are Required:} Some mechanisms for generating dust-density fluctuations, such as the streaming instability, require large local dust-to-gas ratios in the disk midplane $\langle \rhograin \rangle \sim \meanrhogas$ to grow and self-excite turbulence \citep{youdin.goodman:2005.streaming.instability.derivation}. Although this can occur for quite modest vertically-integrated metallicities $\Zgrain \sim Z_{\sun}$ \citep{johansen:2009.particle.clumping.metallicity.dependence}, it has often been incorrectly interpreted to mean that large metallicities are required for any large grain density fluctuations (and has led to a large body of work studying how regions with order-of-magnitude ``enhanced'' metallicities may form). But even if the metallicities are too low to trigger the streaming instability, laboratory experiments \citep{squires:1991.grain.concentration.experiments,rouson:2001.grain.concentration.experiment,gualtieri:2009.anisotropic.grain.clustering.experiments,monchaux:2010.grain.concentration.experiments.voronoi,monchaux:2012.grain.concentration.experiment.review}, simulations \citep{hogan:1999.turb.concentration.sims,yoshimoto:2007.grain.clustering.selfsimilar.inertial.range,carballido:2008.grain.streaming.instab.sims,pan:2011.grain.clustering.midstokes.sims,dittrich:2013.grain.clustering.mri.disk.sims}, and analytic calculations \citep{sigurgeirsson:2002.grain.markovian.concentration.toymodel,bec:2008.markovian.grain.clustering.model,zaichik:2009.grain.clustering.theory.randomfield.review,hopkins:2013.grain.clustering} all find that large grain density fluctuations can still occur (even when $\Zgrain=0$, i.e.\ there is zero back-reaction of grains on gas), {\em provided there is some external source of turbulence}. This may come from the MRI, from Kelvin-Helmholtz or shear instabilities, from gravito-turbulent instabilities if the disk is sufficiently massive, or other objects in the disk. 

Of course (all else being equal), dynamical collapse of pebble piles is easier if the ``initial'' dust-to-gas ratio is larger, since smaller density fluctuations are required, and grains are expected to grow more efficiently to large sizes. However, we again emphasize that since the fluctuation amplitudes can be large under the right conditions, the ``threshold'' for sufficiently large fluctuations depends only weakly (logarithmically) on $\Zgrain$, and can occur at solar metallicities. We do find, however, that it becomes very difficult to reach sufficiently large grain sizes and overdensities if $\Zgrain \lesssim 0.1\,Z_{\sun}$.}

\end{itemize}

\vspace{-0.5cm}
\subsection{Comparison to the Maximum ``Expected'' Grain Sizes}
\label{sec:compare.expected.sizes}

We predict that formation of dynamically collapsing pebble-piles may be possible above a critical $\taustop$ given by Eq.~\ref{eqn:taustop.crit.discussion}. Whether or not disks can produce a large abundance of such grains is a separate question, which is generally beyond the scope of this paper. However, we can make some simple comparisons to speculate on whether this is at all plausible. 

\citet{birnstiel:2012.grain.size.distribution.models} and \citet{draczkowska.2014:coagulation.streaming.trigger,drazkowska.2014:dust.growth.protoplanetary.disk} consider a range of models for the growth and evolution of the dust size distribution in proto-planetary disks, and calibrate these against full numerical simulations; they argue that the upper end of the grain size distribution -- the grains which, in all their models, contain most of the grain mass -- is given by the minimum $\taustop^{\rm max}$ of three criteria: 
\begin{align}
\taustop^{\rm max} &= {\rm MIN}
\begin{cases}
      {\displaystyle \frac{v_{\rm shatter}^{2}}{3\,\alpha\,\cs^{2}} \sim 0.81\,\rau^{3/7}\,\left(\frac{v_{\rm shatter}}{10\,{\rm m\,s^{-1}}} \right)^{2}\,\left(\frac{\alpha}{10^{-4}} \right)^{-1} }\\ 
      \\
      {\displaystyle \frac{v_{\rm shatter}}{\eta\,V_{K}} \sim 0.45\,\rau^{13/14}\,\left(\frac{v_{\rm shatter}}{10\,{\rm m\,s^{-1}}} \right) } \\ 
      \\
      {\displaystyle 0.275\,\frac{Z}{\eta} \sim 7.1\,\rau^{-4/7}\,\left( \frac{Z}{Z_{\sun}} \right)} 
\end{cases}
\end{align}
The first criterion represents turbulent shattering: as $\taustop$ and $\alpha$ increase, so does the rms relative grain-grain velocity; when this exceeds the maximum collision velocity above which grains shatter ($v_{\rm shatter}$), large grains cannot be supported; the second represents shattering by relative velocities induced from radial drift (important only if $\alpha$ is very small); the third represents radial drift depleting grains faster than they can grow (where the metallicity $Z$ enters in the grain growth time). 

If we compare this to Eq.~\ref{eqn:taustop.crit.discussion}, we can check whether we expect grains to grow to large enough sizes to trigger this process. Over the plausible parameter space, we can reduce the complicated expressions in Eq.~\ref{eqn:taustop.crit.discussion} and arrive at two simple, approximate criteria: 
\begin{align}
\Zgrain & \gtrsim 0.15\,\rau^{1/2}\,Z_{\sun} \\ 
\alpha & \lesssim 10^{-4}\,\rau^{1/2}\,\left( \frac{v_{\rm shatter}}{10\,{\rm m\,s^{-1}}} \right)^{2.4}
\end{align}
This ensures that grains grow fast enough (the $Z$ criterion) and avoid shattering (the $\alpha$ criterion) up to the critical $\taustop$ we require. The $Z$ criterion is easily satisfied; the $\alpha$ criterion is more demanding, and depends on the uncertain $v_{\rm shatter}$. Estimates for $v_{\rm shatter}$ for ice-coated grains range from $\sim 10-60\,{\rm m\,s^{-1}}$, while for ``bare'' silicates of $\gtrsim $\,mm size it can be much smaller, $\sim 1\,{\rm m\,s^{-1}}$ \citep[see e.g.][]{wada:2009.dust.growth.collision.conditions}. 

For plausible values of $\alpha$, this suggests that coagulation to large grains and subsequent dynamical collapse may be relatively common beyond the ice line. At $\sim 30\,$au, with the optimistic $v_{\rm shatter}\sim 60\,{\rm m\,s^{-1}}$, we require $\alpha\lesssim 0.04$ (i.e.\ Mach numbers $<0.2$ in the disk). This already extends into the range at which the disk would undergo direct gravito-turbulent fragmentation \citep{gammie:2001.cooling.in.keplerian.disks,hopkins:2013.turb.planet.direct.collapse} -- in other words, essentially all $\alpha$ are (one way or another) prone to dynamical collapse. Inside the ice line, on the other hand, the small shattering velocities represent a serious barrier to this mechanism: at $\sim 1\,$au with $v_{\rm shatter} \sim 1\,{\rm m\,s^{-1}}$, this requires $\alpha\lesssim 4\times10^{-7}$; this is such a low $\alpha$ value that shattering by radial drift starts to dominate, in fact, and we find there is actually {\em no} value of $\alpha$ at $\rau \ll 5$ and $v_{\rm shatter} \lesssim 1\,{\rm m\,s^{-1}}$ which can satisfy all of the relevant criteria. 

In Appendix~\ref{sec:nonsolar}, we consider how these scalings, and the critical $\taustop$ for pebble-pile formation are modified for non-solar type stars. Briefly, because lower-mass stars are cooler, it becomes easier at fixed $\alpha$ and $\Zgrain$ to reach the critical $\taustop$. For a $\sim0.1\,M_{\sun}$ M-dwarf, we find that essentially all radii (outside a few times the stellar radius, where the ice line is located) are expected to support $\approx 1\,$cm-sized grains, which are sufficiently large to reach the critical value in Eq.~\ref{eqn:taustop.crit.discussion}, provided $\Zgrain \gtrsim 0.03\,Z_{\sun}\,(\rau/0.03)^{1/2}$ and $\alpha\lesssim 3\times10^{-4}\,(\rau/0.03)^{1/2}\,(v_{\rm shatter} / 10\,{\rm m\,s^{-1}})^{2}$ (we scale these to $\rau\sim 0.03$, the approximate location of the habitable zone for such a star). On the other hand, for a massive $\sim 10\,M_{\sun}$ O/B type star, we do not expect large enough grains inside the ice line at $\sim 50-100\,$au, and even outside this radius, require very large $v_{\rm shatter} \gtrsim 40\,{\rm m\,s^{-1}}$ (as well as $\Zgrain \gtrsim 0.6\,Z_{\sun}$ and $\alpha \lesssim 2\times10^{-3}\,(v_{\rm shatter}/40\,{\rm m\,s^{-1}})^{2}$) to support large enough ($\sim 10-100\,$cm-sized) boulders to satisfy Eq.~\ref{eqn:taustop.crit.discussion}. 

We caution that this ignores non-equilibrium situations such as the late stages of disk dissipation discussed above. However, it does suggest that formation of planets at small radii from a star may require distinct mechanisms. Or perhaps the formation of inner planets is ``induced'' by dust traps and resonances owing to giant planets which form beyond the ice line  by the mechanisms described here \citep[see e.g.][]{lyra:2009.planet.imf.from.sims}.

\vspace{-0.5cm}
\subsection{Future Work \&\ Areas for Improvement}

This is only a first attempt at constructing a simple semi-analytic model for dynamical collapse of pebble-pile planetesimals. As such there are many approximations we have made which can be improved in future work. Some key areas meriting future study include: 

\begin{itemize}

\item{Multiple grain species/sizes:} We have considered only monolithic, collisionless grain populations here. However the grain clustering statistics can be modified by non-linear interactions between grains of different sizes \citep[for simulations, see e.g.][]{bai:2010.streaming.instability}. At the very least, spreading the grains over a wide size distribution decreases the mass available -- thus the effective $\Zgrain$ and maximum densities reached -- in the largest populations.

\item{Collisions between grains:} Grain clustering can enhance grain collisions, and the grain-grain collision rate will enhance dramatically during the collapse process, which can in turn change the grain size distribution and non-linearly alter the collapse. For example, \citep{johansen:2012.grain.clustering.with.particle.collisions} find, in numerical simulations that collisions can enhance clustering under the right conditions. Ideally, one would self-consistently follow the evolution of the grain size distribution, grain clustering, the generation of turbulence, and self-gravity, in a single simulation.

\item{Effects of intermittency and non-turbulent velocity structures:} The models we use here make some simple assumptions about turbulence, like that it follows a Kolmogorov-type cascade. However, there are diverse range of physical mechanisms which can drive the vorticity field (and therefore grain density fluctuations) in protoplanetary disks, including accretion, magnetic disk winds, the magnetorotational, Kelvin-Helmholtz and Rossby wave, and streaming instabilities. The velocity structures, particularly on the largest scales, are not exactly identical between these regimes. To lowest approximation, this will manifest in the parameters we explicitly include in our model (e.g.\ $\alpha$, $\hgrain$, and the driving/largest eddy scales). But there may be higher-order, more complicated effects; some of these are examined in \citet{hopkins:2013.grain.clustering}, but ultimately they must be directly checked in simulations.

\item{The grain-dominated limit and saturation:} Probably the biggest approximation we make is the extrapolation of the models and simulations described here into the regime where dust grains strongly dominate the local density (i.e.\ where the gas dynamics are dominated by the ``back reaction'' from their collisions with dust). This is discussed in \citet{hopkins:2013.grain.clustering}, and our analytic model makes some crude approximations which provide a reasonable phenomenological fit to the simulated dust density distributions in this regime \citep[compare e.g.][]{hogan:2007.grain.clustering.cascade.model,johansen:2007.streaming.instab.sims,johansen:2012.grain.clustering.with.particle.collisions,bai:2010.streaming.instability,zhu:2014.dust.power.spectra.mri.turbulent.disks}. However, the exact structure of turbulence in this regime (hence, how robust these extrapolations are) is highly uncertain, and must be further explored. 

\item{Non-linear collapse:} The model here allows us to identify ``candidate regions'' for the formation of self-gravitating pebble pile planetesimals: regions which accumulate sufficient grain density to be self-gravitating, linearly unstable, and simultaneously exceed the Roche, Jeans, and Toomre criteria. However, we do not attempt to follow the non-linear evolution of these regions. Simulating in detail the collapse of these pebble piles is extremely important: early work by e.g.\ \citet{wahlberg-jansson:2014.pebble.pile.collapse.process} points out that it is not obvious how the grains will stick or shatter as they collapse (which may modify subsequent collapse). A single region may also fragment into a sub-spectrum of masses: what we identify here is an upper limit (the ``parent region'' mass, not necessarily the mass of a single solid object that will form from the above). Many questions need to be explored to fully link this to planet formation.

\end{itemize}

\vspace{-0.7cm}
\acknowledgments 
We thank Eugene Chiang, Jessie Christiansen, Jeff Cuzzi, Karim Shariff, Kees Dullemond, and Joanna Drazkowska for many helpful discussions during the development of this work. We also thank our anonymous referee, for a very helpful report and a number of excellent suggestions. Support for PFH was provided by NASA through Einstein Postdoctoral Fellowship Award Number PF1-120083 issued by the Chandra X-ray Observatory Center, which is operated by the Smithsonian Astrophysical Observatory for and on behalf of the NASA under contract NAS8-03060.\\

\vspace{-0.1cm}
\bibliography{/Users/phopkins/Dropbox/Public/ms}

\begin{appendix}

\section{Grain Velocity Dispersions}
\label{sec:appendix:graindisp}

Various authors have modeled the statistics of grain velocity dispersions in gas turbulence \citep[see][]{voelk:1980.grain.relative.velocity.calc,markiewicz:1991.grain.relative.velocity.calc,ormel:2007.closed.form.grain.rel.velocities,pan:2013.grain.relative.velocity.calc}. We follow these works to derive an approximate expression for the grain-grain velocity dispersion $\langle v\grain^{2}(k_{i}) \rangle \equiv \alpha\,\cs\,\gturb(\scalevar_{i}\equiv k_{i}^{-1}) = \langle v\gas^{2}(k\rightarrow0)\rangle\,\gturb(\scalevar_{i})$. 

First consider the contribution of eddies larger than the $\scalevar_{i}=k_{i}^{-1}$ of interest. \citet{voelk:1980.grain.relative.velocity.calc} argue that the relative velocity dispersion induced on grains with separation $\ll \scalevar_{i}$, by eddies with size scale $\scalevar>\scalevar_{i}$, can be written
\begin{align}
\langle \Delta V^{2}_{\scalevar>\scalevar_{i}} \rangle = \frac{1}{2}\,{\Bigl(}V_{p,\,i}^{2}+V_{p,\,j}^{2} - 2\,\overline{V_{p,\,i}\,V_{p,\,j}} {\Bigl)} = V_{p}^{2} - V_{c}^{2}
\end{align}
where $V_{p}$ is the inertial-space rms velocity to which all particles are accelerated, and the $\overline{V}$ or $V_{c}$ term is the ``cross term'' -- the component of the velocity imparted on the grains which is {\em coherent} across the scale (since well-coupled grains in large eddies may be accelerated to large absolute velocities by those eddies, but the relative velocity between grains on small scales will be small). The second simplication comes from our adopting a mono-population of grains (so $V_{p,\,i}^{2}=V_{p,\,j}^{2}=V_{p}^{2}$). 
From \citet{ormel:2007.closed.form.grain.rel.velocities}, we have
\begin{align}
V_{p}^{2} &\approx \int_{k_{L}}^{k_{i}}\,{\rm d}k\,2\,E(k)\,{\bigl(}1 - K^{2}{\bigr)} \\
V_{c}^{2} &= \int_{k_{L}}^{k_{i}}\,{\rm d}k\,2\,E(k)\,\phi(k,\,k^{\ast})\,{\bigl(}1 - K^{2} {\bigr)}
\end{align}
with $K\equiv 1/(1 + t_{k}/\tstop)$ where $t_{k}\equiv\Teddy(k)$.\footnote{More generally, we can use
\begin{align}
V_{p}^{2} &= \int_{k_{L}}^{{\rm max}(k^{\ast},\,k_{L})}\,{\rm d}k\,2\,E(k)\,{\bigl(}1-K^{2} {\bigr)} \\
&+ \int_{{\rm max}(k^{\ast},\,k_{L})}^{k_{i}}\,{\rm d}k\,2\,E(k)\,(1-K)\,[g(\chi)+K\,h(\chi)] \nonumber 
\end{align}
where $g(\chi)=\chi^{-1}\,\tan^{-1}(\chi)$ and $h(\chi)=1/(1+\chi^{2})$ with $\chi=K\,t_{k}\,k\,V_{\rm rel}(k)$, $V_{\rm rel}(k)^{2}=\int_{k_{L}}^{k_{i}}\,{\rm d}k\,2\,E(k)\,K^{2}$ from \citet{voelk:1980.grain.relative.velocity.calc}, which must be solved numerically. However as shown in \citet{ormel:2007.closed.form.grain.rel.velocities}, the approximate expression above (which assumes $h(\chi)\approx g(\chi)\approx1$) introduces a negligible error for all particle sizes of interest.}
The $K^{2}$ term in the first integral comes from the ``n=1'' gas velocity autocorrelation function used in \citet{markiewicz:1991.grain.relative.velocity.calc} and \citet{ormel:2007.closed.form.grain.rel.velocities}. 

Here $k^{\ast}$ is the boundary between ``Class I'' eddies (where particles are trapped) and ``Class II'' eddies (where eddies decay before providing more than small perturbations to the particle); formally $k^{\ast}$ is defined by $\tstop^{-1} = t_{k^{\ast}}^{-1} + k^{\ast}\,V_{\rm rel}(k^{\ast})$ \citep{voelk:1980.grain.relative.velocity.calc}. The function $\phi$ is any function which interpolates between $1$ for eddies with $k<k^{\ast}$ and $0$ for eddies with $k>k^{\ast}$. \citet{voelk:1980.grain.relative.velocity.calc} approximate this with a step function at $k=k^{\ast}$; for numerical convenience and slightly improved accuracy, we adopt the simple linear interpolation $\phi = t_{k}/(t_{k} + t_{k}^{\ast})$. We have checked, though, that the difference between this choice and a step function is negligible in our calculations in the text. 

Combining these approximations we have 
\begin{align}
\langle \Delta V^{2}_{\scalevar>\scalevar_{i}} \rangle \approx \int_{k_{L}}^{k_{i}} {\rm d}k\,2\,E(k)\,{\bigl(}1 - K^{2} {\bigr)}{\Bigl(}\frac{1}{1 + t_{k}/t_{k}^{\ast}}{\Bigr)}
\end{align}

At finite scale $\scalevar_{i}>0$, we also need to consider the contribution to grain motion from eddies with smaller sizes. As in the derivation of the above relations, we assume that eddy structure on successive scales is uncorrelated. Thus, the contribution from eddies with $\scalevar<\scalevar_{i}$ is just 
\begin{align}
\langle \Delta V^{2}_{\scalevar<\scalevar_{i}}\rangle &= V_{p}^{2}(\scalevar<\scalevar_{i}) \equiv \int_{k_{i}}^{k_\eta}\,{\rm d}k\,2\,E(k)\,{\bigl(}1-K^{2} {\bigr)}
\end{align}
i.e.\ eddies with internal scale $\scalevar<\scalevar_{i}$ do not contribute to the coherent component $V_{c}$ on scales $\scalevar \ge \scalevar_{i}$. For the cases we study here, we can also take the Kolmogorov scale $k_\eta\rightarrow\infty$ with negligible error.

Thus we obtain 
\begin{align}
\nonumber \Delta V^{2}(k_{i}) &= \int_{k_{L}}^{k_{i}}{\rm d}k\,2\,E(k)\,{\bigl(}1-K^{2} {\bigr)}\,{\Bigl(}\frac{1}{1+t_{k}/t_{k^{\ast}}} {\Bigr)} \\
&+ \int_{k_{i}}^{\infty}{\rm d}k\,2\,E(k)\,{\bigl(} 1-K^{2} {\bigr)}
\end{align}

Determining $k^{\ast}$ is, in general, non-trivial, but \citet{ormel:2007.closed.form.grain.rel.velocities} note that $t_{k^{\ast}}$ can be well-approximated by $t_{k^{\ast}}\approx{\rm MIN}(\phi^{\ast},\,\tstop/t_{k_{L}})$ (with $\phi^{\ast}=(1+\sqrt{5})/2$) or $t_{k^{\ast}}^{-1} \sim (8\,\tstop/5)^{-1} + t_{k_{L}}^{-1}$.

The upper limit $k_{L}$ here represents the driving scale. In \citet{ormel:2007.closed.form.grain.rel.velocities}, this is taken as a fixed value, with $E(k)$ a pure power-law ($\propto k^{-5/3}$) for $k>k_{L}$, so $t_{k}=t_{k_{L}}\,(k/k_{L})^{-2/3}$. In this case, using the definition $\int_{k_{L}}^{\infty}\,{\rm d}k\,E(k) = (1/2)\,\langle v\gas^{2}(k\rightarrow0)\rangle = \alpha\,\cs^{2}/2$, we obtain
\begin{align}
\frac{\Delta V^{2}(\scalevar)}{\alpha\,\cs^{2}} &= \nonumber
\frac{1}{1+y_{k}}\,{\Bigl(} \frac{(y_{k}-y_{L})}{(1+y_{L})\,(\frac{3}{8}\,y_{L}-1)} + \frac{y_{k}^{2}}{y_{L}} {\Bigr)} \\ 
&-\frac{(\frac{5}{8}\,y_{L}+1)}{(\frac{3}{8}\,y_{L}-1)^{2}}\ln{{\Bigl[}\frac{1+y_{k}}{1+y_{L}} {\Bigr]}} \nonumber \\ 
&-\frac{2\,y_{L}\,(\frac{1}{8}\,y_{L}+1)}{(\frac{5}{8}\,y_{L}+1)\,(\frac{3}{8}\,y_{L}-1)^{2}}\ln{{\Bigl[}
\frac{y_{L}\,(\frac{5}{8}\,y_{L}+2)}{y_{L}+y_{k}\,(\frac{5}{8}\,y_{L}+1)}
{\Bigr]}} 
\end{align}
where $y_{L}\equiv t_{k_{L}}/\tstop$ and $y_{k}\equiv t_{k}/\tstop = y_{L}\,(\scalevar/\Lmax)^{2/3}$. This is a tedious expression, but its relevant scalings are clear if we approximate $\phi(k,\,k^{\ast})$ as a step function and $t_{k^{\ast}}\sim{\rm MIN}(\phi^{\ast}\,\tstop/t_{k_{L}})$; then
\begin{align}
\frac{\Delta V^{2}(\scalevar)}{\alpha\,\cs^{2}} &= \gturb(\scalevar/\Lmax) =\frac{(y^{\ast})^{2}}{1+y^{\ast}}\,y_{L}^{-1}\\
y^{\ast} &\equiv {\rm MAX}{\Bigl(} y_{k},\,{\rm MIN}{\Bigl[}\phi^{\ast},\,y_{L}{\Bigr]} {\Bigr)}
\end{align}
For $\tstop\ll t_{L}$ ($y_{L}\gg1$), and $\tstop \ll t_{K}$, this scales as $y_{k}/y_{L} = (\scalevar/\Lmax)^{2/3}$, i.e.\ $\langle v\grain^{2}(\scalevar) \rangle=\langle v\gas^{2}(\scalevar) \rangle$ -- the grain and gas velocities are well-coupled. But on sufficiently small scales where $\tstop\gg t_{k}$, this goes to the constant ($\scalevar$-independent) value $=y_{L}^{-1}=\tstop/t_{L}$ (the turbulent dispersion imparted by eddies with $\Teddy\sim\tstop$).

As noted in the text we can more accurately include the driving-range ($\scalevar>\Lmax$) using a full expression for $E(k)$, and taking $k_{L}\rightarrow \infty$. In this case $\Delta V^{2}(\scalevar)$ can only be evaluated numerically. However, motivated by the form for the turnover of $E(k)$ at $k>k_{L}$, we can approximate the full numerical solution at all $\scalevar$ by simply inserting
\begin{align}
y_{L} &= \frac{\Teddy(\scalevar=\Lmax)}{\tstop} \\
y_{k} &\rightarrow y_{k}^{\rm eff} = y_{L}\,\frac{(\scalevar/\Lmax)^{2/3}}{{\bigl[}1 + (\scalevar/\Lmax)^{7/3} {\bigr]}^{2/7}}
\end{align}
into the expressions above (derived for a sharp cutoff at $\Lmax$). This approximation is accurate to $\sim10\%$, well within the range of uncertainties in our earlier approximations.

\vspace{-0.5cm}
\section{Stability Conditions for a Partially-Coupled Grain-Gas Fluid}
\label{sec:stability.dustgas.fluid}

Here we briefly describe an alternative derivation of a gravitational collapse criterion for grains in a thin disk. Consider a mixture of gas and dust; as in the main text, we introduce the {\em ad hoc} but convenient parameter $\beta$ to describe their coupling. 

The case $\beta=0$ refers to the limit where the dust feels drag, but the gas does not {\em respond} to the dust (for example, the gas does not get compressed by dust motions, leading to higher gas pressure). The equation of motion in this limit should, therefore, correspond to the Euler equation for a collisionless particle fluid (with approximately aligned rotation-dominated particle orbits and negligible dispersion, since we are assuming a thin disk), which can be found in e.g.\ \citet{toomre:Q,julian:disk.stability,binneytremaine}, with addition of the standard drag acceleration $({\bf v}_{\rm dust}-{\bf v}_{\rm gas})/\tstop$ \citep[as appears in various forms in e.g.][]{marble:1970.dust.gas.fluid.dynamics,ward.1976.solar.system.formation,sekiya:1983.secular.grav.instability.pebbles.in.disk,tanga.2004:grav.instability.dust.clustering}.

On the other hand, $\beta=1$ refers to a perfectly-coupled dust-gas mixture, such that the two move together. In this case the explicit drag force must vanish, and since the dust moves with the gas its equation of motion must be that of a razor-thin, single-fluid disk with pressure $p_{\rm gas}$ in cylindrical coordinates \citep[for derivation and detailed discussion of the single-fluid equation, see][]{lau:spiral.wave.dispersion.relations,binneytremaine}. An expression of this form is used to derive in analogous instability criterion in \citet{safronov:1960}, and it is for example the vertically-integrated version of Eqs.~2.12-2.17 in \citet{sekiya:1983.secular.grav.instability.pebbles.in.disk}.

In cylindrical coordinates (radial distance $R$ from the star and azimuthal angle $\phi$), we can write a single set of equations which represents both limits depending on whether we choose $\beta=0$ and $\beta=1$. This has the form: 
%In cylindrical coordinates (radial distance $R$ from the star and azimuthal angle $\phi$), the continuity and Euler equations for such a mixture take the form \citep{toomre:Q,julian:disk.stability,lau:spiral.wave.dispersion.relations,sekiya:1983.secular.grav.instability.pebbles.in.disk}:\footnote{The equations in Eq.~\ref{eqn:Euler} correspond (as they should) for $\beta=1$ and $\beta=0$, respectively, to the well-known Euler equations for a razor-thin disk of either a collisional (gas) fluid or collisionless particle fluid \citep[for derivation and detailed discussion, see][]{binneytremaine}. The $\beta=1$ case is just the Euler equation for a razor-thin single-fluid disk with pressure $p_{\rm gas}$ in cylindrical coordinates \citep[used to derive an analogous instability criterion in e.g.][]{safronov:1960}; it is for example the vertically-integrated version of Eqs.~2.12-2.17 in \citet{sekiya:1983.secular.grav.instability.pebbles.in.disk}. The $\beta=0$ case is just the Euler equation for a collisionless particle fluid with approximately aligned rotation-dominated particle orbits and negligible dispersion \citep[commonly used for e.g.\ stellar galactic disks following][]{toomre:Q,julian:disk.stability}, with a leading-order drag force added, which is straightforward to derive from the Jeans equations \citep{marble:1970.dust.gas.fluid.dynamics,ward.1976.solar.system.formation,sekiya:1983.secular.grav.instability.pebbles.in.disk,tanga.2004:grav.instability.dust.clustering}.}
\begin{align}
\label{eqn:Euler}
\frac{\partial\Sigma}{\partial t} &+ \frac{1}{R}\,\frac{\partial}{\partial R}(R\,\Sigma\,v_{R}) + \frac{1}{R}\,\frac{\partial}{\partial \phi}(\Sigma\,v_{\phi}) = 0\\
\nonumber \frac{\partial v_{R}}{\partial t} & + v_{R}\,\frac{\partial v_{R}}{\partial R} + \frac{v_{\phi}}{R}\,\frac{\partial v_{R}}{\partial \phi} -\frac{v_{\phi}^{2}}{R} = \\
\nonumber & \ \ \ \ \ \ \ \ \ \ \ \ \ \ \ \ \ \ \   -\frac{\partial \Phi}{\partial R} - (1-\beta)\,\frac{(v_{R}-v_{R,\,{\rm gas}})}{\tstop} - \frac{\beta}{\Sigma}\,\frac{\partial p_{\rm gas}}{\partial R}
\\ 
\nonumber \frac{\partial v_{\phi}}{\partial t} & + v_{R}\,\frac{\partial v_{\phi}}{\partial R} + \frac{v_{\phi}}{R}\,\frac{\partial v_{\phi}}{\partial \phi}+\frac{v_{\phi}\,v_{R}}{R} = \\
& \nonumber \ \ \ \ \ \ \ \ \ \ \ \ \ \ \ \ \ \ \  -\frac{1}{R}\,\frac{\partial \Phi}{\partial \phi} - (1-\beta)\,\frac{(v_{\phi}-v_{\phi,\,{\rm gas}})}{\tstop}
- \frac{\beta}{\Sigma\,R}\,\frac{\partial p_{\rm gas}}{\partial \phi}
\end{align}
Here $\Sigma$ is the surface density of the mixture, $\Phi$ the total gravitational potential, and $v_{R}=\dot{R}$, $v_{\phi}=\dot{\phi}$; $p_{\rm gas}$ denotes the pressure of the gaseous component coupled to the dust, and we simplify by taking $\beta$ to be a constant. So, by our definition of $\cs$ and $\rhoratio$, we have $\delta p_{\rm gas} = \cs^{2}\,\delta \Sigma_{\rm gas}^{\rm coupled} = \cs^{2}\,\rhoratio^{-1}\,\delta \Sigma$. 

We can explicitly check that these equations reduce to the limits above for $\beta\rightarrow0$ (gas does not respond to dust; we recover the collisionless/pressure-free Euler equations in with an additional drag term from the gas) and $\beta\rightarrow1$ (gas moves perfectly with dust; the drag vanishes and we recover the collisional/pressurized Euler equations for a single fluid). However, intermediate physical cases do {\em not} necessarily correspond to intermediate $\beta$ (in those cases, one should distinguish the gas and dust velocities; and properly treat the dust velocity distribution function; hence the solutions rapidly become much more complex). In the main text we introduce an interpolation using intermediate values of $\beta$ primarily as a heuristic, ad-hoc matching function between the two (valid) limits.\footnote{Since this paper was submitted, \citet{takahashi.2014:two.component.grav.instability.in.protoplanetary.disks} considered a more detailed derivation of the two-fluid instability criterion in a thin disk, parameterizing the turbulence via a simple effective diffusivity. Their resulting instability criteria (Eqs.~14-16 therein) reduce exactly to ours in \S~\ref{sec:stability.dustgas.fluid} for both the $\beta=0$ ($\taustop\gg1$) and $\beta=1$ ($\taustop\rightarrow0$) limits. For more general cases, they consider only $\rho_{d}/\rho_{g}\ll1$, which leads to the secular limit. However if we start from their Eqs.~6-12, and take $\rho_{d}\gg \rho_{g}$, we can re-derive both $\beta$ limits (and show that the prefactor of the $\Omega^{2}$ term in Eq.~\ref{eqn:dyn.instab.newderiv} is exactly $1$, as used in the text), and if we further expand in $\psi \equiv t_{\rm free-fall}/\tstop \equiv [\tstop\,(2\pi\,G\,\Sigma\,|k|)^{1/2}]^{-1}$, we obtain the approximate scaling of $\beta$ with density used in the text ($\beta \sim \psi$ for $\psi\ll1$, $\sim 1$ for $\psi\gg1$).}

With that caveat in mind, now we assume a perturbation of the form $\Sigma_{1}\propto v_{R,\,1}\propto v_{\phi,\,1}\propto \exp{(\imath\,[m\,\phi + k\,R - \omega\,t ])}$ (to the background equilibrium solution $\Sigma_{0}$, $v_{R,\,0}$, etc.) and linearize the above equations. We also invoke the WKB (local) approximation for the perturbation potential $\Phi_{1}\approx -2\,\pi\,G\,|k|^{-1}\,\Sigma_{1}$; however, for now we retain all terms in the ``unperturbed'' background flow (i.e.\ retain all terms to $\mathcal{O}(|kR|^{-1})$). After some lengthy algebra we obtain the dispersion relation:
\begin{align}
 (\tilde{\omega}+1)\,{\Bigl[} & \tilde{\omega}^{2} + \tilde{\tstop}\,\frac{v_{R,\,0}}{R}\,(1+\eta_{v_{R}})\,\tilde{\omega} \\
\nonumber & + {\tilde{\tstop}^{2}}\,{\Bigl(}\frac{v_{R,\,0}^{2}}{{R^{2}}}\,\eta_{v_{R}} + 2\,\frac{v_{\phi,\,0}^{2}}{{R^{2}}}\,(1+\eta_{v_{\phi}}){\Bigr)} {\Bigr]} = \\
\nonumber -G_{0}
{\Bigl[}&
{m}^{2}\,(\tilde{\omega} + \tilde{\tstop}\,\frac{v_{R,\,0}}{R}) +  {m}\,\tilde{\tstop}\,\frac{v_{\phi,\,0}}{R}\,k\,R\,{\bigl(} 1+\eta_{v_{\phi}} - 2\,{\tilde{k}}{\bigr)} + \\
\nonumber & \tilde{k}\,(k\,R)^{2}\,(\tilde{\omega} +  \tilde{\tstop}\,\frac{v_{R,\,0}}{R}\,\eta_{v_{R}})
{\Bigr]} \\
G_{0} & \equiv \frac{\tilde{\tstop}^{2}}{R^{2}}\,\left[ \frac{2\pi\,G\,\Sigma_{0}}{|k|} - \frac{\beta}{\rhoratio}\,\cs^{2}\right] 
\equiv \frac{\rho_{R}}{\tilde{k}}\,\left(\frac{\tilde{\taustop}}{k\,R}\right)^{2} \\
%\nonumber \tilde{m} & \equiv \frac{m}{k\,R}\\
\nonumber \tilde{k} &\equiv 1 - \imath\,\frac{(1+\eta_{\Sigma})}{k\,R}\\
\nonumber \tilde{\omega} &\equiv \imath\,\tilde{\tstop}\left[\omega - {\Bigl(} m\,\frac{v_{\phi,\,0}}{R} + k\,v_{R,\,0} {\Bigr)} + \imath\,\frac{v_{R,\,0}}{R}\,(1+\eta_{v_{R}})\right]-1
\end{align}
where $\tilde{\tstop} \equiv \tstop / (1-\beta)$, $\tilde{\taustop}\equiv \taustop/(1-\beta)$.

This forms a cubic equation for $\omega$, with three solution branches. Since the interesting parameter space is $\tstop\sim \Omega^{-1}$, and the drift velocity is $\sim \eta\,V_{K}$ with $\eta\ll 1$, we can insert the values for $v_{R,\,0}$ and $v_{\phi,\,0}$ (and the corresponding $\eta_{v_{R}}$, $\eta_{v_{\phi}}$) from the solutions in \citet{nakagawa:1986.grain.drift.solution}, then drop higher-order terms in the drift ($\eta$), and restrict to purely radial modes ($m=0$), to simplify this substantially with negligible effect on the character of the solution. This gives 
\begin{align}
\varpi^{3}\,\tilde{\taustop}^{2} + 2\,\imath\,\varpi^{2}\,\tilde{\taustop} - \varpi\,(1-\tilde{\taustop}^{2}\,(\rho_{R}-1)) + \imath\,\rho_{R}\,\tilde{\taustop} = 0
\end{align}
where $\varpi\equiv {\omega}/\Omega$ and $\rho_{R}\equiv  (2\pi\,G\,\Sigma_{0}\,|k| - \beta\,\cs^{2}\,k^{2}/\rhoratio)\,\tilde{k}\,\Omega^{-2}$. 

First note that if $\rho_{R}\le 0$, then {\em all} solutions for $\varpi$ have imaginary part ${\rm Im}(\varpi)\le 0$, i.e.\ are decaying or stable -- there can be no instability. However, if $\rho_{R}\ge 0$, there is {\em always} a growing mode. If $0<\rho_{R}\ll1$, this mode has $\varpi=\imath\,\rho_{R}\,\tilde{\taustop}$, so grows on a timescale $|\omega|^{-1} = 1/(\rho_{R}\,\tilde{\taustop}\,\Omega)\gg \Omega^{-1}$. This is the ``secular'' sedimentation instability, which grows slowly. This may, in fact, be the mechanism by which planetesimals form \citep[see e.g.][]{cuzzi:2010.planetesimal.masses.from.turbulent.concentration.model,youdin:2011.grain.secular.instabilities.in.turb.disks,shariff:2011.secular.grain.instability}, but it requires a different set of models and collapse criteria, and is outside the scope of this paper (but will be the subject of a future study). 

On the other hand, when $\rho_{R}\gtrsim 1$, then we obtain ${\rm Im}(\varpi) = \rho_{R}^{1/2} - 1/(2\,\tilde{\taustop})$. So growth on the dynamical timescale requires $\rho_{R}>[(1-\beta)/(2\,\taustop)]^{2}$, i.e.\ 
\begin{align}
\label{eqn:dyn.instab.newderiv}
0 >  \Omega^{2}\,{\rm MAX}\left[1,\ \left(\frac{1-\beta}{2\,\taustop}\right)^{2}\right] + \frac{\beta}{\rhoratio}\,\cs^{2}\,k^{2} - 2\pi\,G\,\Sigma_{0}\,|k|
\end{align}
Note that for $\taustop\rightarrow0$, $\beta\rightarrow1$, and for $\beta\ll1$ we expect $\taustop\gtrsim1$, so for the $\taustop\sim1$ of interest, can reasonably take ${\rm MAX}[1,\,(2\,\taustop)^{-2}]\sim1$ and arrive at the dispersion relation used in the text, up to the turbulent terms (with the caveat that additional corrections appear for $\taustop\ll 1$).

It is trivial to see that this satisfies the traditional Toomre, Roche, and Jeans criteria simultaneously. Shear (even the fully non-linear terms) forces are overcome when $\rho>\rho_{\rm Roche}$, and gas pressure and angular momentum are explicitly included. A velocity dispersion term can be added using the approximate methods in \citet{chandrasekhar:1951.turb.jeans.condition,vandervoort:1970.dispersion.relation,bonazzola:1987.turb.jeans.instab}; but to leading order in any of these approaches this is identical to the addition of the $v_{t}(k)$ term in the same manner as $\cs$, as in the text. We consider a more detailed calculation in Appendix~\ref{sec:turb.fluct.fx.on.collapse.criterion}. Still another derivation which arrives at the same criterion for collapse in a coupled dust-gas disk, by treating the turbulence as a diffusion term in the equations of motion, is given in \citet{chavanis:2000.vortex.trapping.disk.instability}. 

As noted in \citet{cuzzi:2008.chondrule.planetesimal.model.secular.sandpiles}, a non-linear term which can suppress collapse is ram pressure from the ``headwind'' encountered by a grain group as it moves through the disk. The relevant criterion for whether the pebble-pile can resist instability in the ram pressure shredding the distribution is the Weber number, the ratio of surface gravity (effectively, ``surface tension'' of the collapsing cloud) $G\,\Sigma^{2}$ to the ram pressure force per unit area $\rhogas\,v_{\rm drift}^{2}$, where $v_{\rm drift} = f(\taustop)\,\eta\,v_{K} \sim (\taustop/(1+\taustop^{2}))\,(c_{s}/v_{K})^{2}\,v_{K}$ is calculated by \citet{nakagawa:1986.grain.drift.solution}. At a radius $r_\ast$ in the disk, with Keplerian velocity $v_{K}$, this is satisfied for all $\taustop$ if $\Sigma \gtrsim (c_{s}/v_{K})^{2}\,f(\taustop)\,Q^{-1/2}\,\Omega^{2}\,r_{\ast}\,G^{-1}$. But it is straightforward to verify that this is automatically satisfied if Eq.~\ref{eqn:dyn.instab.newderiv} is already satisfied. Consider: 
\begin{align}
G\,\Sigma^{2} & > \rhogas\,v_{\rm drift}^{2} = \meanrhogas\,f^{2}(\taustop)\,\eta^{2}\,V_{K}^{2} = \frac{\Sigma_{\rm gas}}{2\,\hgas}\,f^{2}(\taustop)\,\eta^{2}\,V_{K}^{2} 
\end{align}
so that 
\begin{align}
\frac{\Sigma}{\Sigma_{\rm gas}} &> \left( \frac{\pi\,Q}{2} \right)^{1/2}\,f(\taustop)\,\eta\,\frac{V_{K}}{\cs} = \left( \frac{\pi\,Q}{2} \right)^{1/2}\,f(\taustop)\,\Pi
\end{align}
but using the values from \S~\ref{sec:model}, this becomes $\Sigma/\Sigma_{\rm gas} \gtrsim 0.3$, which is always true for an unstable overdensity. More generally, using $\Pi\sim \cs/V_{K}$, with some manipulation we can turn this into 
\begin{align}
2\,\pi\,G\,\Sigma\,|k| \gtrsim \cs^{2}\,k^{2}\,\left(\frac{\scalevar}{R}\,f(\taustop)\,Q^{-1/2} \right) 
\end{align}
Since $Q\gg 1$, $\scalevar \ll R$, $f(\taustop)<1$, this is easily satisfied if $\rho_{R}>0$.

Note that the case of a spherical, non-rotating, constant-density cloud of dust and gas collapsing, as might be appropriate for e.g.\ modes with sufficiently large $|k|\gg \hgrain^{-1}$ near the disk midplane, is described in detail in \citet{shariff:2014.collapse.of.tightly.coupled.dust.gas.mixture}. This essentially amounts to a two-fluid Jeans analysis, for which the key criterion is the effective Jeans number, as discussed in \S~\ref{sec:collapse.crit}. But this is exactly the criterion we obtain if we replace $\Sigma\,|k|$ with $2\,\rho$ in $\rho_{R}$ or the dispersion relation Eq.~\ref{eqn:dyn.instab.newderiv}, as we assumed in \S~\ref{sec:collapse.crit} motivated by the limiting expressions for an exponential vertical profile (where the relevant terms scaled as $2\,\rho\,|k\,\hgrain|/(1+|k\,\hgrain|)$, so $\rightarrow 2\,\rho$ for $|k|\gg \hgrain^{-1}$ and $\rightarrow 2\,\rho\,|k\,\hgrain| = \Sigma\,|k|$ for $|k|\ll \hgrain^{-1}$).

%we replace $\Sigma$ with $\rho$ above and write the $R$, $\phi$ terms instead in terms of the general gradient operator. This becomes a two-fluid Jeans analysis, for which we can simplify the Euler equations to 
%\begin{align}
%\frac{\partial^{2}\rho_{1}}{\partial t^{2}} &= \frac{\beta}{\rhoratio}\,\cs^{2}\,\nabla^{2}\rho_{1} - 4\pi\,G\,\rho\,\rho_{1} + \frac{1}{\tstop}\,\frac{\partial \rho_{1}}{\partial t}\\
%\omega^{2} &= -\imath\,\omega +  \left(\frac{\beta}{\rhoratio}\,\cs^{2}\,k^{2} - 4\pi\,G\,\meanrhogas\,\rhoratio\right)
%\end{align}
%For , it is no longer valid to treat the disk as infinitely thin. Instead, we can consider the opposite limit, of a perturbation in an infinite, homogenous medium (appropriate for e.g.\ the disk midplane) in motion with constant velocity. 

\vspace{-0.5cm}
\section{Accounting for Turbulent Velocity Fluctuations During Collapse}
\label{sec:turb.fluct.fx.on.collapse.criterion}

We now present a derivation of the role of turbulent velocity fluctuations in dynamical collapse, which is simplified but accounts for the fully non-linear turbulent fluctuations (not just their rms value) during collapse.

First assume a grain overdensity exceeds the criterion in \S~\ref{sec:stability.dustgas.fluid} above (Eq.~\ref{eqn:dyn.instab.newderiv}), on some scale, so it can collapse dynamically despite shear and gas pressure effects. If the effects of turbulence were negligible, the collapse timescale $t_{f} = t_{\rm collapse}$ would just be the mode growth timescale $t_{f} = 1/{\rm Im}(\omega)\sim (\rho_{R}^{1/2}\,\Omega)^{-1}$, for the regime of interest. But to survive long enough for this collapse/growth to occur, it must avoid encountering a turbulent gas structure or eddy which induces a shear velocity $>v_{\rm max} \sim v_{\rm collapse} = k^{-1} / t_{\rm collapse}$. This is always less than the ``escape velocity'' ($\sqrt{2\,G\,M\grain(<\scalevar)/\scalevar}$) since that is defined by free-fall from infinite distance; but it is still sufficient to ``reset'' collapse (it will perturb the collapsing region significantly ``away from'' the collapsing state). Define time and velocity in units of the rms eddy turnover time and velocity dispersion on this scale: $\tau\equiv t/\langle\Teddy(k)\rangle$ and $x\equiv v/\langle v_{\rm turb}^{2}(k) \rangle^{1/2}$. Let $\tau_{f}=t_{f}/\langle \Teddy(k) \rangle $ and $B\equiv v_{\rm max}/\langle v_{\rm turb}^{2}(k) \rangle^{1/2}$. Moreover, recall that $\langle \Teddy(k) \rangle\equiv \scalevar/\langle v_{\rm turb}^{2}(k) \rangle^{1/2}$ (where $\scalevar\equiv k^{-1}$), and, for $v_{\rm max}=\scalevar/t_{\rm collapse}$, $\tau_{f}=1/B$. 

In fully-developed turbulence, to lowest order, the distribution of one-dimensional velocities ($v_{x}$, $v_{y}$, $v_{z}$) on a given scale is Gaussian\footnote{In the presence of intermittency, this is not exactly true; however, the effects on the second-order correlation function (which is what matters here) are weak. We can, for example, repeat our derivation using the distribution function predicted by \citet{sheleveque:structure.functions}, and find it gives only a $\sim5\%$-level correction to our calculation.} 
\begin{align}
\label{eqn:P0.x}P_{0}(x\,|\,S_{0}) = \frac{{\rm d}P(<x\,|\,S_{0})}{{\rm d}x} = \frac{1}{\sqrt{2\pi\,S_{0}}}\,\exp{{\Bigl(}-\frac{x^{2}}{2\,S_{0}} {\Bigr)}}
\end{align}
with variance $S_{0,\,v} = \langle v_{\rm turb}^{2}(k) \rangle/3$, or in the units above, $S_{0}\equiv S_{0,\,x} = 1/3$. 

The correlation timescale for $x$ is $\approx \Teddy(k)$ -- this is measured in experiments and simulations \citep{yakhot:2008.lagrangian.structfn,pan:2010.turbulent.mixing.times,konstandin:2012.lagrangian.structfn.turb}, and often is, in fact, how $\Teddy(k)$ is defined. So to lowest order, we can think of the turbulent field as ``refreshed'' or ``resampled'' on a timescale $\Delta t\sim \Teddy(k)$ (or $\Delta \tau \sim 1$). For $\tau_{f}\gg1$, this means we ``draw'' from the distribution in Eq.~\ref{eqn:P0.x} $N\approx \tau_{f}/\Delta \tau=\tau_{f}$ times over the collapse timescale. We require, for each draw, that $|x|<B$, which has probability $P(|x|<B) = {\rm erf}(B\,\sqrt{3/2})$. The probability that all draws are ``successful'' (i.e.\ that the collapsing mode survives) is $P(|x|<B)_{\tau<\tau_{f}} \sim {\rm erf}(B\,\sqrt{3/2})^{\tau_{f}}={\rm erf}(B\,\sqrt{3/2})^{1/B}$. Finally, we note that this was just for one velocity component; we must consider each of three components independently. This gives the probability of survival 
\begin{align}
P(|x|<B)_{\tau<\tau_{f}} \sim {\rm erf}(B\,\sqrt{3/2})^{3\,\tau_{f}}={\rm erf}(B\,\sqrt{3/2})^{3/B}
\end{align}

This is an {\em extremely} steep function of $B$ for $B<1$, approximately $\approx \exp{[3\,B^{-1}\,(\ln{B}+(1/2)\,\ln{(6/\pi)})]}$, and $P\ll1$ for small $B$. So we do not expect ``successful'' collapse to be common for small $B$. Since turbulence is an inherently stochastic process, we cannot deterministically say whether a given region will or will not encounter a large turbulent eddy which would break it up during its collapse. Lacking that, we want our ``collapse criterion'' to identify regions where there is a large (order-unity) probability of ``successful'' collapse (i.e.\ not encountering a too-large turbulent shear/vorticity). We therefore require $P>0.5$ (i.e.\ the probability of survival is larger than that of disruption), which requires $B_{\rm min}>0.8$; this choice of $P$ is is arbitrary but because it is a steep function of $B$, changing the ``threshold'' has weak effects on $B$ (at $B_{\rm min}=0.4$, $P\sim 10^{-3}$, at $B_{\rm min}=1$, $P\sim 0.8$). 

Now recall $v_{\rm max}\approx v_{\rm collapse} = k^{-1}/t_{\rm collapse} = k^{-1}\,{\rm Im}(\omega) \approx k^{-1}\,\Omega\,\rho_{R}^{1/2}$; so this requirement becomes $(\rho_{R}\,\Omega)^{2} > B_{\rm min}^{2}\, \langle v_{\rm turb}^{2}(k) \rangle\,k^{2}$, or 
\begin{align}
0> \Omega^{2} + \frac{\beta}{\rhoratio}\,\cs^{2}\,k^{2} + B_{\rm min}^{2}\,\langle v_{\rm turb}^{2}(k)\rangle\,k^{2}- 2\pi\,G\,\Sigma_{0}\,|k|
\end{align}
Since $B_{\rm min}\sim1$ is somewhat uncertain, we simply adopt $B_{\rm min}=1$ in the text (corresponding to the {\em linear} derivation for a gas fluid in \citealt{chandrasekhar:1951.turb.jeans.condition}); however, the difference between this and $B_{\rm min}=0.8$ is negligible for all of our results. We simply note that the choice ($B_{\rm min}=1$) in the text also applies to non-linear, fluctuating turbulent velocity fields during collapse, and corresponds to a probability $P>0.8$ that the region will ``successfully'' collapse in the limit where turbulence is the dominant source of support (compared to rotation and shear). 

Still another approach to calculating the effects of turbulence on the collapse is given in \citet{hogan:2007.grain.clustering.cascade.model,cuzzi:2010.planetesimal.masses.from.turbulent.concentration.model}, who explicitly model a bivariate probability distribution of particle concentration ($\rho$ or $\Sigma$) and enstrophy density ($|\nabla\times{\bf v}|^{2} \sim v_{\rm turb}^{2}(k)\,k^{2}$). This has the advantage of accounting directly for the variation in $v_{\rm turb}^{2}$ from one location to another, where in some regions there can be less (or more) support versus collapse; however it requires a numerical model for the bivariate cascade.

\vspace{-0.5cm}
\section{Scalings for Non-Solar Type Stars}
\label{sec:nonsolar}

Here we briefly note how the scalings used in this paper are modified for stars which differ significantly from solar-type.

First, we repeat our calculation of basic disk properties in \S~\ref{sec:disk.model}. Our definition of $\Sigmagas(r)$ was already generalized for any surface density, so we only need to correct for the properties of the star. 
\begin{align}
\Omega &= \sqrt{\frac{G\,M_{\ast}}{\Rstar^{3}}} \approx 6.3\,\rau^{-3/2}\,{\rm yr^{-1}}\,\Minsolar^{1/2} \\ 
\Sigmagas &= \SigmaMMSN\,1000\,\rau^{-3/2}\,{\rm g\,cm^{-2}} \\ 
T_{\rm eff,\,\ast} &= {\Bigl(}\frac{(0.05\,\rau^{2/7})\,R_{\ast}^{2}}{4\,\Rstar^{2}} {\Bigr)}^{1/4}\,T_{\ast} \approx 140\,\rau^{-3/7}\,\Linsolar^{1/4}\,{\rm K} 
\end{align}
where we have used the fact that $L_{\ast} \propto R_{\ast}^{2}\,T_{\ast}^{4}$ (by definition), and defined the mass and luminosity of the star relative to solar: 
\begin{align}
\Minsolar & \equiv \frac{M_{\ast}}{M_{\sun}}\ \ \ , \ \ \ \ \ \Linsolar \equiv \frac{L_{\ast}}{L_{\sun}}
\end{align}

Inserting these into the same definitions we used in the text (following \citealt{chiang:2010.planetesimal.formation.review}) we obtain
\begin{align}
\cs &= \sqrt{\frac{k_{B}\,T_{\rm mid}}{\mu\,m_{p}}} \approx 0.64\,\rau^{-3/14}\,\Linsolar^{1/8}\,{\rm km\,s^{-1}}\\
\frac{\hgas}{\Rstar} &= \frac{\cs}{V_{K}} \approx 0.022\,\rau^{2/7}\,\Minsolar^{-1/2}\,\Linsolar^{1/8}\\
\meanrhogas &= \frac{\Sigmagas}{2\,\hgas} \approx 1.5\times10^{-9}\,\SigmaMMSN\,\rau^{-39/14}\,\Minsolar^{1/2}\,\Linsolar^{-1/8}\,{\rm g\,cm^{-3}}\\ 
Q &= \frac{\cs\,\Omega}{\pi\,G\,\Sigmagas} \approx 61\,\SigmaMMSN^{-1}\,\rau^{-3/14}\,\Minsolar^{1/2}\,\Linsolar^{1/8} \\ 
\Pi &= \frac{1}{2\,\meanrhogas\,V_{K}\,\cs}\frac{\partial (\meanrhogas\,\cs^{2})}{\partial \ln{r}} \approx 0.035\,\rau^{2/7}\,\Minsolar^{-1/2}\,\Linsolar^{1/8} \\
\scalevar_\sigma &= \frac{1}{n\gas\,\sigma(H_{2})} \approx \frac{1.2\,\SigmaMMSN^{-1}\,\rau^{39/14}\,\Minsolar^{-1/2}\,\Linsolar^{1/8}}{1+\Linsolar^{-1/4}\,(\rau/3.2)^{3/7}}\,{\rm cm}\\
\taustop &\approx {\rm MAX}
\begin{cases}
      {\displaystyle 0.004\,\SigmaMMSN^{-1}\,\rau^{3/2}\,\Rgraincm} \\ 
      \\
      {\displaystyle 0.0014\,\Rgraincm^{2}\,\frac{\rau^{-9/7}\,\Minsolar^{1/2}\,\Linsolar^{-1/8}}{1+\Linsolar^{-1/4}\,(\rau/3.2)^{3/7}}}
\end{cases}
\end{align}

In \S~\ref{sec:approx.largegrain.criterion}, we derived an approximate collapse criterion for large grains -- a critical $\taustop$ above which fluctuations becomes sufficiently large such that pebble-pile formation is likely in our model. Note that our derivation was cast in terms of quantities like $\taustop$, $Q$ and $\cs$, so we do not need to change it in those terms -- only the relation of those terms to quantities like the location in the disk and absolute size of the grains will be altered. So we retain the ``threshold'' criteria that the ``effective'' Jeans number $J < 1$ (Eq.~\ref{eqn:jeans.number}):  
\begin{align}
\rhoratio_{\rm crit}(\scalevar) &\sim \frac{\cs}{\scalevar\,\sqrt{G\,\meanrhogas}} \sim Q^{1/2}\, \frac{\cs}{\scalevar\,\Omega} \\ 
\end{align}
and consequent (Eq.~\ref{eqn:taustop.crit.discussion})
\begin{align}
\taustop \gtrsim \taustop^{\rm crit} \approx 0.05\,\frac{\ln(Q^{1/2}/\Zgrain)}{\ln{(1+10\,\alpha^{1/4})}} \approx 0.4\,\psi(Q,\,\Zgrain,\,\alpha)
\end{align}

In \S~\ref{sec:compare.expected.sizes}, we noted the maximum grain sizes, in terms of $\taustop$, expected to contain significant mass at a given location in a protoplanetary disk (as estimated by \citealt{birnstiel:2012.grain.size.distribution.models}). We can use the revised values above to correct these: the grains containing most of the grain mass have $\taustop$ given by the minimum $\taustop^{\rm max}$ of three criteria: 
\begin{align}
\taustop^{\rm max} &= {\rm MIN}
\begin{cases}
      {\displaystyle \frac{v_{\rm shatter}^{2}}{3\,\alpha\,\cs^{2}} \sim 0.81\,\rau^{3/7}\,\left(\frac{v_{\rm shatter}}{10\,{\rm m\,s^{-1}}} \right)^{2}\,\left(\frac{\alpha}{10^{-4}} \right)^{-1}\,\Linsolar^{-1/4} }\\ 
      \\
      {\displaystyle \frac{v_{\rm shatter}}{\eta\,V_{K}} \sim 0.45\,\rau^{13/14}\,\left(\frac{v_{\rm shatter}}{10\,{\rm m\,s^{-1}}} \right)\,\Minsolar^{1/2}\,\Linsolar^{-1/4} } \\ 
      \\
      {\displaystyle 0.275\,\frac{Z}{\eta} \sim 7.1\,\rau^{-4/7}\,\left( \frac{Z}{Z_{\sun}} \right)\,\Minsolar\,\Linsolar^{-1/4}} 
\end{cases}
\end{align}

If we now insert numbers into these scalings, it is straightforward to see that we predict ``easier'' pebble-pile formation around low-mass stars. Consider an M-dwarf with $M_{\ast}=0.1\,M_{\sun}$ and typical luminosity $L_{\ast}\approx 5\times10^{-4}\,L_{\sun}$ ($T_{\ast}\approx 2900\,$K, $R_{\ast}\approx 0.1\,R_{\sun}$). The disk is cool, and the ice line lies at just a few times the stellar radius ($\sim 4\,R_{\ast} \sim 0.002\,$au). Thus we expect $v_{\rm shatter}\gtrsim 10\,{\rm m\,s^{-1}}$ throughout. Plugging in these values suggests that the critical $\taustop^{\rm crit}$ is modest ($\sim 0.3-0.6$) and ``expected'' ($\taustop^{\rm crit} < \taustop^{\rm max}$) at almost all radii. For example at $\sim 0.03\,$au (the approximate location of the habitable zone after the disk is evaporated), we have $\taustop^{\rm crit}\approx 0.5$, with $\taustop^{\rm max} = {\rm MIN}(1.2\times10^{-4}\,v_{s,\,10}^{2}/\alpha,\,1.2\,v_{s,\,10},\,35\,\Zgrain/Z_{\sun})$ (where $v_{s,\,10} \equiv v_{\rm shatter} / 10\,{\rm m\,s^{-1}}$).  More generally, we obtain a metallicity and $\alpha$ criterion as in \S~\ref{sec:compare.expected.sizes}, with $\Zgrain \gtrsim 0.15\,\rau^{1/2}\,Z_{\sun}$ as before (except now the relevant radii are smaller, so at $\sim 0.03\,$au we require only $\Zgrain \gtrsim 0.03\,Z_{\sun}$), and $\alpha \lesssim 1.5\times10^{-3}\,\rau^{1/2}\,v_{s,\,10}^{2}$ ($\alpha\lesssim 3\times10^{-4}\,v_{s,\,10}^{2}$ at $\sim 0.03\,$au). The implied critical grains/pebble size at this radius is modest, $\Rgraincm\approx 1.5$ (i.e.\ cm-size). 

On the other hand, now consider a massive zero-age main sequence O/B star with $M_{\ast} = 12\,M_{\sun}$ and $L_{\ast} \approx 8800\,L_{\sun}$ ($T_{\ast}\approx 28000\,$K, $R_{\ast}\approx 4.3\,R_{\sun}$). The disk is hot, so the ice line lies at $\sim 50-100\,$au. Inside this radius, assuming $v_{\rm shatter}\sim 1\,{\rm m\,s^{-1}}$, we expect the maximum grain sizes to be well below those required to reach $\taustop^{\rm crit}$. Outside this radius, it is still quiet challenging: at e.g.\ $50\,$au we expect $\taustop^{\rm crit}\approx 0.5$, with $\taustop^{\rm max} = {\rm MIN}(0.45\times10^{-4}\,v_{s,\,10}^{2}/\alpha,\,0.12\,v_{s,\,10},\,0.9\,\Zgrain/Z_{\sun})$ -- so unless $v_{\rm shatter}\gtrsim 40-50\,{\rm m\,s^{-1}}$ (the highest values estimated), this is a serious challenge. Even if this is satisfied, we {\em also} require $\Zgrain \gtrsim 0.6\,Z_{\sun}$, and $\alpha\lesssim 2\times10^{-3}\,(v_{\rm shatter} / 40\,{\rm m\,s^{-1}})^{2}$. The critical pebble sizes would be $\Rgraincm \gtrsim 10\,(\SigmaMMSN/30)$ -- approaching the ``boulder'' range.

Qualitatively, if we assume $\Linsolar \approx \Minsolar^{4}$ over the main-sequence stellar mass range, we see that the various values for $\taustop^{\rm max}$ scale inversely or not at all with stellar mass (the turbulent and drift-based shattering criteria scale as $\Minsolar^{-1}$ and $\Minsolar^{-1/2}$, respectively, while the radial drift/dust growth criterion is independent of $\Minsolar$), while $\taustop^{\rm crit}$ is only weakly (logarithmically) sensitive to the stellar mass. We can similarly explore the effects of stellar metallicity, but this is even weaker: if we assume, following observations, that $L_{\ast} \propto Z_{\ast}^{-1/3}$ at fixed mass \citep[see e.g.][]{kotoneva:2002.luminosity.metallicity.relation.stars}, then we obtain small corrections to all the above. Assuming the stellar $Z_{\ast}$ is the same as the gas disk, we find the correction to the minimum metallicities required are small over the plausible range $\sim 0.1 \lesssim Z_{\ast}/Z_{\sun} \lesssim 10$.

\end{appendix}

\end{document}